\documentclass[11pt,times, 3p, reqno]{amsart}
\usepackage[T1]{fontenc}
\usepackage{a4wide}

\usepackage{eucal}
\usepackage{exscale}

\usepackage{hyperref}
\usepackage{enumerate,fullpage}
\usepackage{amssymb,amsmath,graphicx,amsthm}
\usepackage{color}

\usepackage{amsfonts,amsmath,amsthm,mathrsfs}
\usepackage{amssymb}
\usepackage{color}
\usepackage{enumerate}
\usepackage{subfloat,subfigure}
\usepackage{graphics,verbatim,url}
\usepackage{graphicx}
\usepackage{epsfig}
\usepackage{epstopdf}
\include{graphics}

\DeclareMathOperator{\sech}{sech}

\usepackage{index}

\newcommand{\be}{\begin{eqnarray*}}
    \newcommand{\en}{\end{eqnarray*}}
\newcommand{\bes}{\begin{eqnarray}}
    \newcommand{\ens}{\end{eqnarray}}

\usepackage{epic, curves}

\def\bq{\begin{equation}}
    \def\eq{\end{equation}}
\def\bqq{\begin{eqnarray*}}
    \def\eqq{\end{eqnarray*}}

    \title[ Frequency shifting for solitons based on transformations on the Fourier domain and applications ]{ Frequency shifting for solitons based on transformations on the Fourier domain and applications}

        \author[Q.M. Nguyen]{ Quan M. Nguyen }
        \address[Q.M. Nguyen]{Department of Mathematics, International University, Vietnam National University-HCMC, Ho Chi Minh City, Vietnam }
        \email{quannm@hcmiu.edu.vn }

    \author[T.T. Huynh]{Toan T. Huynh}
    \address[T.T. Huynh]{Department of Mathematics, University of Science, Vietnam National University-HCMC, Ho Chi Minh City, Vietnam}
    \address{Department of Mathematics, University of Medicine and Pharmacy at Ho Chi Minh City, Ho Chi Minh City,
Vietnam} 
    
    \email{huynhthanhtoan@ump.edu.vn}

\begin{document}
        
   \begin{abstract}

             We develop the theoretical procedures for shifting the frequency of a single soliton and of a sequence of solitons of the nonlinear Schr\"odinger equation. The procedures are based on simple transformations of the soliton pattern in the Fourier domain and on the shape-preserving property of solitons.  
These theoretical frequency shifting procedures are verified by numerical simulations 
with the nonlinear Schr\"odinger equation using the split-step Fourier method. In order to demonstrate the use of the frequency shifting procedures, two important applications are presented:
(1) stabilization of the propagation of solitons in waveguides with frequency dependent linear gain-loss; 
(2) induction of repeated soliton collisions in waveguides with weak cubic loss. 
The results of numerical simulations with the nonlinear Schr\"odinger model are in very good agreement with the theoretical predictions.
            
   \end{abstract}
       
        \maketitle
    
%        \tableofcontents
    

%\maketitle

\section{Introduction}
\label{intro}

Solitons are stable shape preserving traveling-wave solutions of a class of nonlinear partial differential equations such as the nonlinear Schr\"odinger (NLS) equation, the Ginzburg-Landau equation, the sine-Gordon equation, and the Korteweg-de Vries equation
\cite{Ablowitz11, Tao06}. In nonlinear dispersive media, the perfect balance between nonlinearity and dispersion forms solitons. Solitons appear in a wide range of fields, including hydrodynamics
\cite{Zakharov84},
condensed matter physics
\cite{Malomed89}, optics \cite{Agrawal2001, Mollenauer2006}, and plasma physics \cite{Horton96}.
One of the most fundamental properties of solitons is their shape-preserving property in a soliton collision, that is, a soliton collision is elastic \cite{Agrawal2001}. In 1973, Hasegawa and Tappert showed the existence of solitons in optical waveguides, which can be described by the NLS equation
\cite{Hasegawa1973}.
Due to the integrability of the NLS equation and the stability of optical solitons, they are ideal candidates for information transmission
and processing in broadband waveguide systems \cite{Agrawal2001, Mollenauer2006}. In the optical fiber transmission technology, temporal solitons of the NLS equation can be used as bits of information
\cite{Agrawal2001, Mollenauer2006}.
This has lead to the explosion of high-speed communication technologies based on solitons transmission in the last three decades. 

A soliton of the NLS equation (NLS soliton) is characterized by four parameters: its amplitude ($\eta$), phase ($\alpha$), position ($y$), and frequency ($\beta$). The group velocity $v_g$ of a soliton is double its central frequency, i.e., $v_{g}=2\beta$. In broadband waveguide systems, many sequences of solitons can propagate through the same waveguide. The solitons in each sequence
(each ``frequency channel'') propagate with the same group
velocity, but the group velocity differs for solitons from different
sequences. The frequency shift process is a process that changes the frequency  of a soliton (or, of a sequence of solitons), and so changes its velocity and, as a consequence, changes the bit rate in transmission. In this process, the frequency of solitons is shifted to a new frequency while other parameters of the solitons are kept intact. In fiber-optics technology, a few techniques of frequency shifting have been developed such as methods of wavelength conversion \cite{Yoo1996} and of coherent detection \cite{Ip08} in which one can extract any phase, amplitude, and frequency carried by a transmitted electric field.
A physical process which affects the frequency of optical solitons is the Raman self frequency shift, which is a continuous downshift of the soliton frequency due to energy transferring from high-frequency components of the soliton to lower ones \cite{Agrawal2001, Mitschke86}.
The analytic expression for the Raman frequency shift on soliton collisions in the presence of delayed Raman response was found in \cite{CP2005}.
The implementation of frequency shifting, in particular, for a sequence of solitons, is challenging since it is difficult to shift only the frequency by an arbitrary large value while keeping other parameters unchanged. As shown in section \ref{2.2.2} of the current paper, a ``naive'' frequency shifting procedure for a sequence of solitons might lead to a change of the phase of new pulses. Recently, in \cite {PC2018B}, the authors have found a way to introduce a large frequency shift to solitons  along a waveguide using the delayed Raman response or guiding filters with a varying central frequency. However, this method is also limited since it shifts the frequency only of the soliton and not of the accompanying radiation. In addition, the soliton's frequency shift developed in \cite {PC2018B} might not be applied for shifting an arbitrary frequency at a certain propagation distance. So far, to the best of our knowledge, a complete mathematical procedure for frequency shifting of any given frequency at a given propagation distance is still lacking.  

In the current paper, this important and challenging task will be addressed. We develop the theoretical procedures for shifting the frequency of a single NLS soliton and of a sequence of NLS solitons, and validate them by simulations. The procedures of frequency shifting are based on simple transformations of the Fourier transform (FT) of the wave field in the Fourier domain (or the frequency domain). In particular, the {\it decomposition method} is developed to perform frequency shifting for a sequence of solitons. Furthermore, two major applications for the use of soliton frequency shifting are also presented. More specifically, the frequency shifting procedures are applied to:
(1) stabilize the propagation of solitons in waveguides with frequency dependent linear gain-loss;
(2) enable repeated two-soliton collisions in the presence of weak cubic loss. In the first application, we show that the dynamics of the soliton amplitude in waveguides with frequency dependent linear gain-loss can be described by an ordinary differential equation (ODE).
We then use this ODE model to study the robust transmission stabilization of a single soliton and a sequence of solitons experiencing frequency dependent linear gain-loss in multiple periodic frequency shifting. It is found that
under the multiple periodic frequency shifting, the pulse shape of the sequence of solitons is preserved over intermediate to long distances and the amplitude fits  well with the theoretical prediction.
Therefore, the soliton dynamics studied in this work can be used for controlling and switching soliton sequences in broadband waveguides. Furthermore, in experiments and simulations, it is often difficult to measure a very small quantity such as a collision-induced amplitude shift due to the effect of weak perturbation. In the second application, we suggest a solution for this problem by the use of the frequency shifting to repeat two-soliton collisions in the presence of weak cubic loss. As a result, the theoretical prediction for the accumulative amplitude shift from several repeated collisions, which is a much larger value, can be measured.

The use of frequency dependent linear gain-loss for the transmission stability of NLS solitons was investigated in several earlier studies such as
\cite{PC2018B, CPN2016, PNH2017}. In these papers, it has been shown that the use of frequency dependent linear gain-loss can suppress radiative effects and stabilize multi-sequence soliton propagation at long distances. The main factor for the stabilization propagation is due to the fact that the use of frequency dependent linear gain-loss is efficient in suppressing the resonant part of the emitted radiation \cite{PC2018B}. In \cite{CPN2016}, the authors found that the introduction of frequency dependent loss leads to significant enhancement of
transmission stability via decay of
radiative sidebands, which can be described as a dynamic phase
transition. Furthermore, the waveguide's cubic loss can be a result of two-photon absorption (TPA) or gain and loss saturation in a silicon waveguide. TPA has been received considerable attention in recent years due to the importance of
TPA in silicon nanowaveguides, which are expected to play
a crucial role in optical processing applications in optoelectronic devices, including pulse switching and compression,
wavelength conversion, regeneration, etc \cite{Husko2013, PNC2010}. It has been shown that the presence of weak cubic loss strongly affects soliton parameters and, in particular, lead to a downshift of the amplitude and frequency of solitons in a single two-soliton collision. The analytic expressions for the amplitude and frequency shift in a single two-soliton collision in the presence of weak cubic loss were found in \cite{PNC2010}. Soliton propagation of the NLS equation with weak cubic loss has also been investigated in \cite{NPT2015} for robust transmission stabilization and dynamic switching in hybrid waveguides.

The rest of the paper is organized as follows. In section \ref{Procedure}, the frequency shifting procedures for a single NLS soliton and for a sequence of NLS solitons are developed. In section \ref{Applications}, applications of the frequency shifting for a single soliton and a sequence of solitons are theoretically analyzed as follows: (1) stabilizing soliton propagation in waveguides with frequency dependent linear gain-loss; (2) enabling repeated two-soliton collisions in the presence of weak cubic loss and then measuring the theoretical prediction for the accumulative amplitude shift. In section \ref{Num}, the theoretical results presented in section \ref{Procedure} and section \ref{Applications} are verified by numerical simulations with the NLS equation. Section \ref{conc} is reserved for conclusions.

\section{The theoretical procedures for frequency shifting}
\label{Procedure}
\subsection{Model, solitons and their Fourier transforms}

In this section, we describe the NLS solitons and their Fourier transforms. The propagation of solitons through a nonlinear optical waveguide 
can be described by the unperturbed NLS equation \cite{Ablowitz11}:
\begin{eqnarray} &&
i\partial_{z}\psi  + \partial_{t}^2\psi  + 2\left|\psi\right|^{2}\psi  = 0,
\label{single1}
%Equation 1
\end{eqnarray}    
where $\psi$ is proportional to the envelope of the electric field, 
$z$ is the normalized propagation distance, and $t$ is the time \cite{dimensions}.
The second term of Eq. (\ref{single1}) 
describes the effects of temporal second-order dispersion 
while the third term describes the effects of Kerr nonlinearity.  
Equation (\ref{single1}) has the fundamental soliton solution which is given by
%\cite{ACN1985, Tao09, Agrawal2001, Hirota73}:
\cite{Ablowitz11}:
%\begin{eqnarray} &&
\[\psi_{0}(t,z)=\eta \frac{\exp(i\chi)}{\cosh(x)},
\]%\label{single1a}
%\end{eqnarray}    
where $x=\eta (t-y_{0}-2\beta z)$, $\chi=\alpha+\beta(t-y_{0})+ (\eta^{2}-\beta^{2})z$, 
and $\eta$, $\beta$, $y_{0}$, and $\alpha$ 
are the soliton amplitude, frequency, initial position, and phase, respectively.

First, we calculate the Fourier transform of a single soliton.
The initial condition is given by
\begin{eqnarray} &&
\psi_{1} (t,0) =\eta(0)
\frac{\exp\left\{i[\alpha(0)+\beta(0)(t-y(0))]\right\}}
{\cosh \left\{\eta(0)[t-y(0)]\right\}},
\label{IC1}
\end{eqnarray}
where $\eta(0)$, $\beta(0)$, $y(0)$, and $\alpha(0)$ 
are the initial soliton's amplitude, frequency, position, and phase, respectively.  
In a typical optical waveguide system, solitons propagate 
in the presence of additional weak physical perturbations, such as linear 
or nonlinear loss, Raman scattering, etc. \cite{Agrawal2001}.
In the current work, it is assumed that the effects of these perturbation processes 
on the soliton propagation are weak. Under this assumption, one can employ 
the standard adiabatic perturbation theory for the NLS soliton,
which was developed by Kaup \cite{Kaup76, Kaup76B}, 
and can write the total electric field at a distance $z$ as: 
$\psi(t,z)=\psi_{s1}(t,z)+v_{r1}(t,z)$, 
where $\psi_{s1}(t,z)$ is the soliton part and $v_{r1}(t,z)$ is the 
very small radiation part \cite{Hasegawa95}. 
The soliton part is expressed as 
\begin{eqnarray} &&
\psi_{s1} (t,z) =\eta(z)
\frac{\exp\left\{i[\beta(z)\left(t-y(z)\right)+\theta(z)]\right\}}
{\cosh \left\{\eta(z)[t-y(z)]\right\}},
\label{single2}
\end{eqnarray} 
where $\eta(z)$, $\beta(z)$, $y(z)$, 
and $\theta(z)$ are the $z$-dependent soliton's
amplitude, frequency, position, and overall phase, respectively.
The FT of $\psi_{s1}(t,z)$ 
with respect to $t$ is defined by
%\begin{eqnarray} &&
\[\hat \psi_{s1} (\omega ,z) = \frac{1}{(2\pi)^{ 1/2}}
\int_{ - \infty }^{ \infty } {\psi_{s1}(t,z)e^{-i\omega t} }dt,
\]
%\label{single2b} 
%\end{eqnarray} 
where $\omega$ is the frequency. Thus,
%\begin{eqnarray} &&
\[\hat \psi_{s1} (\omega ,z) = \frac{\eta(z)} {(2\pi)^{1/2}}
e^{i\theta(z)} \int_{ - \infty }^{ \infty }
\frac {e^{i\beta(z)[t-y(z)]} e^{-i\omega t} } {\cosh \left\{\eta(z)[t-y(z)]\right\}} dt.
\]%\label{single2c} 
%\end{eqnarray} 
By the Residue theorem, one can obtain $\int_{-\infty}^{\infty} \dfrac{e^{-ibx}}{\cosh (ax)}  dx  = \dfrac{\pi/ a} {\cosh[\pi b /(2a)]}$, where $a$ and $b$ are constants, $a>0$ and $b \in \mathbb{R}$. Therefore, the pattern of a single soliton in the frequency domain is given by
\begin{eqnarray} &&
\hat \psi_{s1}(\omega,z) = 
\left( \frac{\pi}{2} \right)^{1/2}
\frac{\exp[i\theta \left( z \right) - i\omega y( z )]}
{\cosh\left\{\pi \left[\omega - \beta \left( z \right)\right]/
\left[2\eta \left( z \right)\right]\right\}}.
\label{single3} 
\end{eqnarray}

\begin{figure}[ptb]
\begin{tabular}{cc}
\epsfxsize=6.5cm  \epsffile{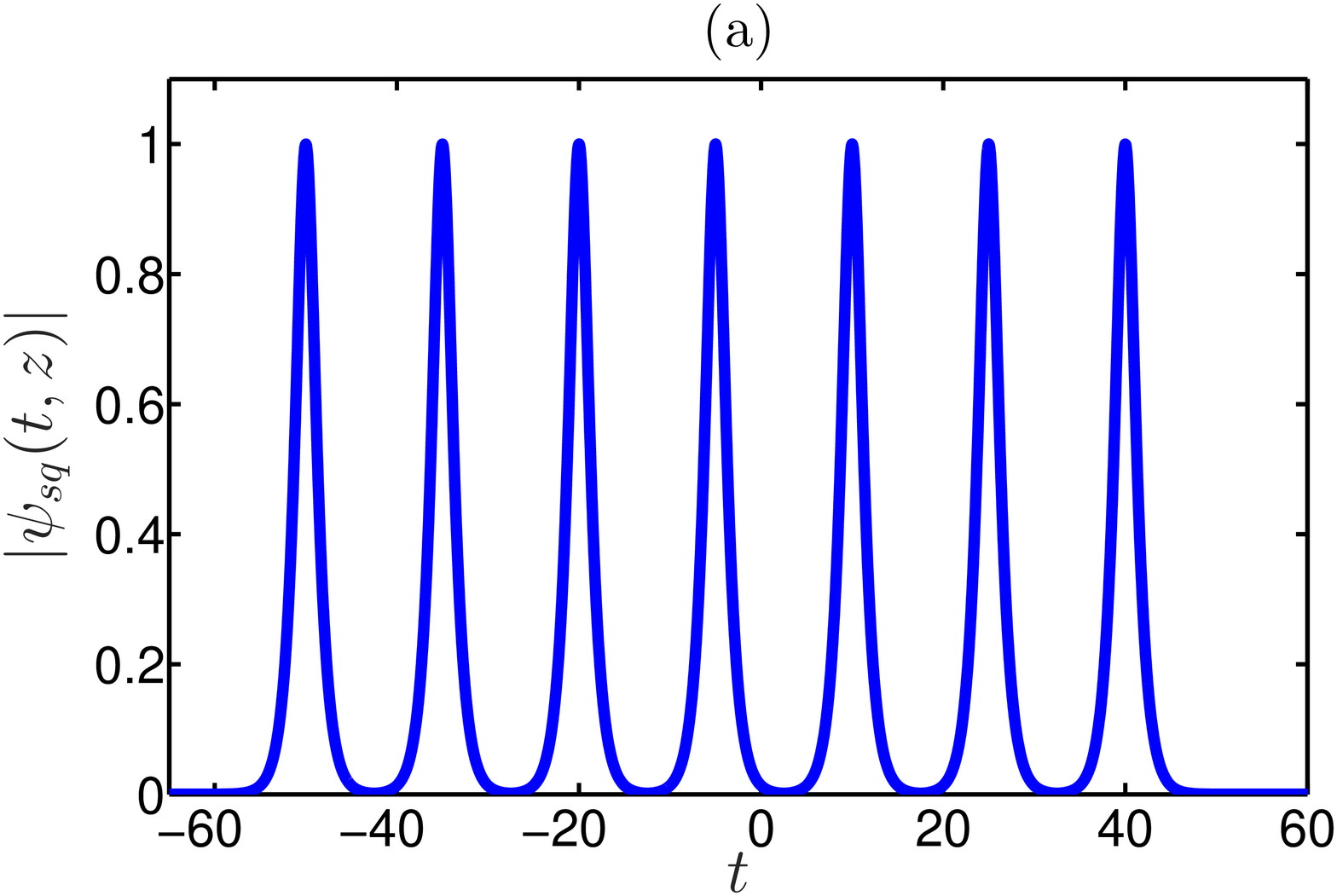} &
\epsfxsize=6.5cm  \epsffile{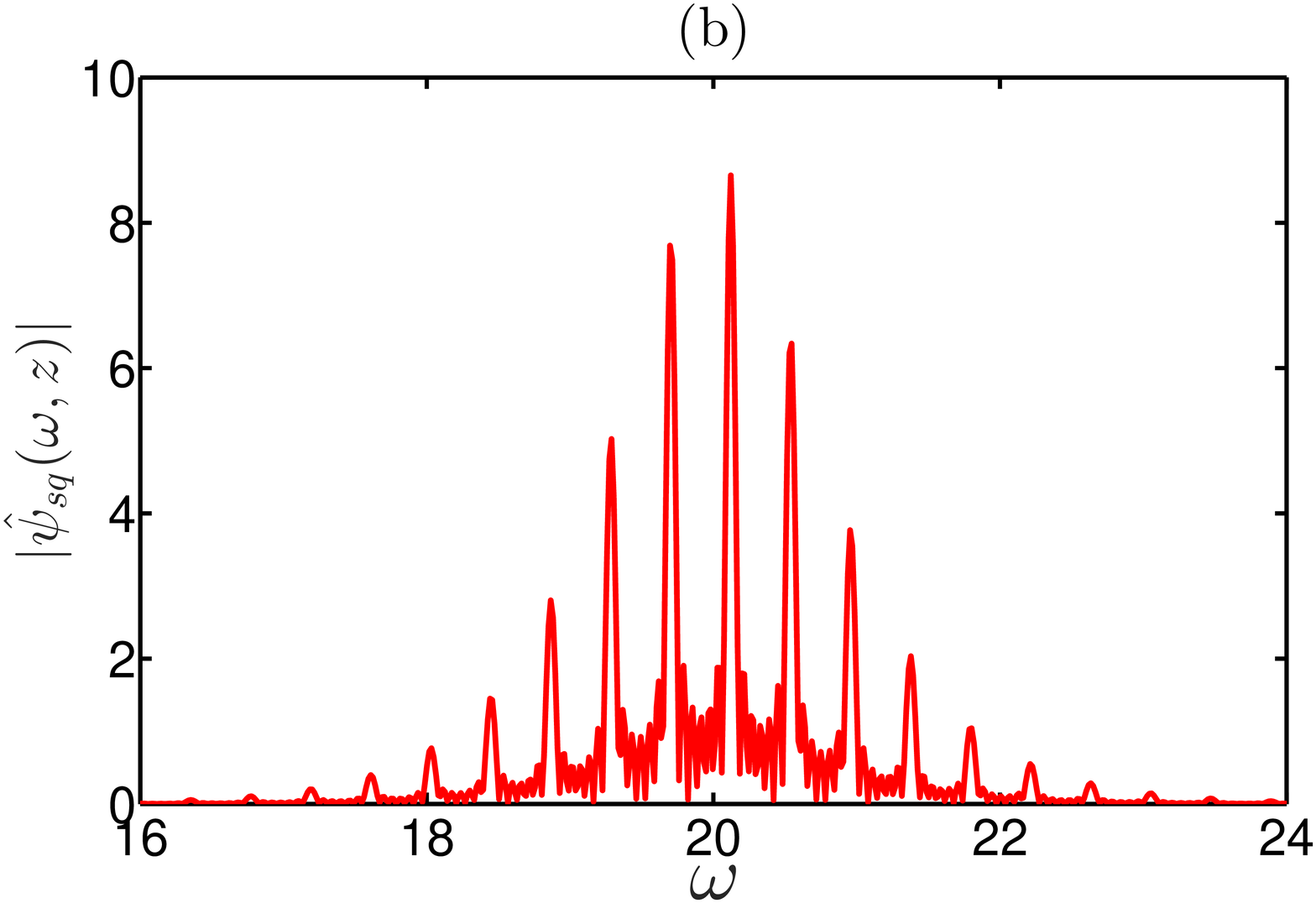} \\
\epsfxsize=6.5cm  \epsffile{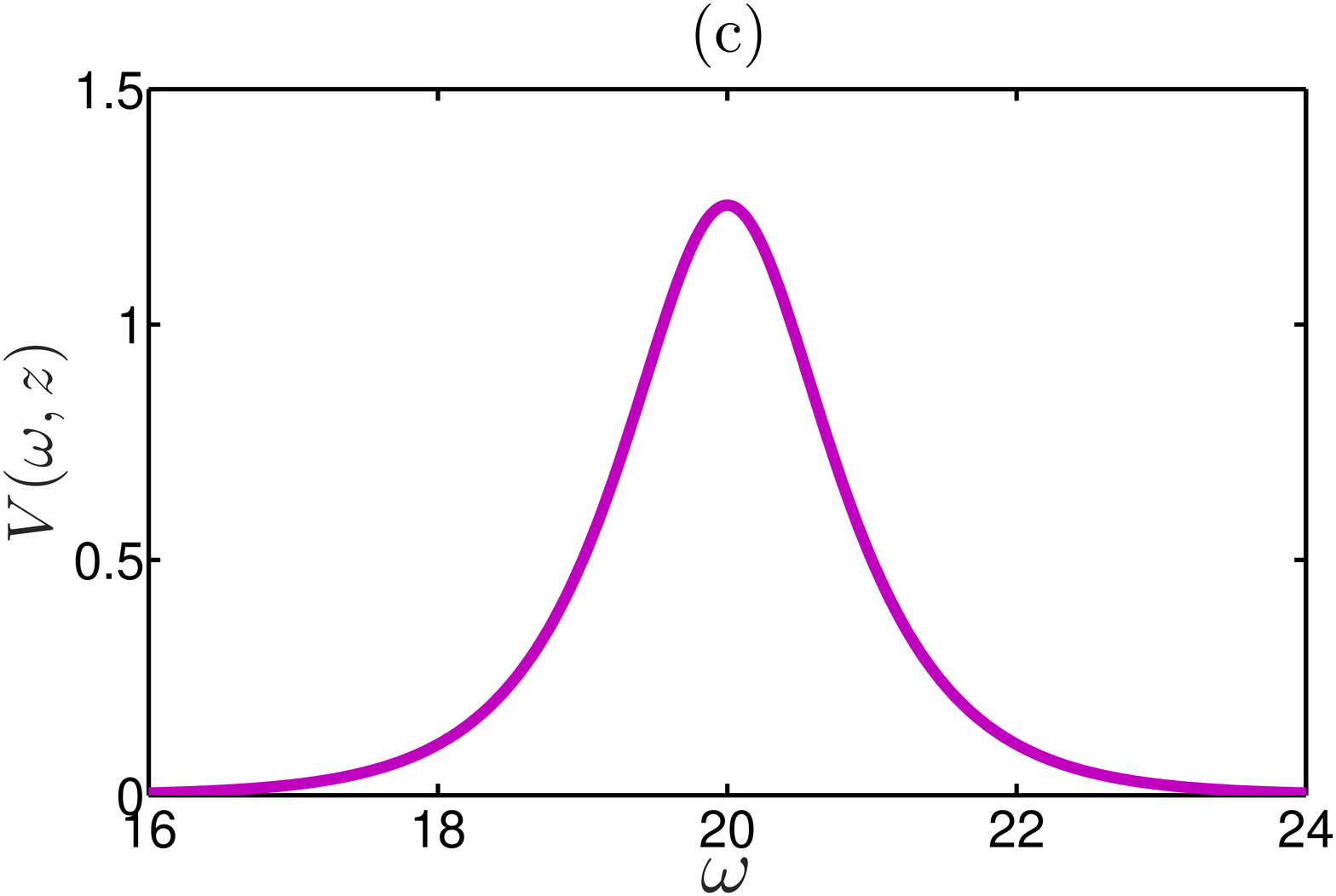} &
\epsfxsize=6.5cm  \epsffile{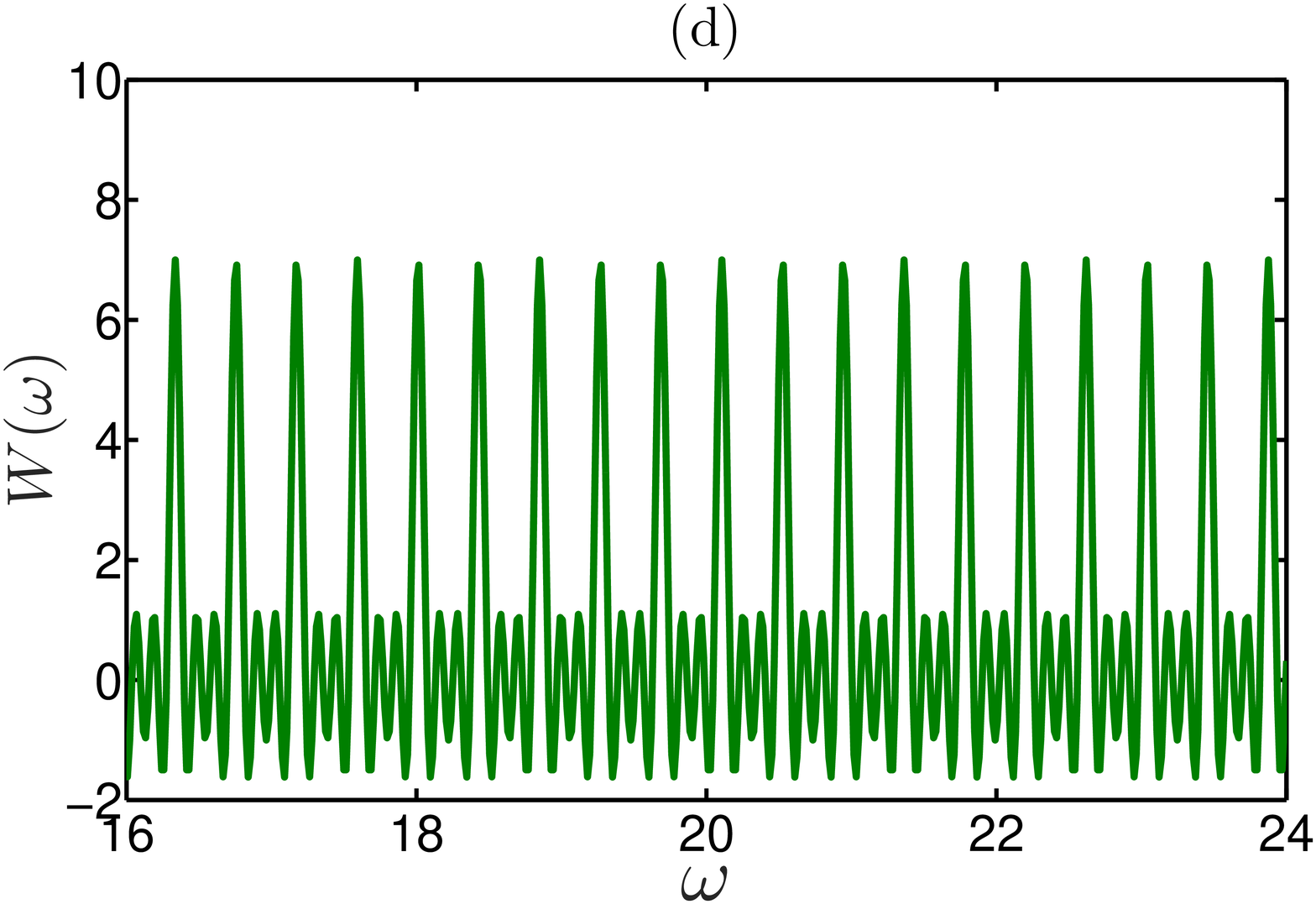} \\
\end{tabular}
\caption{(Color online) The soliton pattern $|\psi_{sq}(t,z)|$, its FT pattern $|\hat\psi_{sq}(\omega,z)|$, and the functions $V(\omega,z)$ and $W(\omega)$, 
given by Eqs. (\ref{sequence1})-(\ref{sequence4}) with $J=3$, $\eta (z)=1$, $\beta(z)=20$, 
$y(z)=-5$, and $T=15$.  
(a) The solid blue curve represents $|\psi_{sq}(t,z)|$ obtained by Eq. (\ref{sequence1}). 
(b) The solid red curve represents the FT pattern $|\hat\psi_{sq}(\omega,z)|$. 
(c) The solid purple curve represents $V(\omega,z)$ 
obtained by Eq. (\ref{sequence3}). 
(d) The solid green curve represents $W(\omega)$ obtained by Eq. (\ref{sequence4}).}   
% The parameter values are: 
% $t_{max}=150,\,J=3,\,\omega_{max}=150,\,y(z)=-5,\,
% \beta(z)=20,\, \theta (z)=0,\,\eta (z)=1,\,T=15$. 
\label{add_fig1}
\end{figure}

Second, we turn to calculate the FT of a sequence of solitons. It is assumed that the envelope 
of the electric field at the waveguide's entrance 
consists of $2J+1$ solitons with equal initial amplitudes, 
frequencies, and phases. This sequence of solitons can be used in a finite waveguide link or in a closed waveguide loop. Thus, the initial soliton sequence $\psi_{sq}(t,0)$ is given by
\begin{eqnarray} &&
\psi_{sq}(t,0)\!=\!\sum_{k=-J}^{J}
\frac{\eta(0)\exp\left\{i[\alpha(0)+\beta(0)(t-y(0)-kT)]\right\}}
{\cosh[\eta(0)(t-y(0)-kT)]}, 
\label{IC2}
\end{eqnarray}
where $\eta(0)$, $\beta(0)$, and $\alpha(0)$ are the 
common initial amplitude, frequency, and phase,   
$y(0)$ is the initial position of the central soliton,  
and $T$ is the temporal separation between adjacent solitons, 
i.e., the time-slot width. 
Assuming the soliton sequence propagates in the presence of 
weak physical perturbations, the total electric field at a propagation distance $z$ 
can be written as: $\psi(t,z)=\psi_{sq}(t,z)+v_{rq}(t,z)$, 
where $\psi_{sq}(t,z)$ is the soliton sequence and $v_{rq}(t,z)$ 
is the very small radiation part \cite{Hasegawa95}.
The soliton sequence $\psi_{sq}(t,z)$ can be expressed as 
\begin{eqnarray} &&
\psi_{sq}(t,z) = \eta(z)e^{i\theta(z)}
\sum_{k = -J}^{J}\frac{\exp\{ i\beta(z) 
\left[ t-y(z)-kT \right] \}}
{\cosh\{\eta(z)\left[t-y(z)-kT\right]\}}, 
\label{sequence1}
\end{eqnarray} 
where $\eta(z)$, $\beta(z)$, $y(z)$, 
and $\theta(z)$ are the common amplitude, frequency, 
overall position shift, and overall phase of solitons at the distance $z$, respectively.  
The FT of $\psi_{sq}(t,z)$ 
with respect to $t$ is given by 
\begin{eqnarray} &&
\!\!\!\!\!\!\!\!\!\!\!\!\!\!\!\!\!\!\!\!\!
\hat\psi_{sq}(\omega,z) = 
\left( \frac{\pi}{2} \right)^{1/2}
\frac{\exp[i\theta(z) - i\omega y(z)]}
{\cosh\left\{\pi \left[\omega - \beta(z)\right]/
\left[2\eta(z)\right]\right\}}
\sum\limits_{k=-J}^{J} e^{-ikT\omega}.
\label{sequence1b}
\end{eqnarray} 
This can be written as
\begin{eqnarray} &&
\hat \psi_{sq}(\omega,z) = 
V(\omega,z) W(\omega)
e^{iU(\omega,z)},
\label{sequence2}
\end{eqnarray}
where
\begin{eqnarray} &&
V(\omega,z)=
\left(\frac{\pi}{2}\right)^{1/2}
\sech\left\{\frac{\pi\left[\omega - \beta(z)\right]}{2\eta(z)}\right\}, 
\label{sequence3}
\end{eqnarray}             
\begin{eqnarray} &&
W(\omega)=
\sum\limits_{k =  - J}^{J} {e^{ -ikT\omega }}=
1 + 2\sum\limits_{k = 1}^{J} {\cos (kT\omega )}, 
\label{sequence4}
\end{eqnarray}    
and 
\begin{eqnarray} &&
U(\omega,z)=
\theta(z) - \omega y(z). 
\label{sequence5}
\end{eqnarray} 
The function $V(\omega,z)$ describes the spectral shape of a single soliton in the frequency domain, 
$W(\omega)$ is related to the solitons' positions relative to 
one another, while $U(\omega,z)$ is the overall common phase. 
Figure \ref{add_fig1} illustrates a typical example for a sequence of solitons $|\psi_{sq}(t,z)|$, its FT $|\hat\psi_{sq}(\omega,z)|$, $V(\omega,z)$, and $W(\omega)$. It is seen that $V(\omega,z)$  
is related to an envelope function for $|\hat\psi_{sq}(\omega,z)|$, 
while $W(\omega)$ corresponds to oscillations of the pulse pattern
$|\hat\psi_{sq}(\omega,z)|$ inside $V(\omega,z)$.

%fig1
              
\subsection{The frequency shifting procedures}

\subsubsection{Frequency shifting for a single soliton}
\label{2.2.1}

The frequency shifting procedure for a single soliton is simply based on shifting the pulse pattern in the frequency domain by a given value $\Delta\beta$. 
If the FT of a soliton before performing the
frequency shifting is centered at $\beta(z)$, then it will be centered at the new frequency 
$\beta_{n}(z)=\beta(z)+\Delta\beta$ after employing the frequency shifting. 
Formally, the frequency shifting is implemented by 
employing the transformation $\omega  \to \omega - \Delta \beta$ 
in the Fourier domain. 
From Eq. (\ref{single3}), the FT of the soliton after performing the frequency shifting is then given by        
\begin{eqnarray} &&
\hat\psi_{s1,n}(\omega,z) = 
\left( \frac{\pi}{2} \right)^{1/2}
\frac{\exp[i\theta_{n}(z) - i\omega y(z)]}
{\cosh\left\{\pi[\omega - \beta_{n}(z)]/
[2\eta(z)]\right\}},
\label{single4} 
\end{eqnarray} 
where $\theta_{n}(z) = \theta (z) + \Delta \beta y(z)$ is the {\it new} overall phase, the sub-index $n$ denotes the {\it new} parameter of the pulse pattern after employing the frequency shifting. 
Taking the inverse FT of $\hat\psi_{s1,n}(\omega,z)$ with respect to $\omega$, it yields the {\it new} (frequency shifted) soliton: 
\begin{eqnarray} 
\psi_{s1,n}(t,z)=\eta(z)
\frac{\exp\left\{i[\beta_{n}(z)\left(t-y(z)\right)+\theta_{n}(z)]\right\}}
{\cosh \left\{\eta(z)[t-y(z)]\right\}}.
\label{single5}
\end{eqnarray}   
Comparing Eq. (\ref{single5}) with Eq. (\ref{single2}), we conclude that after performing the frequency shifting one can obtain the soliton with the new frequency $\beta_{n}(z)=\beta(z)+\Delta\beta$.

\subsubsection{Frequency shifting for a sequence of solitons}
\label{2.2.2}

In this section, the frequency shifting procedure for a sequence of solitons propagating through 
a nonlinear optical waveguide will be presented.
Assuming that the soliton sequence propagates in the presence of weak physical perturbations,
the soliton part and its FT are then given by Eq. (\ref{sequence1}) and Eq. (\ref{sequence2}), respectively.
It is emphasized that the current frequency shifting procedure can be used in the following setups: (i) transmission through a finite-length optical waveguide link; (ii) propagation in a closed optical waveguide loop. The frequency shift method for a sequence of solitons is more complicated than for a single soliton. Before describing the decomposition procedure in the procedure III, we present and discuss the first two problematic procedures:

{\it Procedure I - a ``naive'' frequency shifting procedure.}
First, it is emphasized that a frequency shifting procedure, which is based on a ``naive'' attempt 
to extend the procedure used for a single soliton to a sequence of solitons, 
is generally not applicable. In this ``naive'' procedure, the transformation 
$\omega  \to \omega - \Delta \beta$ of the FT of the soliton sequence is performed in the frequency domain. Indeed, by performing this transformation to the FT of 
a soliton sequence, which is given by Eq. (\ref{sequence1b}), 
one can arrive at the following expression for the FT 
of the soliton sequence after shifting the frequency :           
%\begin{eqnarray} &&
%\!\!\!\!\!\!\!\!\!\!\!\!\!\!\!\!\!\!\!\!\!
\[\hat\psi_{sq,n}(\omega,z) = 
\left( \frac{\pi}{2} \right)^{1/2}
\frac{\exp[i\theta_{n}(z) - i\omega y(z)]}
{\cosh\left\{\pi \left[\omega - \beta_{n}(z)\right]/
\left[2\eta(z)\right]\right\}}
\sum\limits_{k=-J}^{J} e^{-ikT\omega}e^{ikT\Delta\beta},
\]
%\label{sequence6}
%\end{eqnarray} 
where $\theta_{n}(z) = \theta (z) + \Delta \beta y(z)$ is the {\it new} overall phase and $\beta_{n}(z)=\beta(z)+\Delta\beta$. 
Taking the inverse FT of $\hat\psi_{sq,n}(\omega,z)$ with respect to $\omega$, 
it implies the new sequence of pulses: 
\begin{eqnarray} &&
\!\!\!\!\!\!\!\!\!\!\!\!\!\!\!\!\!\!\!\!\!
\psi_{sq,n}(t,z) = \eta(z)e^{i\theta_{n}(z)}
\sum_{k=-J}^{J}\frac{\exp\{i\beta_{n}(z)\left[t-y(z)-kT\right]+ikT\Delta\beta\}}
{\cosh\{\eta(z)\left[t-y(z)-kT\right]\}}.
\label{sequence7}
\end{eqnarray} 
Comparing Eq. (\ref{sequence7})  with Eq. (\ref{sequence1}),
it can be seen that the new pulse sequence does not have the form expected for a sequence 
of solitons due to the multiplicative phase factor $\exp(ikT\Delta\beta)$ in Eq. (\ref{sequence7}). Note that if $\Delta\beta=2m\pi/T$, where $m \in \mathbb{Z}$, then $\exp(ikT\Delta\beta)=1$ for $-J \le k \le J$. 
As a result, in this specific case, Eq. (\ref{sequence7}) has the same form with Eq. (\ref{sequence1}). 
Therefore, the procedure I is only applicable  when $\Delta\beta$ is an integer multiple of $2\pi/T$. This problem will be fixed in the decomposition procedure described in the procedure III.

{\it Procedure II - the frequency shifting of $V(\omega,z)$ and $U(\omega,z)$.}
We present another theoretical approach to perform the frequency shift for a sequence of solitons. A promising theoretical procedure for employing the frequency shifting by a value $\Delta\beta$ for a sequence of solitons in form of Eq. (\ref{sequence1}), which is called the ``shifting envelope'' procedure, is described as follows. First, employing the transformation $\omega \to \omega-\Delta\beta$ for $V\left( {\omega ,z} \right)$ and $U\left( {\omega ,z} \right)$, which are given by
Eq. (\ref{sequence3}) and Eq. (\ref{sequence5}), respectively, it yields:
\begin{eqnarray} &&
U_n(\omega,z) = U(\omega  - \Delta\beta,z), \,\,
V_n(\omega,z) = V(\omega - \Delta\beta,z).
\label{sequence8}%Equation 15
\end{eqnarray}   
Second, one can define the FT of the new pulse pattern after employing the frequency shifting:
\begin{eqnarray} &&
\hat \psi_{sq,n}(\omega ,z)
=V_{n}(\omega,z) W(\omega) e^{i U_{n}(\omega,z)}.
\label{sequence9}%Equation 16
\end{eqnarray}
Thus, in Eq. (\ref{sequence9}), the frequency shift is performed in the normalized ``envelope function'' $V(\omega,z)$ and the phase $U(\omega,z)$, while $W(\omega)$ is kept to be fixed during transformations. Eq. (\ref{sequence9}) can be written as
%\begin{eqnarray} &&
\[\!\!\!\!\!\!\!\!\!\!\!\!\!\!\!\!\!\!\!\!\!
\hat\psi_{sq,n}(\omega,z) = 
\left( \frac{\pi}{2} \right)^{1/2}
\frac{\exp[i\theta_{n}(z) - i\omega y(z)]}
{\cosh\left\{\pi \left[\omega - \beta_{n}(z)\right]/
\left[2\eta(z)\right]\right\}}
\sum\limits_{k=-J}^{J} e^{-ikT\omega},
\]%\label{sequence10}
%\end{eqnarray}
%%%%%%%%%%%%
where $\theta_{n}(z) = \theta (z) + \Delta \beta y(z)$ and $\beta_{n}(z)=\beta(z)+\Delta\beta$. Third, taking the inverse FT of $\hat\psi_{sq,n}(\omega,z)$, it implies
\begin{eqnarray} &&
\!\!\!\!\!\!\!\!\!\!\!\!\!\!\!\!\!\!\!\!\!
\psi_{sq,n}(t,z) = \eta(z)e^{i\theta_{n}(z)}
\sum_{k=-J}^{J}\frac{\exp\{i\beta_{n}(z)\left[t-y(z)-kT\right]\}}
{\cosh\{\eta(z)\left[t-y(z)-kT\right]\}}.
\label{sequence11}
\end{eqnarray} 
Equation (\ref{sequence11}) has the required form of the new theory prediction for a sequence of solitons, which is the form as in Eq. (\ref{sequence1}). Therefore, it preserves the relative phase between solitons.

{\it Shortcomings of the procedure II.} In theory, the procedure II is applicable since Eq. (\ref{sequence11}) has the same form with Eq. (\ref{sequence1}). However, it is important to emphasize the important shortcomings of the second 
frequency shifting procedure in numerical implementations. To explain this, we note that the pulse pattern in the frequency domain after performing the frequency shifting 
$\hat\psi^{(num)}_{sq,n}(\omega,z)$ must be centered at the {\it new} numerical frequency value $\beta_{n}^{(num)}(z)$, where 
$\beta_{n}^{(num)}(z)=\beta^{(num)}(z) + \Delta\beta^{(num)}$ with $\Delta\beta^{(num)}$ is the number of grid points shifted in the frequency domain. 
Therefore, the high accuracy of the numerical measurements and implementations  for
$\beta^{(num) }(z)$, $\Delta\beta^{(num)}$, ${V^{(num)}_n(\omega, z)}$, $W^{(num)}(\omega)$, 
and $U_{n}^{(num)}(\omega,z)$ in the neighborhood of $\beta_{n}^{(num)}(z)$ is extremely important. 
These values can be determined 
by the procedures described as in \ref{App_B}. In this Appendix, 
the values of $W^{(num)}(\omega)$ are determined by the extrapolations for   
$\omega<\beta^{(num)}(z)-L/2$ and $\omega>\beta^{(num)}(z)+L/2$,
where $15 \leq L \leq 40$. Consequently, if $|\Delta\beta^{(num)}|$ is large, $|\Delta\beta^{(num)}| \simeq L/2$ or $|\Delta\beta^{(num)}| > L/2$, 
then the numerical data points used for calculating  
$W^{(num)}(\omega)$ of the main body of $\hat\psi^{(num)}_{sq,n}(\omega,z)$, i.e., in the neighborhood of $\beta_{n}^{(num)}(z)$, are obtained by extrapolations, see Fig. \ref{fig_appB1} in \ref{App_B}. These extrapolations in the main body of the new soliton pattern might lead to inaccuracies for calculating $\hat\psi^{(num)}_{sq,n}(\omega,z)$ and, especially, change the physical effect of the soliton sequence, and as a result, 
to the breakdown of the frequency shifting procedure.  
Therefore, the procedure II can only 
be accurately implemented for a small value of the frequency shift: $|\Delta\beta| \ll L/2$.

{\it Procedure III - the decomposition procedure.} This method can completely fix the mentioned impairments of both previous procedures. This new method, which is called the decomposition method, is based on the following key ideas: (1) The ``naive'' procedure I works perfectly for a frequency shift value of $\Delta\beta=2m\pi/T$, $m \in \mathbb{Z}$. (2) The impairment of the ``shifting envelope'' procedure in the procedure II can be fixed by shifting a small frequency value $\Delta\beta$: $|\Delta\beta| \ll L/2$. The theoretical procedure of the decomposition method can be summarized as follows: 
\begin{enumerate}
\item Decompose $\Delta \beta=\Delta \beta_{1}+\Delta \beta_{2}$, where $\Delta\beta_{1}=2m\pi/T$, $m \in \mathbb{Z}$, and $-2\pi/T < \Delta\beta_{2} < 2\pi/T$. 
\item Define $\hat \psi_{sq,n1}(\omega,z) = \hat \psi_{sq}(\omega  - \Delta \beta_{1},z)$ by the ``naive'' method as the {\it procedure I}.
\item Employ the frequency shift by $\Delta \beta_{2}$ for $\hat \psi_{sq,n1} (\omega,z)$
by using the {\it procedure II}.
That is, $\hat \psi_{sq,n}(\omega,z) = V_{n}(\omega,z)W(\omega)e^{iU_{n}(\omega,z)}$
as in Eqs. (\ref{sequence8})-(\ref{sequence9}).
\item Define $\psi_{sq,n}(t,z)=\mathcal {F}^{-1}\left(\hat\psi_{sq,n}(\omega,z)\right)$, where $\mathcal {F}^{-1}$ is the inverse FT.
\end{enumerate}

%fig 3
\begin{figure}[ptb]
\begin{tabular}{cc}
\epsfxsize=7.0cm  \epsffile{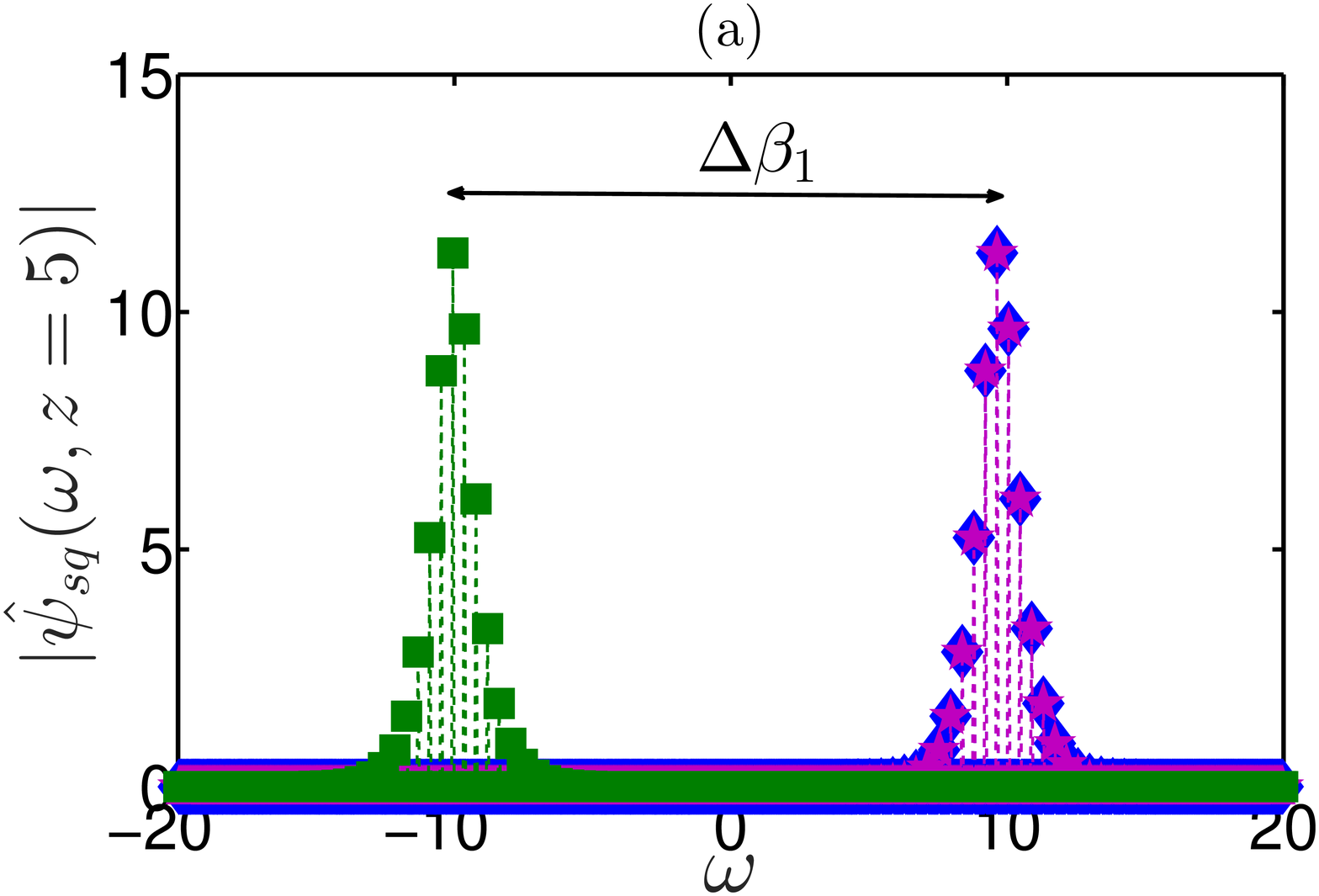} &
\epsfxsize=7.0cm  \epsffile{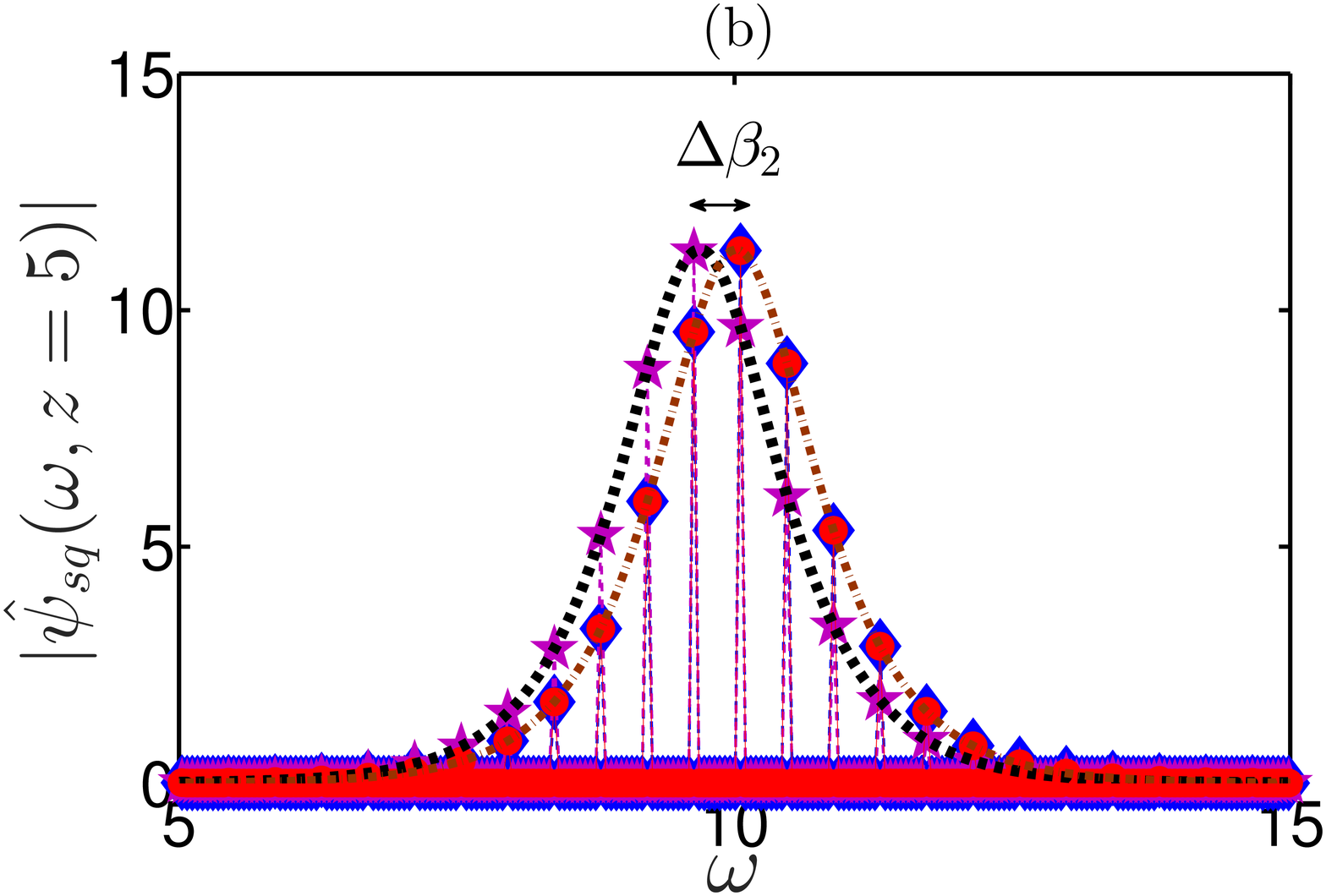} \\
\epsfxsize=7.0cm  \epsffile{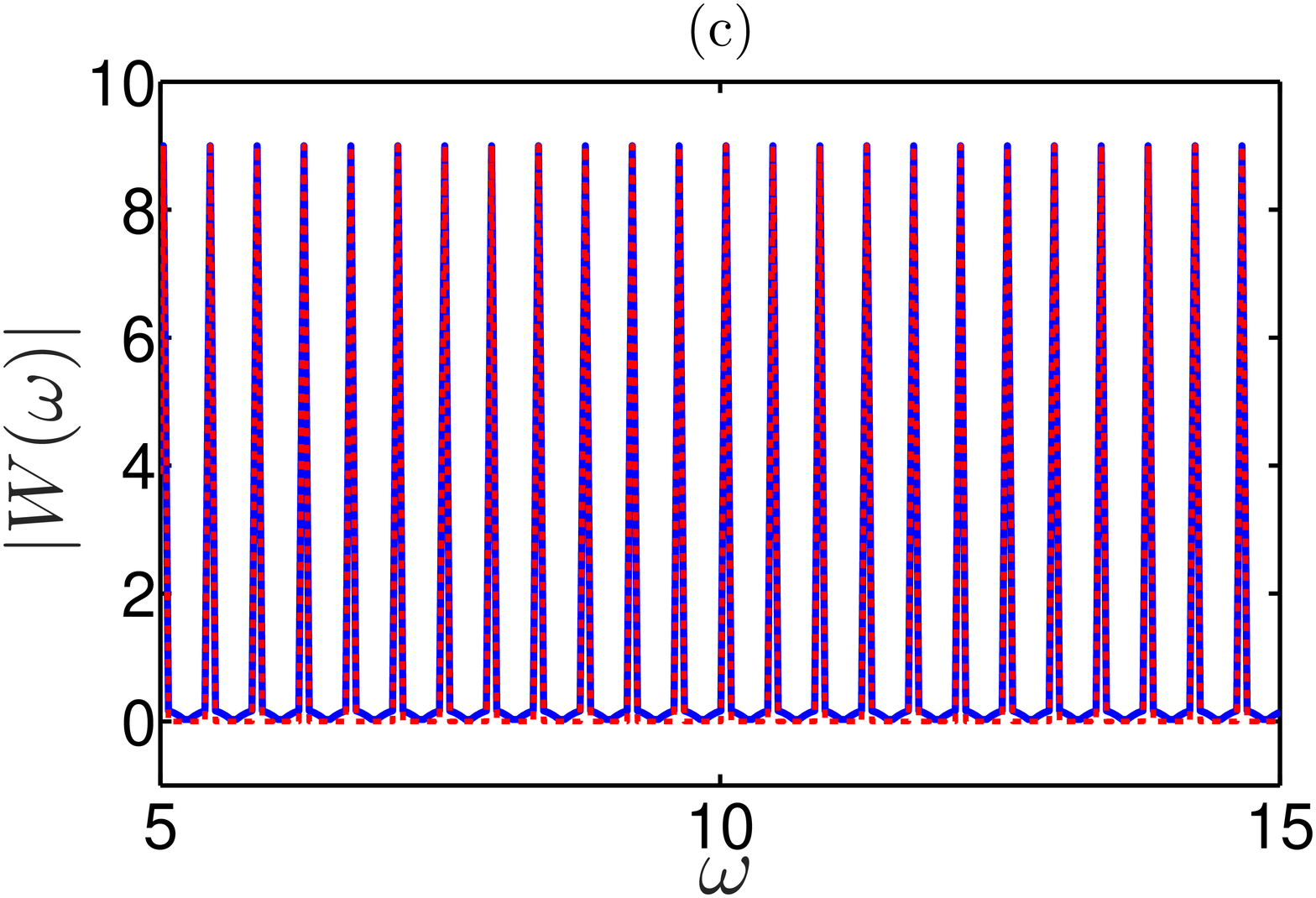} &
\epsfxsize=7.0cm  \epsffile{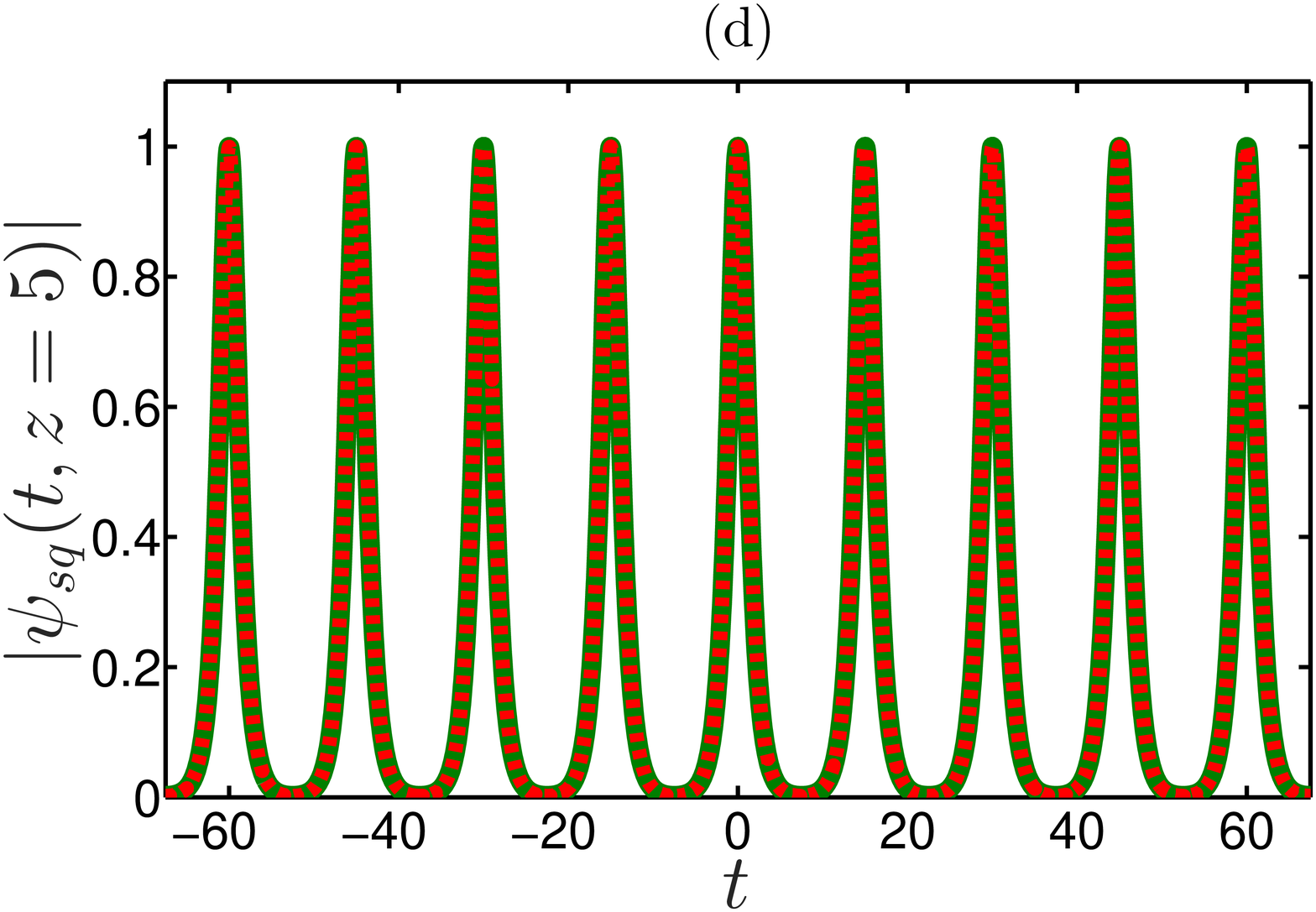} 
\end{tabular}
\caption{(Color online) The frequency shifting procedure by the decomposition method with {\bf setup 1}. (a)-(b) The pulse patterns in the frequency domain before and after shifting the frequency by $\Delta\beta_{1}$ and $\Delta\beta_{2}$, respectively. (a) The green squares represent $|\hat \psi^{(num)}_{sq} (\omega ,z) |$ before shifting the frequency. The purple stars and the blue diamonds correspond to $|\hat \psi^{(num)}_{sq,n1} (\omega ,z)|$ after shifting the frequency and its theoretical prediction $|\hat \psi^{(th)}_{sq,n1} (\omega ,z)|$, respectively. (b) The purple stars are the same as in (a). The red circles and the blue diamonds represent $|\hat \psi^{(num)}_{sq,n} (\omega ,z)|$ after shifting the frequency and its theoretical prediction $| \hat \psi^{(th)}_{sq,n} (\omega ,z)|$, respectively. The black dashed and brown dashed-dotted curves represent $\tilde V^{(num)}(\omega,z)$ and $\tilde V_{n}^{(num)}(\omega,z)$, which are described as in \ref{App_B}, before and after shifting the frequency by $\Delta\beta_2$, respectively. (c) The form of fast oscillation function $|W(\omega)|$. The red dashed and blue solid curves represent $| W^{(num)}(\omega)|$ measured by Eq. (\ref{App_B2}) and its theoretical prediction $| W^{(th)}(\omega)|$, respectively. (d) The soliton patterns in the time domain before and after shifting the frequency. The green solid and red dashed curves correspond to $| \psi^{(num)}_{sq} (t ,z) |$, measured by the simulation with Eq. (\ref{single1}), and $|\psi^{(num)}_{sq,n} (t ,z)|$ measured from the decomposition method, respectively. }
 \label{fig2}
\end{figure}

To validate the accuracy for implementing the procedure, one can calculate the relative error in measuring the pulse pattern at the propagation distance $z$ by using the following normalized integral:
\[\!\!\!\!\!\!\!\!\!\!\!\!\!\!\!\!\!\!\!\!\!
E(z)=\left\{\int\limits_{t_{\min}}^{t_{\max}} \left[|\psi^{(num)}(t,z)| - |\psi^{(th)}(t,z)|\right]^{2}dt\right\}^{1/2}/
\left\{\int\limits_{t_{\min}}^{t_{\max}} |\psi^{(th)}(t,z)|^{2}dt\right\}^{1/2},
\]
where the pulse patterns $\left|\psi^{(num)}(t,z)\right|$ and $\left|\psi^{(th)}(t,z)\right|$ are measured from the simulation and the theoretical prediction, respectively, and $[t_{\min},t_{\max}]$ is the computational time domain
\cite{trefethen2000,Yang2010}.
In measuring the theoretical prediction for $\psi_{s1}(t,z)$ 
from Eq. (\ref{single2})
and for $\psi_{sq}(t,z)$
from Eq. (\ref{sequence1}), 
the soliton parameters $\eta(z)$, $\beta(z)$, $y(z)$, and $\theta(z)$ are calculated from simulations.
Note that the relative error in measuring the pulse pattern $|\hat\psi(\omega,z)|$ in the frequency domain can be calculated similarly.

To illustrate the decomposition method, we perform the frequency shifting procedure for a soliton sequence propagating periodically in a waveguide loop with Eq. (\ref{single1}) and the initial condition as in Eq. (\ref{IC2}).
The NLS equation (\ref{single1}) is numerically solved by using the split-step Fourier method with periodic boundary conditions (see Appendix \ref{App_C}).
The use of periodic boundary conditions means that the numerical simulations describe soliton dynamics in a closed waveguide loop. It is useful to consider the numerical {\bf setup 1}
with parameters as follows: $\Delta\beta  = 20$, $L=15$, $\eta(0) = 1$, $\beta(0) =  - 10$, $y(0) =  - 5$, $\alpha(0) =  0$, $t_{\max}=67.5$, $t_{\min}=-t_{\max}$, $T=15$, and $J=4$.
Figure \ref{fig2} presents the frequency shifting procedure at the propagation distance $z=5$. One can measure $\Delta\beta^{(num)}= 20.0131$ ($n_{shift}=430$) that consists $\Delta\beta_1=94\pi/T=19.6873$ ($n_{shift}=423$) and $\Delta\beta_{2}=0.3258$ ($n_{shift}=7$), where $n_{shift}$ is the number of grid points shifted in
the frequency domain. 
The numerical measurements for the frequency before and after performing the frequency shifting are $\beta^{(num)}(z=5)=-10$ and $\beta_{n}^{(num)}(z=5)=10.0131$, respectively.
Figure \ref{fig2}(a) illustrates the first step of the procedure with shifting the frequency by $\Delta\beta_{1}$, that is, one can obtain $\hat \psi^{(num)}_{sq,n1}(\omega,z)$ from $\hat \psi^{(num)}_{sq}(\omega,z)$ by the procedure I.
Figure \ref{fig2}(b) represents the second step of the procedure with shifting the frequency by $\Delta\beta_{2}$, that is, one can obtain $\hat \psi^{(num)}_{sq,n}(\omega,z)$ from $\hat \psi^{(num)}_{sq,n1}(\omega,z)$ by the procedure II. Figure \ref{fig2}(c) shows the comparison of the oscillation function $\left|W^{(num)}(\omega)\right|$ measured by Eq. (\ref{App_B2}) and its theoretical prediction $\left|W^{(th)}(\omega)\right|$ as in Eq.(\ref{sequence4}). Figure \ref{fig2}(d) shows the pulse patterns $\left|\psi^{(num)}_{sq} (t ,z) \right|$ and $\left|\psi^{(num)}_{sq,n}(t ,z) \right|$ in the time domain before and after shifting the frequency by the decomposition method, respectively. As can be seen, there are very good agreements between the numerical measurements and the theory predictions in Figs. \ref{fig2}[(a)-(c)], as well as between the pulse patterns before and after employing the frequency shifting in Fig. \ref{fig2}(d).
In fact, the relative errors in measuring the soliton patterns $|\hat\psi_{sq,n}(\omega,z)|$ in Fig. \ref{fig2}(b)
and $|\psi_{sq,n}(t,z)|$
in Fig. \ref{fig2}(d)
are $1.5\times 10^{-5}$ and $1.13\times 10^{-5}$, respectively.
These very small relative errors and the simulation results demonstrate that the decomposition method is validated and implemented successfully by numerical simulations for a sequence of solitons.

{\it Discussion on the robustness of the procedure III.} 
It is worthy to emphasize that the decomposition method is simple to implement. Indeed, the frequency can be perfectly shifted for a value of $\Delta\beta_{1}$ by the ``naive'' method without adjusting or approximating the value of $\Delta\beta_{1}$ to fit with the grid size in the frequency domain. That is, $\Delta\beta_{1}^{(num)}=\Delta\beta_{1}$ in simulations. In fact, we recall the following relation between the wavenumber separation in the frequency domain and the length of the time domain $\Delta\omega=2 \pi/L_{t}$, where $L_{t}$ is the length of the computational time domain
\cite{trefethen2000,Yang2010}.
Thus, in a closed waveguide loop setup, one has $\Delta\omega=2 \pi/[(2J+1)T]$. As a result, the value of $\Delta\beta_{1}$, which is in form of $2m \pi/T$, is exactly a multiple of the wavenumber separation $\Delta\omega$ in the frequency domain. That is, there exists a positive integer $n=(2J+1)m$ such that $\Delta\beta_{1}=n\Delta\omega$. Therefore, one can perfectly shift $\Delta\beta_{1}$ by the procedure I without finding the nearest integer $n$ such that $\Delta\beta_{1}^{(num)}=n\Delta\omega$. 

\section{Applications of frequency shifting procedures}
\label{Applications}

\subsection{Stabilization of the propagation of solitons in waveguides with frequency dependent linear gain-loss} 
\label{App-stabi}
%\cite{CPN2016, PNH2017, PC2018} 

\begin{figure}[ptb]
\begin{tabular}{cc}
\epsfxsize=6.8cm  \epsffile{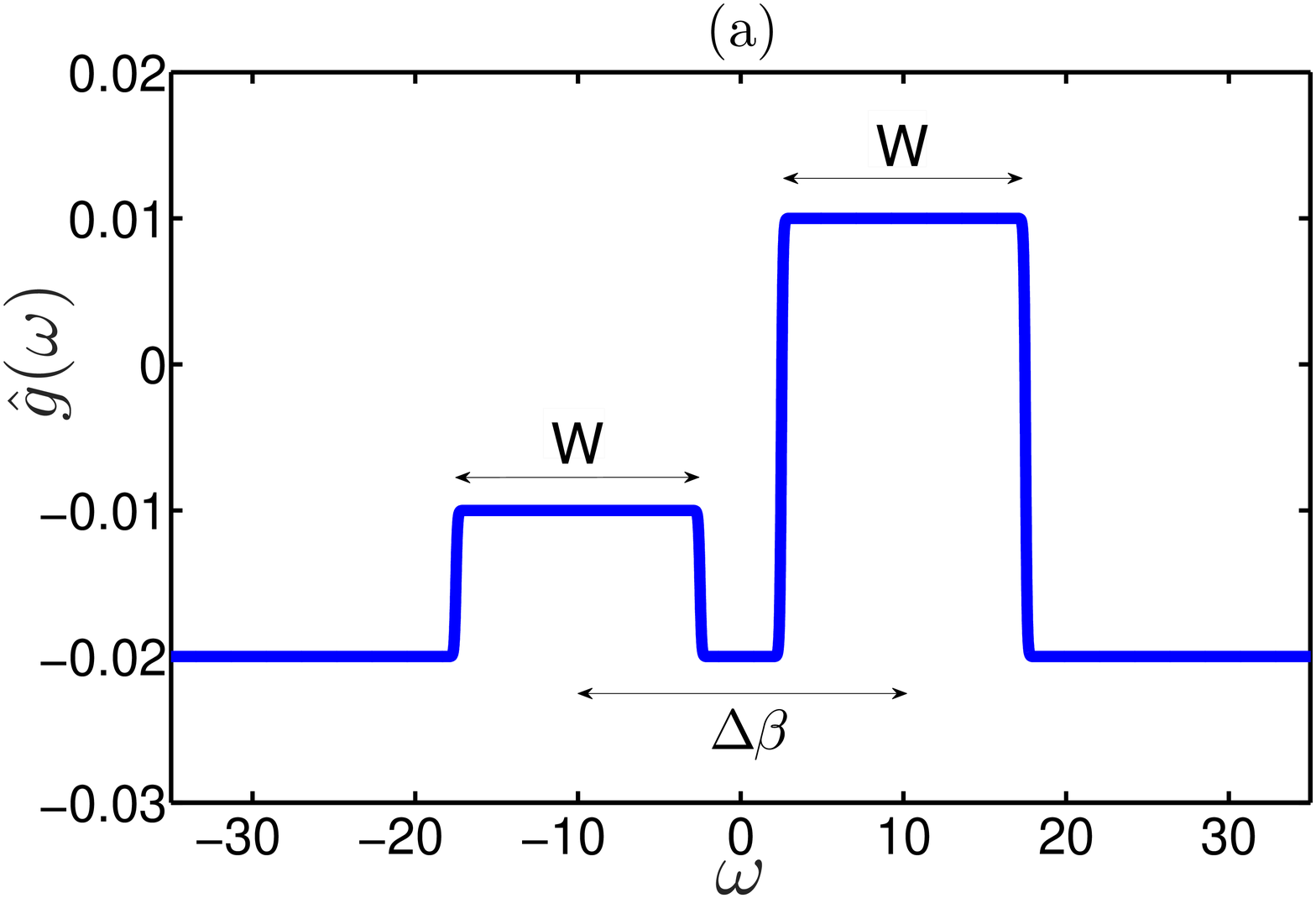}&
\epsfxsize=6.8cm  \epsffile{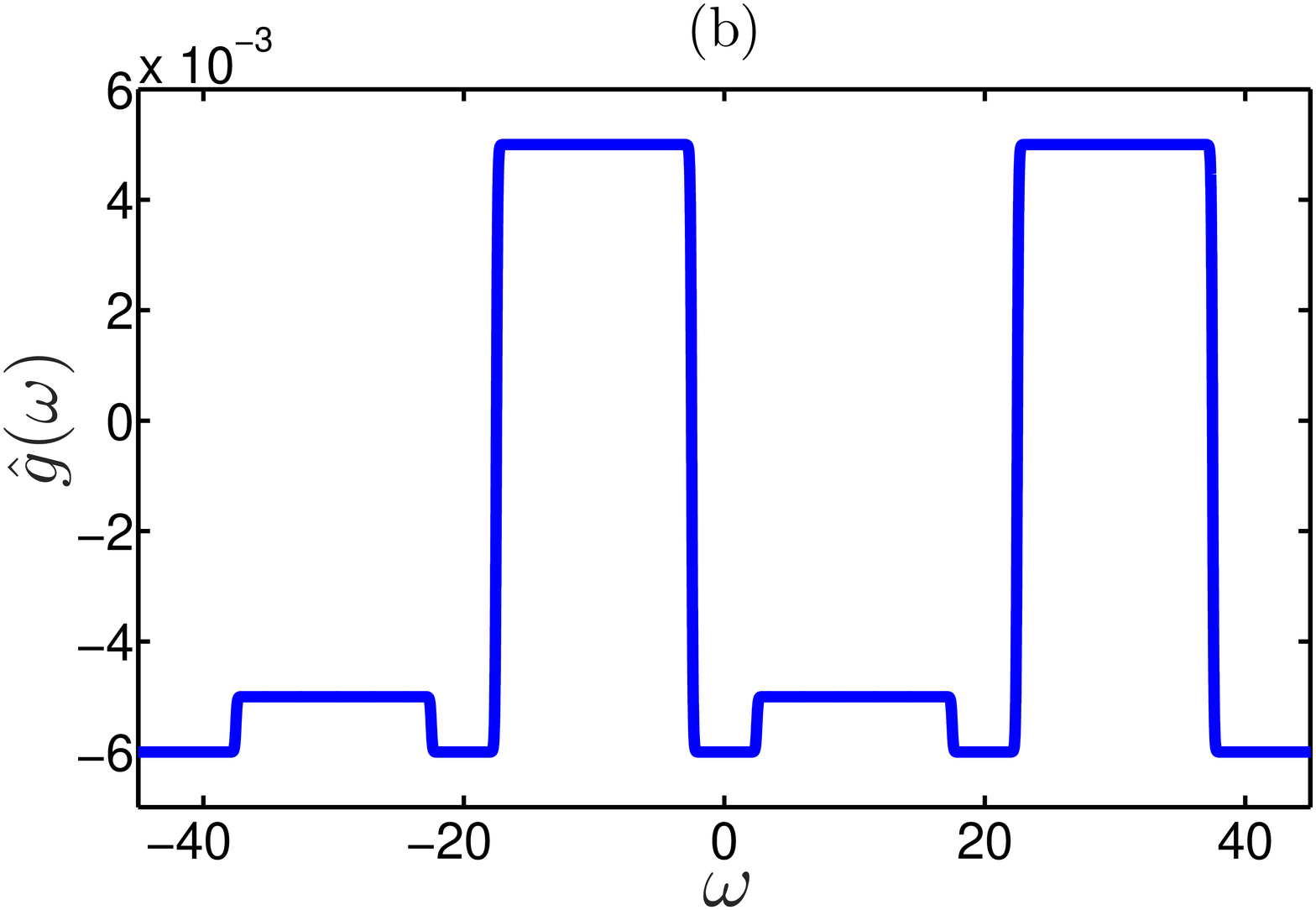} \\
\end{tabular}
\caption{(Color online) The profile for $\hat g(\omega )$ with $\kappa=1$ (a) and $\kappa=2$ (b). (a) $\hat g(\omega)$ vs $\omega$ in Eq. (\ref{App_gl12b}) with parameters used in {\bf setup 4} and used for Fig. \ref{fig6}. (b) $\hat{g}(\omega)$ vs $\omega$ in Eq. (\ref{App_gl2}) with parameters used in {\bf setup 5a} and used for Fig. \ref{fig9}.
}
\label{fig5}
\end{figure}

We consider the propagation of solitons in the presence of frequency dependent linear gain-loss described by the following NLS model: 
\begin{eqnarray} &&
i\partial_{z}\psi  + \partial_{t}^{2}\psi  + 2|\psi|^{2}\psi
={i}\mathcal{F}^{-1}\left[\hat g(\omega)\hat \psi\right]/{2}, 
\label{App_gl1}  %App_gain_loss_1
\end{eqnarray}        
where $\hat g(\omega)$ is the linear gain-loss, $\hat\psi$ is the FT of $\psi$ with respect to time, $\mathcal{F}^{-1}$ is the inverse FT \cite{dimensions}.
It is emphasized that similar forms of $\hat g(\omega )$, which are step-size functions, have been studied in Refs. \cite{CPN2016,PNH2017}.
In these papers, it has been shown that the use of frequency dependent linear gain-loss can suppress the radiative effects and stabilize soliton transmission at long distances instead of using constant gain-loss coefficients. In Eq. (\ref{App_gl1}), the form of ${\hat g}(\omega )$ is defined such that radiation emission effects are mitigated by the use of a negative loss in term of $-g_{L}$ in the frequency nearby the central frequency $\beta(z)$. For this purpose, one can use the following form with experiencing loss-gain $\kappa$ times for $\hat g(\omega )$:
\begin{eqnarray} &&
\begin{array}{l}
\hat g(\omega ) =  - g_{L}
+ 0.5\sum\limits_{j = 0}^{2\kappa  - 1} {\left\{ \left[(-1)^{j} g_{0} + g_{L} \right]
\sum\limits_{k = 0}^{1} {(-1)^{k}{g_{j,k}}(\omega )} \right\}},
\end{array}
\label{App_gl2}%equation 34
\end{eqnarray}
where $g_{j,k}(\omega)=\tanh \left[\rho \left(\omega  + (2j - 2\kappa + 1) \Delta \beta/2 + (-1)^{k} W/2 \right) \right]$, $0< g_{0} \ll 1$, $g_{L} \ge 0$, $\rho \gg 1$, and the spectral width $W$ satisfies $1 \ll W \le \Delta \beta $. The form of ${\hat g}(\omega )$ defined in Eq. (\ref{App_gl2}) can be considered as the continuous version of the step gain-loss function ${\hat g}(\omega )$ used in Refs. \cite{CPN2016,PNH2017}. Physically, the value of $-g_{L}$ is the loss required for suppressing radiation emission, while $g_{0}$ is the gain required for maintaining stable propagation against decaying soliton amplitudes experiencing by the linear loss $-g_{0}$. The multiple frequency shifting is performed such that the solitons experience the linear gain or loss in the new shifted frequency and then continue propagating along the waveguide.
As an example, Fig. \ref{fig5} depicts a graph for $\hat g(\omega )$ with $\kappa=1$ in Fig. \ref{fig5}(a) and $\kappa=2$ in Fig. \ref{fig5}(b), respectively. Let us describe the frequency shifting procedure for the use of $\hat g(\omega )$ as in Fig. \ref{fig5}(a).
Assuming that a sequence of solitons is propagating at the central frequency $\beta=-10$ in the presence of the gain-loss function $\hat g(\omega )$ as in Fig. \ref{fig5}(a), solitons then experience the linear loss at the central frequency $\beta=-10$ and the amplitude will be decreasing due to the linear loss.
At the propagation distance $z=z_{s}>0$, one can perform shifting the current frequency $\beta=-10$ by $\Delta \beta =20$, then the solitons will turn to experience the linear gain at the new central frequency of $\beta_{n}=10$. By simulations in section {\ref{Num}},
it is demonstrated that the propagation of a sequence of solitons described by Eq. (\ref{App_gl1}) is stable under performing the frequency shifting, even with periodic shifting frequency multiple times as the use of $\hat g(\omega )$ in Fig. \ref{fig5}(b). 

%%Fig. 4 example for \hat{g}(\omega) vs. \omega
%fig4

In the rest of the current section, we theoretically derive the ODE model describing the amplitude dynamics of a single soliton and of a soliton sequence in the presence of frequency dependent linear gain-loss in Eq. (\ref{App_gl1}). This ODE can be used to verify the soliton dynamics in section \ref{Num}.

First, we derive the equation for the amplitude dynamics of a single soliton described by Eq. (\ref{App_gl1}) with the initial condition as in Eq. (\ref{IC1}).
It is useful to denote by $z^{*}$ the complex conjugate of a complex number $z$. From Eq. (\ref{App_gl1}), by deriving the energy balance, that is, by simplifying $[\psi^{*}$Eq. (\ref{App_gl1}) $- \psi$ Eq. \ref{App_gl1}$]$ and integrating over $z$, it implies
\begin{eqnarray} &&
\partial_{z}\left[\int_{-\infty}^{\infty} |\psi(t,z)|^{2}dt \right] = I_{1} + I_{2},
\label{App_gl5}  
\end{eqnarray}
where
%\begin{eqnarray} &&
\[I_{1} = \frac{1}{2} \int_{-\infty}^{\infty} \psi^{*}(t,z)
\mathcal{F}^{-1}\left[\hat g(\omega)\hat \psi(\omega,z)\right]dt,
\]
%\label{App_gl6}  
%\end{eqnarray}
and
%\begin{eqnarray} &&
\[
I_{2} = \frac{1}{2} \int_{-\infty}^{\infty} \psi(t,z)
\mathcal{F}^{-1}\left[\hat g^{*}(\omega)\hat \psi^{*}(\omega,z)\right]dt.
\]
%\label{App_gl7}  
%\end{eqnarray}
%In order to solve Eq. (\ref{App_gl5}, one can employ the adiabatic perturbation method, which was developed by Kaup \cite{Kaup76, Kaup76B}, and has been extensively used to study soliton dynamics, see, for example, Refs. \cite{CP2005, PNC2010, NPT2015, Hasegawa95}.
In order to solve Eq. (\ref{App_gl5}), by the adiabatic perturbation method,  
one can determine the effects of small dissipations on the evolution of parameters by the exactly solvable NLS \cite{Hasegawa95},
that is, by using Eq. ({\ref{single2}}) for the soliton part. This calculation yields
\begin{eqnarray} &&
2\frac{d\eta}{dz} = I_{1} + I_{2},
\label{App_gl14b}  
\end{eqnarray}
where $I_1$ and $I_2$, by the convolution theorem for the inverse FT, are simplified by
%\begin{eqnarray} &&
\[I_{1} = \frac{\eta^{2}}{2(2\pi)^{1/2}} \int_{-\infty}^{\infty} \frac{dt}{\cosh\left[\eta(t-y)\right] }
\int_{-\infty}^{\infty} \frac{g(s)e^{-i\beta s} ds} {\cosh\left[\eta(t-s-y)\right]},
\]%\label{App_gl9}  
%\end{eqnarray}
and 
%\begin{eqnarray} &&
\[
I_{2} = \frac{\eta^{2}}{2(2\pi)^{1/2}} \int_{-\infty}^{\infty} \frac{dt}{\cosh\left[\eta(t-y)\right] }
\int_{-\infty}^{\infty} \frac{g^{*}(s) e^{i\beta s} ds} {\cosh\left[\eta(t-s-y)\right]}.
\]%\label{App_gl10}  
%\end{eqnarray}
From Eq. (\ref{App_gl14b}), it arrives at
\begin{eqnarray} &&
\frac{d\eta}{dz} = \frac{\eta} {4(2\pi)^{1/2}}
\int_{-\infty}^{\infty} \frac{dv}{\cosh(v) }
\int_{-\infty}^{\infty} \frac{g(s)e^{-i\beta s} + g^{*}(s) e^{i\beta s}}
{\cosh(v - \eta s)} ds,
\label{App_gl12}  
\end{eqnarray}
where $v=\eta (t - y)$. Equation (\ref{App_gl12}) describes the amplitude dynamics of a single soliton of Eq. (\ref{App_gl1}). 

We now simplify Eq. (\ref{App_gl12}) for a specific form of $\hat g(\omega)$.
For simple, it is useful to consider $\kappa=1$ and $\beta(0)=-\Delta \beta/2$. Then $\hat g(\omega)$ has the form
\begin{eqnarray} &&
\hat g(\omega) =  - g_{L}
+ 0.5(-g_{0} + g_{L})\left\{ \tanh \left[\rho(\omega  + b) \right] - \tanh \left[\rho (\omega  + a) \right] \right\}
\nonumber \\ &&
\,\,\,\,\,\,\,\,\,\,\,\,\,\,\,\,\,\,+ 0.5(g_{0} + g_{L})\left\{ \tanh \left[\rho(\omega - a) \right] - \tanh \left[\rho (\omega  - b) \right] \right\},
\label{App_gl12b}
\end{eqnarray}   
where $a=(\Delta\beta - W)/2$ and $b=(\Delta\beta + W)/2$ (see, for example, Fig. \ref{fig5} (a)).
Noting that in the limit as $\rho \gg 1$, $\hat g(\omega)$ can be approximated by the following function:
\begin{eqnarray} &&
\hat g(\omega ) = \left\{ \begin{gathered}
-g_{0}, \text{if} \, -\Delta\beta/2 - W/2 < \omega < -\Delta\beta/2 + W/2, \hfill \\
\,\,\,\,g_{0}, \text{if} \,\,\,\, \Delta\beta/2 - W/2 < \omega < \Delta\beta/2 + W/2, \hfill \\
-g_{L}, \text{elsewhere}. \hfill \\ 
\end{gathered}  \right.
\label{App_gl12c}
\end{eqnarray}
Without loss of generality, it is assumed that the current frequency of the soliton is $\beta \simeq \beta(0) = -\Delta \beta/2$. That is, the soliton is currently experienced the linear loss before shifting the frequency, and thus the frequency of the soliton after shifting the frequency is $\beta + \Delta \beta  = \Delta \beta/2$.
From Eq. (\ref{App_gl12c}), by calculating $\mathcal{F}^{-1}(\hat g(\omega ))$, it yields 
\begin{eqnarray} &&
g(t) =  - (2\pi )^{1/2}{g_L}\delta (t)
+ (2/\pi )^{1/2} (g_{L} - g_{0}) e^{ - i\Delta \beta t/2 } \sin (W t/2) / t
\nonumber \\&&
\,\,\,\,\,\,\,\,\,\,\,\,\,\,\,\,\,\,+ (2/\pi )^{1/2} (g_{L} + g_{0}) e^{ i\Delta \beta t/2 } \sin (W t/2) / t,
\label{App_gl14}
\end{eqnarray}
where $\delta (t)$ is the Dirac delta function. Substituting Eq. (\ref{App_gl14}) into Eq. (\ref{App_gl12}) and simplifying, the ODE for the amplitude dynamics is then given by:
\begin{eqnarray} &&
\frac{d\eta} {dz} =
 - \frac{g_{L}\eta}{2}J_{1}
 + \frac{(g_{L} - g_{0})\eta} {2\pi} J_{2}
 + \frac{(g_{L} + g_{0})\eta} {2\pi} J_{3}, 
\label{App_gl15}
\end{eqnarray}
where
%\begin{eqnarray} &&
\[
J_{1}=\int_{-\infty}^{\infty} \frac{dv}{\cosh(v) }
\int_{-\infty}^{\infty} \frac{\delta(s)\cos(\beta s)} {\cosh(v - \eta s)} ds
=\int_{-\infty}^{\infty} \frac{dv}{\cosh^{2}(v) } = 2,
\]
%\label{App_gl16}  
%\end{eqnarray}
%%%%%%%%%%%%%%%
%\begin{eqnarray} &&
\[
J_{2} = \int_{-\infty}^{\infty} \frac{dv}{\cosh(v) }
\int_{-\infty}^{\infty} \frac{\sin(W s/2)ds}
{\cosh (v-\eta s) s}
=\int_{-\infty}^{\infty} \frac{\sin(W s/2) ds}{ s}
\int_{-\infty}^{\infty} \frac{dv}{\cosh(v) \cosh (v-\eta s) },
\]
%\label{App_gl17}  
%\end{eqnarray}
and
%\begin{eqnarray} &&
\[
J_{3} = \int_{-\infty}^{\infty} \frac{dv}{\cosh(v) }
\int_{-\infty}^{\infty} \frac{ \cos(\Delta \beta s) \sin(W s/2) }
{\cosh(v - \eta s) s} ds.
\]
%\label{App_gl18}  
%\end{eqnarray}
%We now calculate $J_2$ and $J_3$. We note that
By using identity 2.444(1) and 3.981(1) in \cite{Grad_Ryz}, it yields
%\begin{eqnarray} &&
\[J_{2} = 2\eta \int_{-\infty}^{\infty} \frac{\sin(W s/2)} {\sinh(\eta s) } ds
= 2\pi \tanh \left(\frac{\pi W}{4\eta}\right).
\]%\label{App_gl21}  
%\end{eqnarray}
On the other hand, by using identity 2.444(1) in \cite{Grad_Ryz}, one can arrive at
%\begin{eqnarray} &&
\[
J_{3} = 2\eta \int_{-\infty}^{\infty} \frac{ \cos(\Delta \beta s) \sin(W s/2) } {\sinh(\eta s)} ds.
\]%\label{App_gl23}  
%\end{eqnarray}
It implies
%\begin{eqnarray} &&
\[
J_{3} = 2\eta \int_{0}^{\infty} \frac{\sin\left[(\Delta \beta + W/2) s \right]}{\sinh(\eta s)} ds
-2\eta \int_{0}^{\infty} \frac{\sin\left[(\Delta \beta - W/2) s \right]}{\sinh(\eta s)} ds.
\]%\label{App_gl24}  
%\end{eqnarray}
By using identity 3.981(1) in \cite{Grad_Ryz}, one can get
%\begin{eqnarray} &&
\[
J_{3} = \pi \left[\tanh \left(\frac{\pi \left( 2\Delta \beta  + W \right)} {4\eta} \right)
 - \tanh \left(\frac{\pi \left(2\Delta \beta  - W \right)} {4\eta} \right) \right].
\]%\label{App_gl25}
%\end{eqnarray}
Substituting $J_{1}$, $J_{2}$, and $J_{3}$ into Eq. (\ref{App_gl15}), the ODE for the amplitude dynamics is:
\begin{eqnarray} &&
\frac{d\eta}{dz}= 
 - g_{L} \eta
 + (-g_{0} + g_{L}) \eta \tanh \left(\frac{\pi W} {4\eta } \right) 
\nonumber \\&&
+ \frac{\eta}{2}(g_{0} + g_{L})
\left[\tanh \left(\frac{\pi \left( 2\Delta \beta  + W \right)} {4\eta} \right)
 - \tanh \left(\frac{\pi \left(2\Delta \beta  - W \right)} {4\eta} \right) \right].
\label{App_gl26}
\end{eqnarray}
This approximate ODE model describes the amplitude dynamics of a single soliton experiencing the linear loss before shifting the frequency. If $2\Delta \beta  + W \gg 1$ and $2\Delta \beta  - W \gg 1$, then one can neglect the 3rd and 4th terms of the RHS of Eq. (\ref{App_gl26}). Therefore,
\begin{eqnarray} &&
\frac{{d\eta }}
{{dz}} =  - g_L \eta  + \left( { - g_{0} + g_L } \right)\eta \tanh \left( {\frac{{\pi W}}
{{4\eta }}} \right).
\label{App_gl27}
\end{eqnarray}
Similarly, an approximate ODE for the amplitude dynamics of a soliton experiencing the linear gain after shifting the frequency can be found as follows:
\begin{eqnarray} &&
\frac{{d\eta }}
{{dz}} =  - g_L \eta  + \left( {  g_{0} + g_L } \right)\eta \tanh \left( {\frac{{\pi W}}
{{4\eta }}} \right).
\label{App_gl28}
\end{eqnarray}

Second, we derive the equation for the amplitude dynamics of a {\it sequence} of solitons propagating in the presence of frequency dependent linear gain-loss in Eq. (\ref{App_gl1}) with the initial condition as in Eq. (\ref{IC2}). By the adiabatic perturbation method, we turn to consider the soliton sequence solution of Eq. (\ref{App_gl1}) in form of Eq. (\ref{sequence1}). The $k^{\text{th}}$ soliton of this sequence of solitons is
%\begin{eqnarray} &&
\[\psi^{(k)}_{sq}(t,z) = \eta(z)e^{i\theta(z)}
\frac{\exp\{ i\beta(z) 
\left[ t-y(z)-kT \right] \}}
{\cosh\{\eta(z)\left[t-y(z)-kT\right]\}}.
\]%\label{App_gl29}
%\end{eqnarray} 
Similarly to the derivation of the amplitude dynamics for a single soliton, one can obtain an approximate ODE for the amplitude dynamics of the $k^{\text{th}}$ soliton experiencing the linear loss as in Eq. (\ref{App_gl27}) and experiencing the linear gain as in Eq. (\ref{App_gl28}). Therefore, an approximate ODE model for the amplitude dynamics of a sequence of solitons experiencing the linear loss and experiencing the linear gain can be obtained in form of Eq. (\ref{App_gl27}) and Eq. (\ref{App_gl28}), respectively. The ODE model for amplitude dynamics can be used to verify the robust propagation of a sequence of solitons in section \ref{Num}.

\subsection{Repeated soliton collisions in waveguides in the presence of weak cubic loss}
\label{App-coll}

In this section, we demonstrate the use of the frequency shift to enable repeated two-soliton collisions in the presence of weak cubic loss and measure the theoretical prediction for the accumulative amplitude shift. The propagation model in terms of coupled NLS equation is \cite{PNH2017}
\begin{eqnarray} &&
i\partial_{z}\psi_{j}  + \partial_{t}^{2}\psi_{j}  + 2|\psi_{j}|^{2}\psi_{j} + 4|\psi_{k}|^{2}\psi_{j}
= - i\epsilon_{3} |\psi_{j}|^{2}\psi_{j} - 2i\epsilon_{3} |\psi_{k}|^{2}\psi_{j},
\label{App_coll_2}  
\end{eqnarray}
where $\epsilon_{3}$ is the weak cubic loss coefficient ($0 < \epsilon_{3} \ll 1$), $\psi_{j}$ is the electric field's envelope for the $j^{\text{th}}$ soliton, $j=1,2$. The third and the fourth terms on the left hand side of Eq. (\ref{App_coll_2}) describe the effects of intrasequence and intersequence interaction due to Kerr nonlinearity, respectively. The first and the second terms on the right hand side of Eq. (\ref{App_coll_2}) 
describe intrasequence and intersequence interaction due to cubic loss, respectively.  
The initial conditions are considered as follows:
\begin{eqnarray} &&
\psi_{1}(t,0) = \frac{\eta_{1}(0)} {\cosh \left[\eta_{1}(0)t \right]},\,
\psi_{2}(t,0) = \frac{\eta_{2}(0)\exp{[i\beta(0)t]}} {\cosh \left[\eta_{2}(0)\left(t - y_{2}(0) \right) \right]}, 
\label{App_coll_3}
\end{eqnarray}
where $\beta(0)>0$ and $y_{2}(0)<0$ such that two solitons are initially well separated, for example, $y_{2}(0)=-20$. That is, soliton 1 is a standing wave and soliton 2 with the group velocity of $v_g=2\beta(0)>0$ is located at $y_{2}(0)<0$ on the left. Therefore, there will be a two-soliton collision at the propagation distance $z_{c_{1}} = -y_{2}(0)/[2\beta(0)]$. 

The idea of using the frequency shift for repeating soliton collisions is to change their group velocity. First, we employ the frequency shift procedure for solitons as follows.
After the $1^{\text{st}}$ collision, soliton 2 is going to the right and two solitons are then well separated at the propagation distance $z_{s_{1}} = 2z_{c_{1}}$. Now one can abruptly change their group velocity by employing the $1^{\text{st}}$ frequency shifting of the value $\Delta\beta=\beta(0)$ at the propagation distance $z_{s_{1}} $ such that
%\begin{eqnarray} &&
\[\hat \psi_{1}(\omega,z_{s_{1}}) \to \hat \psi_{n1}(\omega  - \beta (0),z_{s_{1}}),\,
\hat \psi_{2}(\omega,z_{s_{1}}) \to \hat \psi_{n2}(\omega  + \beta(0),z_{s_{1}}).\]
%\label{App_coll_6}
%\end{eqnarray}
%and
%\begin{eqnarray} &&
%\label{App_coll_7}
%\end{eqnarray}
Therefore, after the $1^{\text{st}}$ frequency shifting, soliton 1 is going to the right and going to collide with soliton 2, which currently becomes a standing wave. Let $z_{c_{j}}$ be the propagation distance of the $j^{\text{th}}$ collision. The propagation distance of the $2^{\text{nd}}$ collision thus is
%\begin{eqnarray} &&
\[
z_{c_{2}} = - 3y_{2}(0)/[2\beta(0)].
\]
%\label{App_coll_8}
%\end{eqnarray}
Similarly, one continues employing the frequency shifting of size $\beta(0)$ at $z_{s_{2}} = 2z_{s_{1}}$ such that
%\begin{eqnarray} &&
\[
\hat \psi_{1}(\omega,z_{s_{2}}) \to \hat \psi_{n1}(\omega  + \beta (0),z_{s_{2}}),
\,
\hat \psi_{2}(\omega,z_{s_{2}}) \to \hat \psi_{n2}(\omega  - \beta(0),z_{s_{2}}).
\]
%\label{App_coll_9}
%\end{eqnarray}
%and
%\begin{eqnarray} &&
%\label{App_coll_10}
%\end{eqnarray}
Repeating this process $n$ times, the $n^{\text{th}}$ collision distance is thus located at
%\begin{eqnarray} &&
\[z_{c_{n}} =  - (2n - 1) y_{2}(0)/[2\beta (0)],
\]%\label{App_coll_11}
%\end{eqnarray}
and the $n^{\text{th}}$ frequency shifting distance is
%\begin{eqnarray} &&
\[z_{s_{n}} = n z_{s_{1}}=- n y_{2}(0)/\beta (0).
\]
%\label{App_coll_12}
%\end{eqnarray}
In Figs. \ref{fig11} and \ref{fig12}, an example for repeating soliton collision by employing the frequency shift is illustrated. 

Second, we calculate the theoretical collision-induced amplitude shift. Neglecting the frequency shift due to weak perturbations, it can be assumed that $\beta(z) = \beta \simeq \beta(0)$. We recall that the dynamics of solitons in the presence of weak cubic loss is given by \cite{PNC2010}
\begin{eqnarray} &&
\eta_{j}(z) = \eta_{j}(z_{0}) \left[ 1 + 8\epsilon_{3} \eta_{j}^{2}(z_{0})(z-z_{0})/3 \right]^{-1/2},
\label{App_coll_13}
\end{eqnarray}
for $z>z_{0}$, and the collision-induced amplitude shift on the $j^{\text{th}}$ soliton at the propagation distance $z_{c}$ is \cite{PNC2010}
\begin{eqnarray} &&
\Delta \eta_{j}^{(c)} =  - 4\epsilon_{3} \eta_{j}(z_{c}) \eta_{k}(z_{c})/|\beta|.
\label{App_coll_14}
\end{eqnarray}
%%%%%%%%%%%%%%%
Thus, the $k^{\text{th}}$ collision-induced amplitude shift is calculated by
\begin{eqnarray} &&
\Delta \eta_{j}^{(c_{k})} =  - 4\epsilon_{3} \eta_{1}(z_{c_{k}}^{-}) \eta_{2}(z_{c_{k}}^{-})/|\beta|,
\label{App_coll_18}
\end{eqnarray}
where the soliton amplitude before the $k^{\text{th}}$ collision $\eta_{j}(z_{c_{k}}^{-} )$ is defined by Eq. (\ref{App_coll_13}) with $\eta_{j}(z_{c_{1}}^{-})$ is calculated from $\eta_{j}(0)$ on
$[0, z_{c_{1}}^{-}]$ and $\eta_{j}(z_{c_{k}}^{-} )$ is calculated from $\eta_{j}(z_{c_{k-1}}^{+} )$ on
$[z_{c_{k-1}}^{+}, z_{c_{k}}^{-}]$ by
%\begin{eqnarray} &&
\[
\eta_{j}(z_{c_{k}}^{-} ) = \eta_{j}(z_{c_{k-1}}^{+} )
\left[ 1 + 8\epsilon_{3} \eta_{j}^{2}(z_{c_{k-1}}^{+})(z_{c_{k}}-z_{c_{k-1}})/3 \right]^{-1/2},
\]%\label{App_coll_17}
%\end{eqnarray}
for $k \geq 2$, where $\eta_{j}(z_{c_{k-1}}^{+} ) = \eta_{j}(z_{c_{k-1}}^{-} ) + \Delta \eta_{j}^{(c_{k-1})}$ is the soliton amplitude after the $(k-1)^{\text{th}}$ collision. Therefore, the total of collision-induced amplitude shift after $n$ collisions $\Delta \eta_{j} ^{(c)(th)} $ can be theoretically measured as follows:
\begin{eqnarray} &&
\Delta \eta_{j}^{(c)(th)} =  \sum\limits_{k=1}^{n} {\Delta \eta_{j}^{(c_{k})}}.
\label{App_coll_18a}
\end{eqnarray}

Third, we describe the numerical implementation for measuring the value of the accumulative collision-induced amplitude shift in simulations. The $k^{\text{th}}$ collision-induced amplitude shift can be numerically calculated by
\begin{eqnarray} &&
\Delta \eta_{j}^{(c_{k})(num)} = \eta_{j}(z_{c_{k}}^{+} ) - \eta_{j}(z_{c_{k}}^{-} ),
\label{App_coll_19}
\end{eqnarray}
where $\eta_{j}(z_{c_{k}}^{-} )$ and $\eta_{j}(z_{c_{k}}^{+} )$ are measured from the simulation of Eq. (\ref{App_coll_2}) and from Eq. (\ref{App_coll_13}).
% on $[0,z^{-}_{c_{1}}]$ or $[z{s_{k-1}}, z^{-}_{c_{k}}]$ for $k \ge 2$ and $[z^{+}_{c_{k}}, z{s_{k}}]$, respectively
That is, $\eta_{j}(z_{c_{1}}^{-} )$ is calculated from $\eta_{j}(0)$ on $[0,z^{-}_{c_{1}}]$, $\eta_{j}(z_{c_{k}}^{-} )$ is calculated from $\eta_{j}(z_{s_{k-1}})$ for $k \ge 2$ on $[z_{s_{k-1}}, z^{-}_{c_{k}}]$ by 
%\begin{eqnarray} &&
\[
\eta_{j}(z_{c_{k}}^{-} )= \eta_{j}(z_{s_{k-1}})
\left[ 1 + 8\epsilon_{3} \eta_{j}^{2}(z_{s_{k-1}})( z_{c_{k}}- z_{s_{k - 1}})/3 \right]^{-1/2},
\]%\label{App_coll_20}
%\end{eqnarray}
and $\eta_{j}(z_{c_{k}}^{+} )$ is calculated from $\eta_{j}(z_{s_{k}})$ for $k \ge 1$ on $[z^{+}_{c_{k}}, z_{s_{k}}]$ by  
%\begin{eqnarray} &&
\[\eta_{j}(z_{c_{k}}^{+}) = \eta_{j}(z_{s_{k}})
\left[ 1 - 8\epsilon_{3} \eta_{j}^{2}(z_{s_{k}})( z_{s_{k}}- z_{c_{k}})/3 \right]^{-1/2},
\]%\label{App_coll_21}
%\end{eqnarray}
where $ \eta_{j}(z_{s_{k}})$ is measured from the numerical simulation of Eq. (\ref{App_coll_2}). Thus, the numerical value of the accumulative collision-induced amplitude shift after $n$ collisions $\Delta \eta_{j} ^{(c)(num)} $ is calculated as follows:
\begin{eqnarray} &&
\Delta \eta_{j}^{(c)(num)} =  \sum\limits_{k=1}^{n} {\Delta \eta_{j}^{(c_{k})(num)}}.
\label{App_coll_22}
\end{eqnarray}

\section{Numerical simulations}
\label{Num}

In this section, the simulation results for frequency shifting procedures of a single soliton and of a sequence of solitons and their applications are presented.
The numerical simulations are implemented in a large time domain with initial condition as in Eq.(\ref{IC1}) for a single soliton and in a closed waveguide loop with initial condition as in Eq.(\ref{IC2}) for a sequence of solitons. 
The NLS equation Eq. (\ref{single1}), Eq. (\ref{App_gl1}), and Eq. (\ref{App_coll_2}) are numerically solved by using the split-step Fourier method (see Appendix \ref{App_C}).

\subsection{ Numerical simulations for frequency shifting procedures} 

\begin{figure}[ptb]
\begin{tabular}{cc}
\epsfxsize=6.8cm  \epsffile{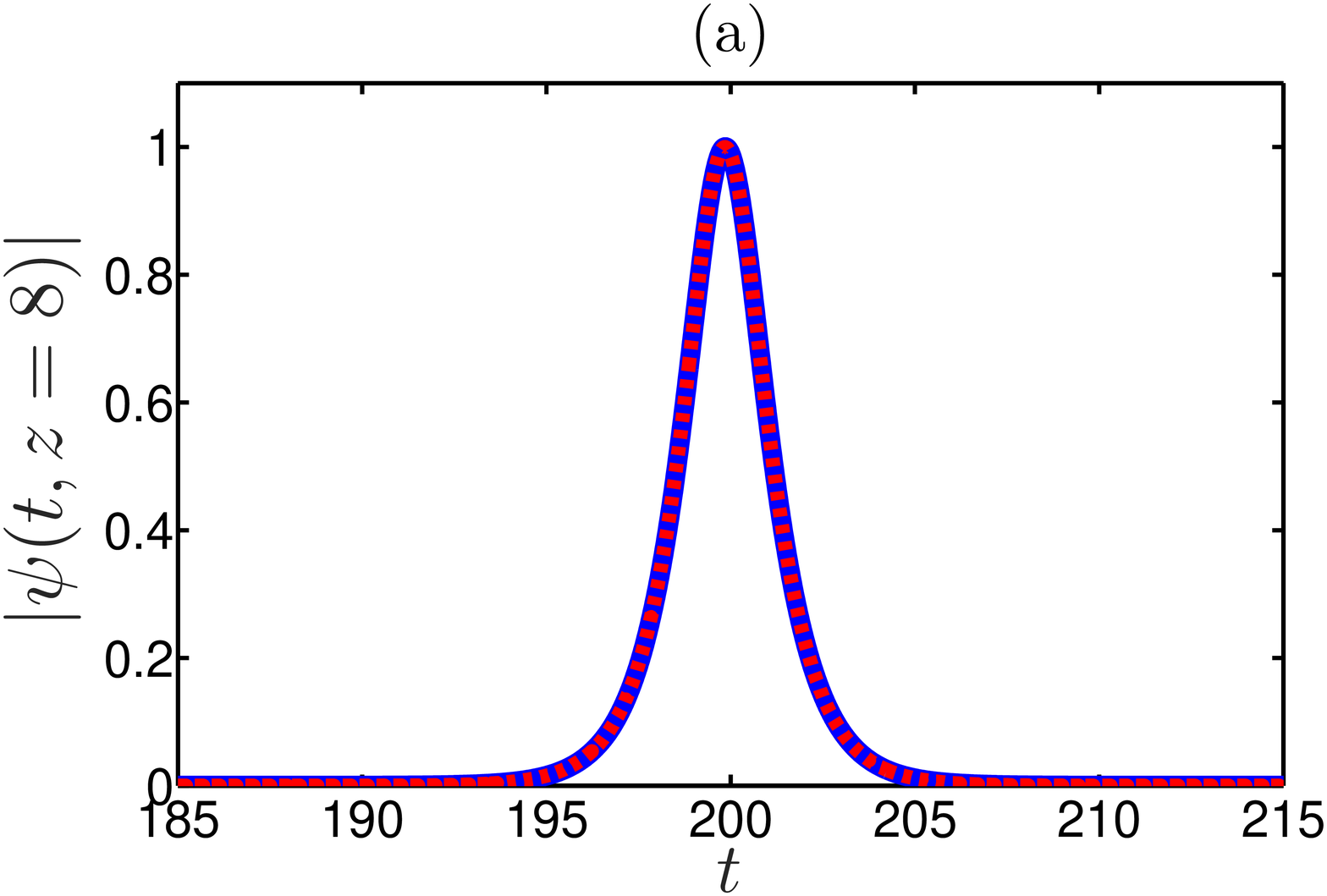} &
\epsfxsize=6.8cm  \epsffile{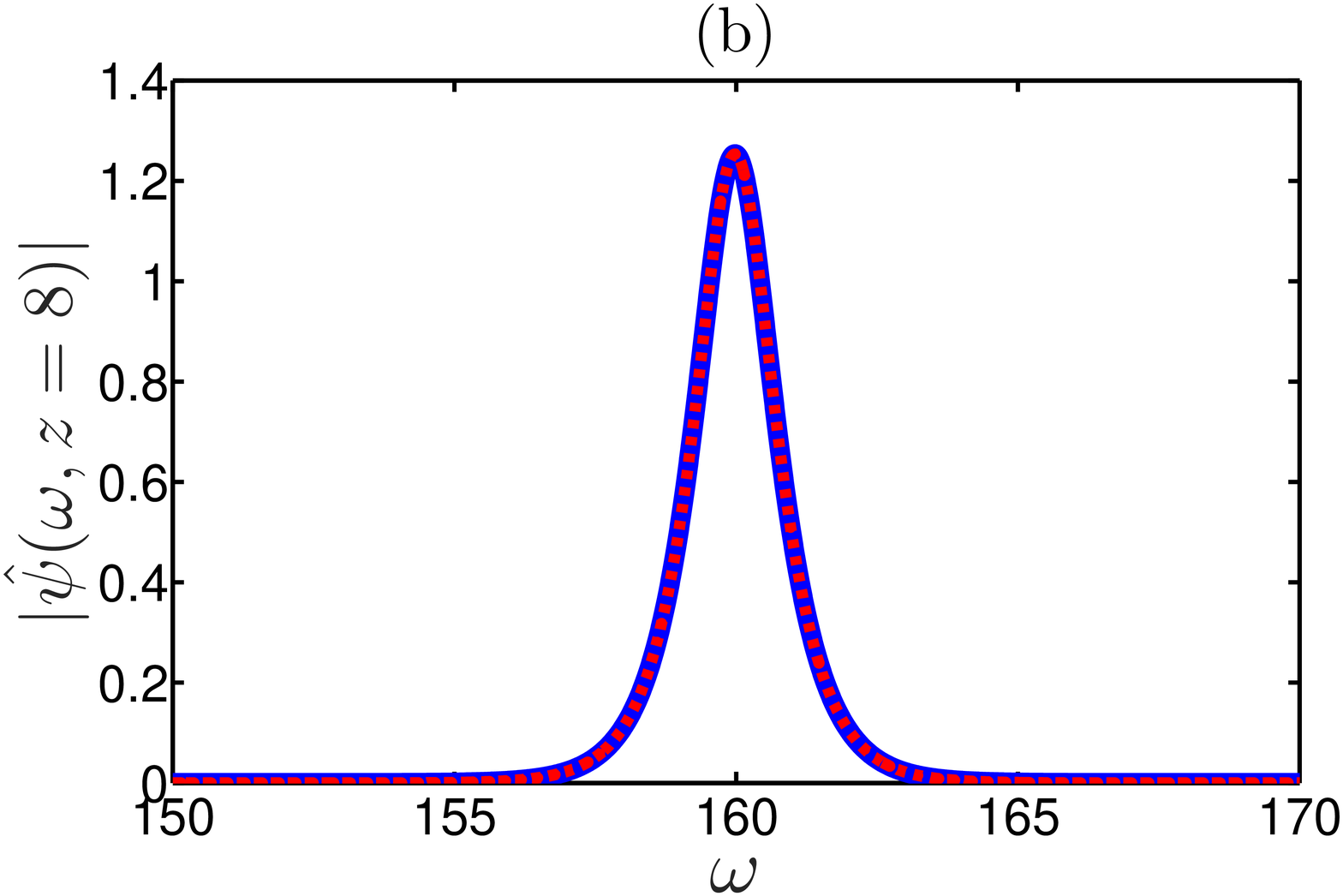} \\
\end{tabular}
\caption{(Color online) Multiple frequency shifting for a single soliton in a large time domain. The pulse patterns are in the time domain (a) and in the frequency domain (b) at the distance $z=8$ after performing 16 shifts of the frequency with parameters in {\bf setup 2}.
The red dashed and blue solid curves correspond to $\left| {{\psi}(t,z)} \right|$ (a) or $\left| {{\hat\psi}(\omega,z)} \right|$ (b) obtained by the numerical simulation with Eq. (\ref{single1}) and its theoretical prediction, respectively.
}
%(a) The red dashed and blue solid curves correspond to $\left| {{\psi}(t,z)} \right|$ obtained by the numerical simulation with Eq. (\ref{single1}) and its theoretical prediction, respectively.
% (b) The red dashed and and blue solid curves represent  $\left| {{\hat\psi}(\omega,z)} \right|$ measured from simulation and its theoretical prediction, respectively.
\label{fig3}
\end{figure}

We first present the numerical simulation with Eq. (\ref{single1}) for the frequency shifting procedure of a single soliton and then compare the numerical results with the theoretical predictions. It is useful to consider the numerical {\bf setup 2} with following parameters: $\Delta \beta  = 20$, $\beta(0) =  - 160,\,y(0) = 360,\,\eta(0) = 1,\alpha(0)=0,\,\Delta z = 0.0001,\,\Delta t = 0.018,\,t_{\max} = 380,\,t_{\min}=-t_{\max}$, and $z_{s} = 0.5$. One then has the numerical measurements: $\Delta \omega = 0.00826735,\,\Delta\beta^{(num)}=19.9987$. The $k^{\text{th}}$ frequency shift is implemented at the propagation distances $z_{s_{k}}=kz_{s}=0.5k$, where $1 \leq k \leq 16$. Figures \ref{fig3}(a) and \ref{fig3}(b) represent the pulse patterns in the time domain and in the frequency domain, respectively, after the 16$^{\text{th}}$ frequency shifting ($k=16$).
The relative errors in measuring the soliton patterns $|\psi_{s1}(t,z)|$ and $|\hat \psi_{s1}(\omega,z)|$ at $z=z_{f}=8$ are $5.2\times 10^{-3}$ and $6.32\times 10^{-7}$, respectively.
The numerical measurement for the frequency of the soliton is $\beta^{(num)}(z=8)=158.979$.

%fig 5

%%Simulations for a sequence

\begin{figure}[ptb]
\begin{tabular}{cc}
\epsfxsize=6.8cm  \epsffile{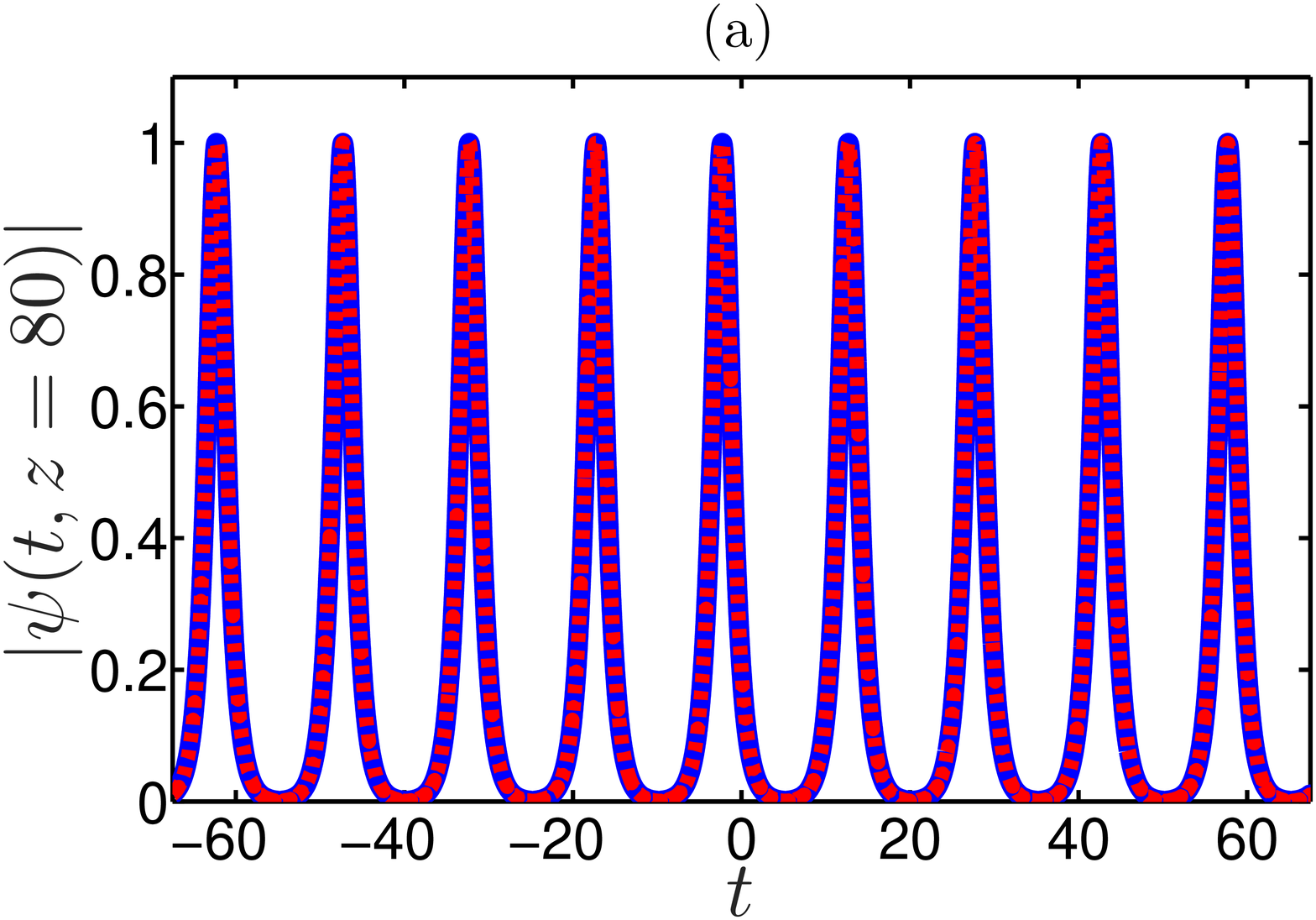} &
\epsfxsize=6.8cm  \epsffile{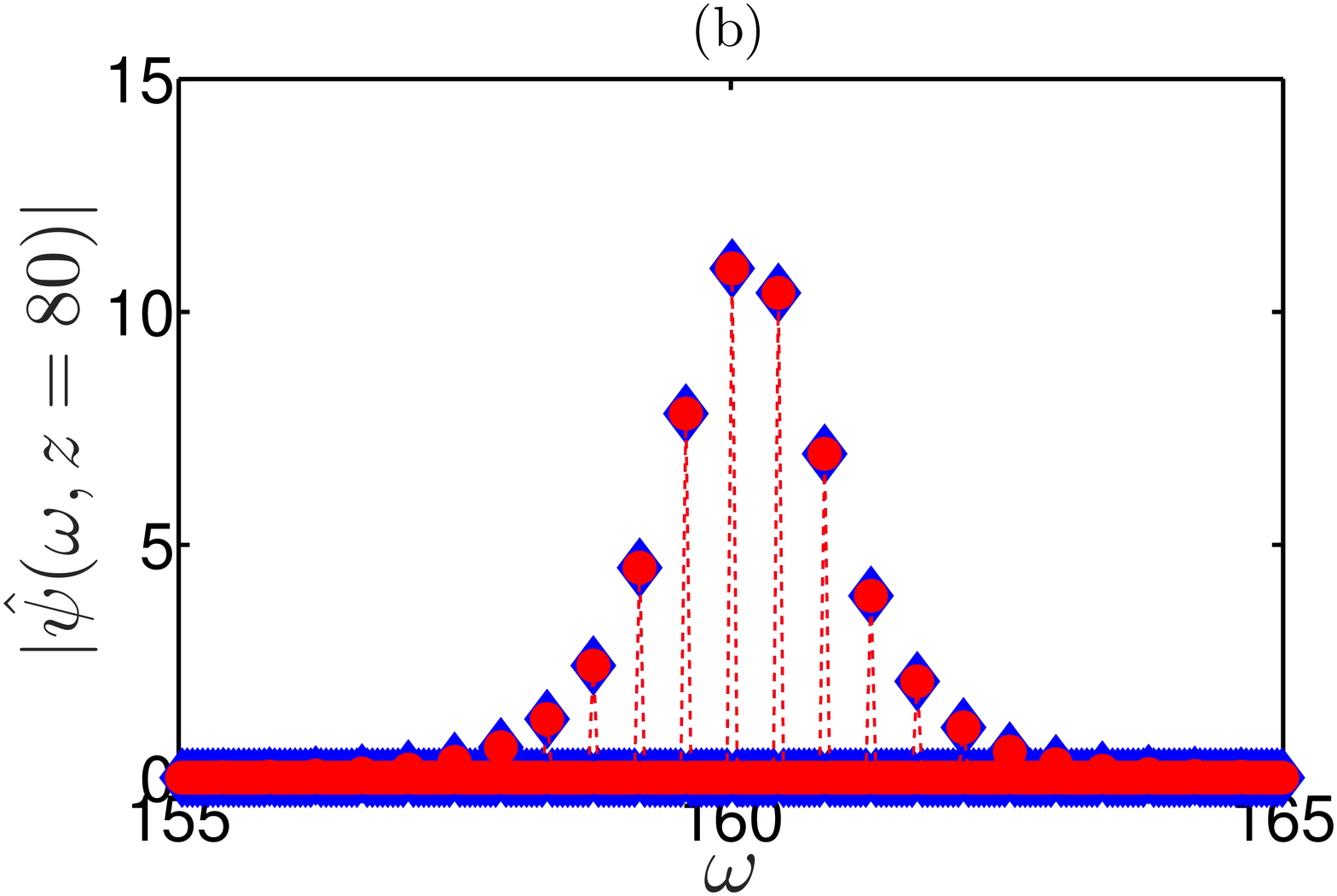} \\
\end{tabular}
\caption{(Color online) Multiple frequency shifting for a sequence of solitons by the decomposition method. The soliton patterns are in the time domain (a) and in the frequency domain (b) at the distance $z=80$ after performing 16 shifts of the frequency with parameters in {\bf setup 3}. (a) The red dashed and blue solid curves correspond to $|\psi(t,z)|$ obtained by the simulation with Eq. (\ref{single1}) and its theoretical prediction. (b) The red circles and blue diamonds represent $|\hat\psi(\omega,z)|$ obtained by the simulation and its theoretical prediction.
}
 \label{fig4}
\end{figure}

Second, we present the simulation with Eq. (\ref{single1}) for the frequency shifting procedure of a sequence of solitons with the decomposition method and then compare the numerical results with the theoretical predictions.
%Here the simulation for the propagation of a soliton sequence in a closed waveguide loop with the initial condition in the form of a periodic sequence of $2J+1$ solitons as in Eq. (\ref{IC2}) is chosen to illustrate the procedure.
The initial condition is in the form of a periodic sequence of $2J+1$ solitons as in Eq. (\ref{IC2}).
It is useful to consider the numerical {\bf setup 3} with following parameters: 
$\Delta\beta  = 20$, $L=15$, $\eta(0) = 1$, $\beta(0) =  - 160$, $y(0) =  - 5$, $\alpha(0) =  0$, $T=15$, $J=4$, $\Delta z = 0.0001,\,\Delta t = 0.018,\,t_{\max } = 67.5,\,t_{\min}=-t_{\max}$, and $z_{s} = 5$.
It then yields $\Delta\omega = 0.0465$ and $\Delta\beta^{(num)}= 20.0131$ that consists $\Delta\beta_1=94\pi/T=19.6873$ and $\Delta\beta_{2}=0.3258$. 
The $k^{\text{th}}$ frequency shifting is implemented at the propagation distances $z_{s_k}=kz_{s}=5k$, where $1 \leq k \leq 16$. Figures \ref{fig4}(a) and \ref{fig4}(b) represent the soliton patterns in the time domain and in the frequency domain, respectively, after the 16$^{\text{th}}$ frequency shifting.
The relative errors in measuring the soliton patterns $|\psi_{sq}(t,z)|$ and $|\hat \psi_{sq}(\omega,z)|$ at $z=z_{f}=80$ are $2.7\times 10^{-3}$ and $2\times 10^{-4}$, respectively.
The numerical measurement for the frequency of the soliton sequence is $\beta^{(num)}(z=80)=160.172$. 
The very good agreement between the results of the numerical simulations and the theoretical predictions after a multiple of frequency shifting confirms the robustness of the frequency shifting procedure for a sequence of solitons.

%fig 6
%Frequency shifting for sequence of solitons 

\subsection{Numerical simulations for soliton propagations with frequency dependent linear gain-loss}

%fig 7
\begin{figure}[ptb]
\begin{tabular}{cc}
\epsfxsize=7.4cm  \epsffile{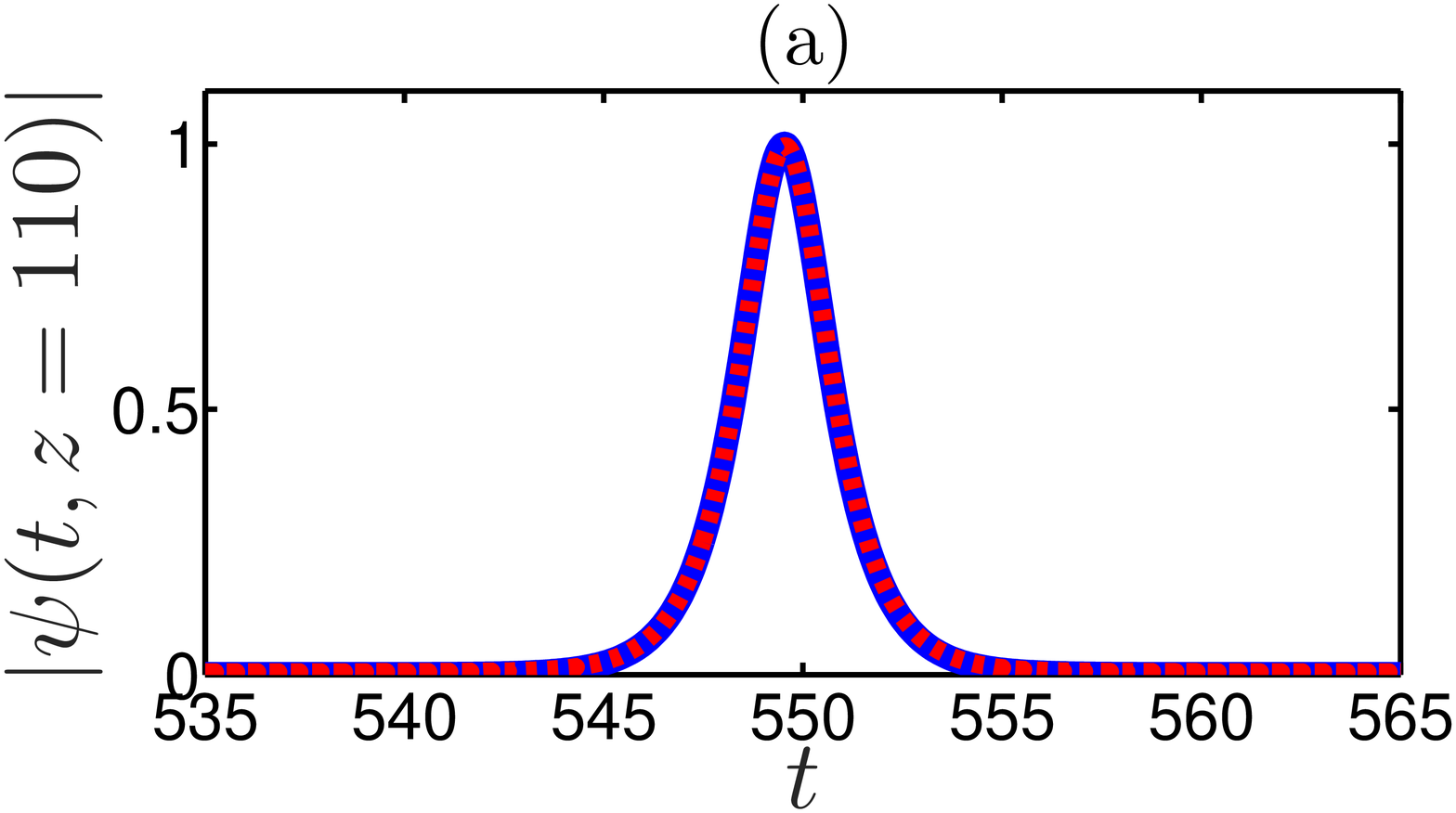} &
\epsfxsize=7.2cm  \epsffile{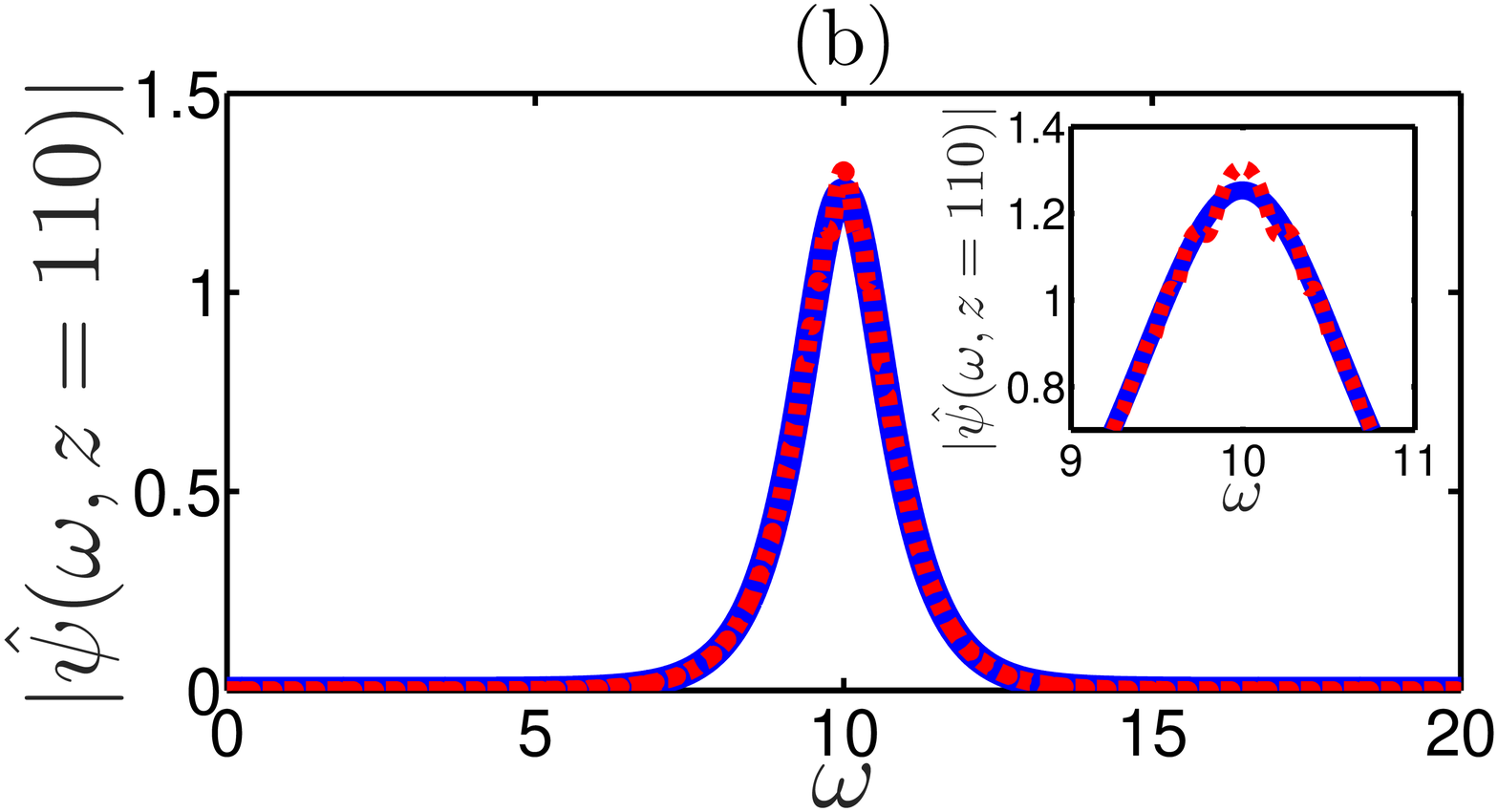} \\
\epsfxsize=7.6cm  \epsffile{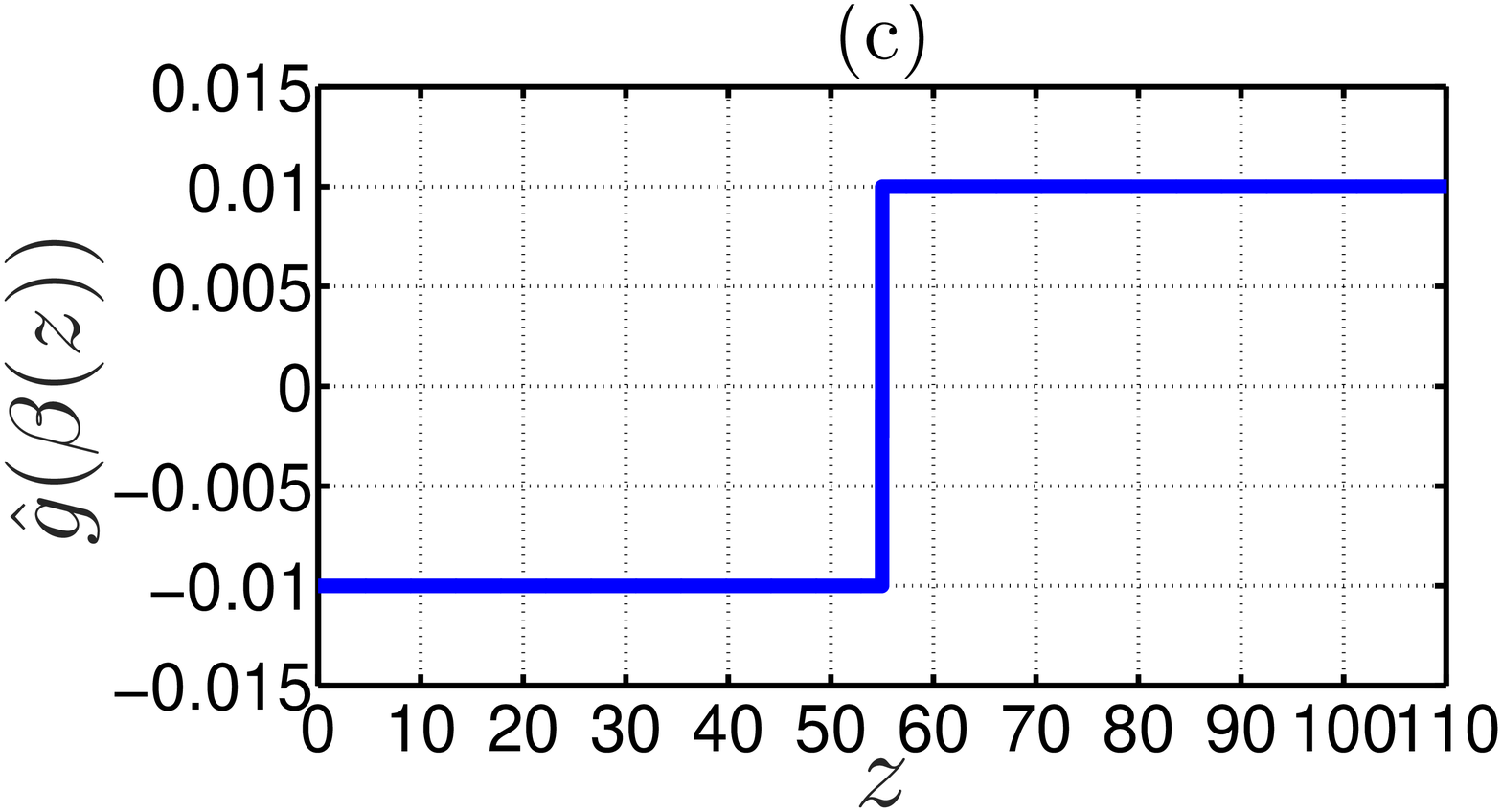}&
\epsfxsize=7.7cm  \epsffile{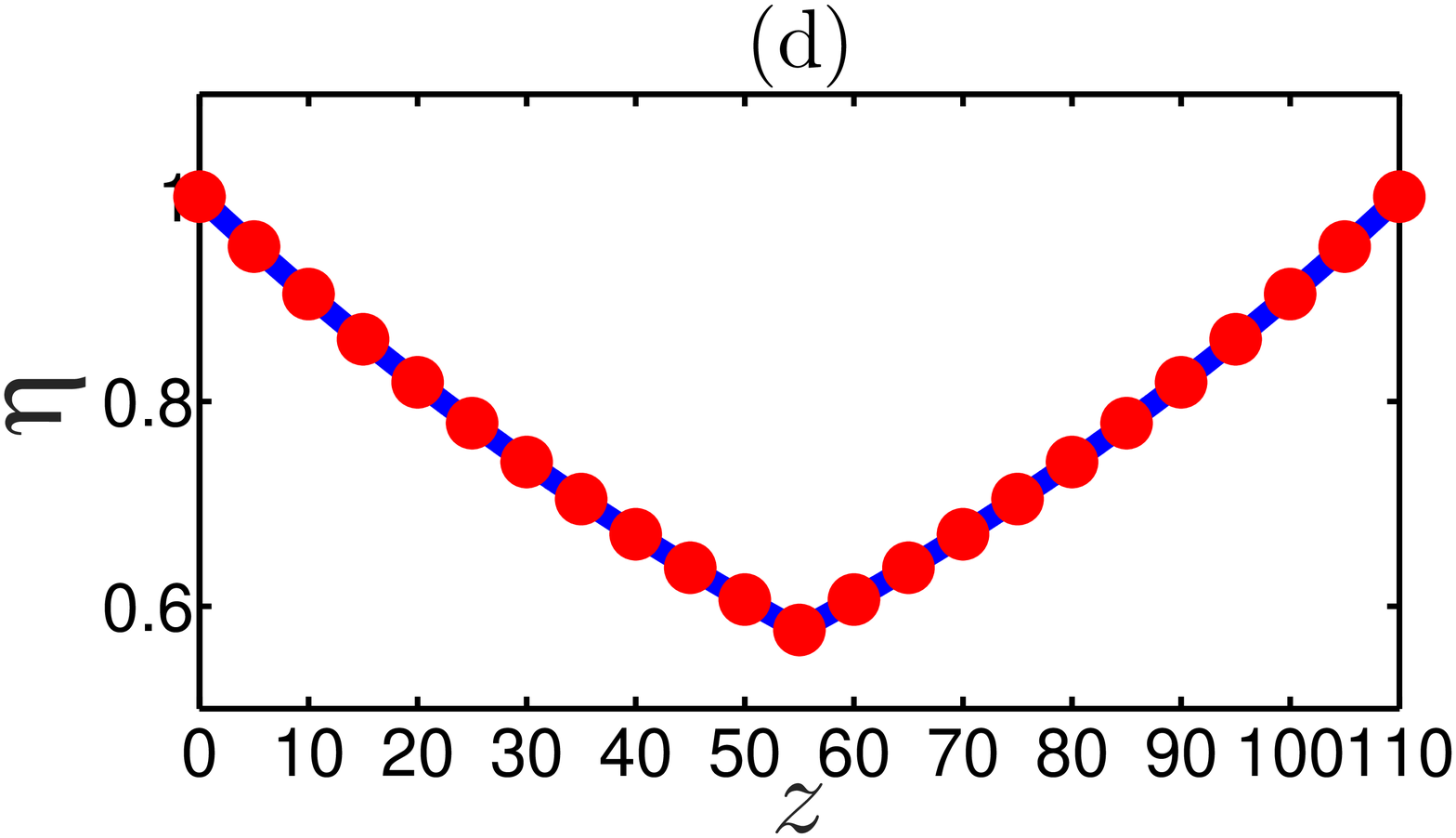} \\
\end{tabular}
\caption{(Color online) Single frequency shifting in the presence of frequency dependent linear gain-loss for a single soliton ($\kappa=1$) with parameters in {\bf setup 4}.
(a)-(b) The soliton patterns in the time domain and in the frequency domain at the final propagation distance $z_{f}=110$, respectively.
The red dashed and blue solid curves represent numerical results obtained from the simulation with Eq. (\ref{App_gl1}) and Eq. (\ref{App_gl12b})
and its theoretical prediction, respectively.
The inset at the upper right of (b) is a magnified version of the pulse pattern on the top.
(c) The $z$ dependence of $\hat g(\beta(z))$ in Eq. (\ref{App_gl12b}).
(d) The $z$ dependence of the soliton amplitude. The red circles and blue solid curve represent the amplitude $\eta(z)$ obtained by Eq. (\ref{App_gl1}) and Eq. (\ref{App_gl12b}) and its theoretical prediction obtained by Eq. (\ref{App_gl27}) and Eq. (\ref{App_gl28}).
% (a) The soliton patterns in the time domain at the final propagation distance $z=110$. The red dashed and blue solid curves represent $|\psi(t,z)|$ obtained by the simulation with Eq. (\ref{App_gl1}) and Eq. (\ref{App_gl12b}) and its theoretical prediction. (b) The soliton patterns in the frequency domain at $z=110$. The red dashed and blue solid curves represent $|\hat\psi(\omega,z)|$ obtained by the simulation and its theoretical prediction, respectively. The inset at the upper right of (b) is a magnified version of the pulse pattern on the top. (c) The $z$ dependence of $\hat g(\beta(z))$ in Eq. (\ref{App_gl12b}). (d) The $z$ dependence of the soliton amplitude. The red circles and blue solid curve represent the amplitude $\eta(z)$ obtained by Eq. (\ref{App_gl1}) and Eq. (\ref{App_gl12b}) and its theoretical prediction obtained by Eq. (\ref{App_gl27}) and Eq. (\ref{App_gl28}).
}
 \label{fig6}
\end{figure}

First, we present the simulations for the single frequency shifting ($\kappa=1$) in the presence of frequency dependent linear gain-loss for a single soliton with Eq. (\ref{App_gl1}) and Eq. (\ref{App_gl12b}). That is, the frequency shift is implemented such that the soliton will experience gain at the new frequency. It is useful to consider the numerical {\bf setup 4} with following parameters for $\hat g(\omega)$: $g_{0}=0.01$, $g_{L}=0.02$, $\Delta\beta=20$, $W=15$, and $\rho=10$. This function $\hat g(\omega)$ is presented in Fig. \ref{fig5}(a). Other parameters are as follows:
$\beta(0)=-\Delta\beta /2 = -10,\,y_{0} = 550,\,\eta_{0} = 1,\,\alpha(0)=0$,
$\Delta z = 0.001,\,\Delta t = 0.0588$,
$t_{\max} = 570,\,t_{\min}=-t_{\max}$, and $z_{s}=55$. The single frequency shifting is implemented at the propagation distance $z_s=55$ by $\Delta\beta$. Therefore, the soliton is experienced the linear loss over the propagation distance $[0, \,z_{s}]=[0, \,55]$ and experienced the linear gain over the propagation distance $[z_{s}, \,z_{f}]$.
The results are presented in Fig. \ref{fig6}.
Figures \ref{fig6}(a) and \ref{fig6}(b) show the soliton patterns in the time domain and in the frequency domain at the final propagation distance $z_{f}=110$ (end of the gain period), respectively.
The inset at the upper right corner of Fig. \ref{fig6}(b) shows a magnified version of the pulse pattern on the top.
The relative errors in measuring the soliton patterns $|\psi_{s1}(t,z)|$ and $|\hat \psi_{s1}(\omega,z)|$ at $z=z_{f}=110$ are $2.79\times 10^{-2}$ and $2.04\times 10^{-2}$, respectively.
Figure \ref{fig6}(c) presents the profile of $\hat{g}(\beta(z))$ vs. $z$ to illustrate the propagation of solitons experienced different linear gain and loss at their central frequencies, as they propagate along the waveguide, while Fig. \ref{fig6}(d) presents the amplitude dynamics $\eta(z)$.
The relative errors in calculating the amplitude dynamics, which are defined by $|\eta^{(num)}(z) - \eta^{(th)}(z)|/\eta^{(th)}(z)$, are less than $5.5\times 10^{-7}$ over the propagation distance $0 \leq z \leq z_{f}$.
We observe that there is a small instability of the pulse pattern in the frequency domain slightly arising at the final propagation distance $z=z_{f}$ with the relative error of $2.04\times 10^{-2}$ as showing in the inset at the upper right corner of
Fig. \ref{fig6}(b).
Moreover, by extensively implementing the numerical simulations for different values of $g_{0}$, it is observed that the instability of the pulse pattern at the distance $z=z_{f}$ depends on the magnitude of $g_{0}$ and the amplitude at the end of loss period $\eta(z_{s})$.
More specifically, if $0 < g_{0} \leq 0.01$, then the relative error in measuring $\left|\psi_{s1}(t,z_{f})\right|$ is less than $1.1\times 10^{-2}$ for $0.8 \leq \eta(z_{s}) \leq 1$ and less than $5.04\times 10^{-2}$ for $0.4 \leq \eta(z_{s}) < 0.8$. 
If $0.01 < g_{0} \leq 0.05$, then the relative error in measuring $\left|\psi_{s1}(t,z_{f})\right|$ is less than $5.5\times 10^{-2}$ for $0.8 \leq \eta(z_{s}) \leq 1$, less than $7.4\times 10^{-2}$ for $0.6 \leq \eta(z_{s}) < 0.8$, and less than  $23.5\times 10^{-2}$ for $0.4 \leq \eta(z_{s}) < 0.6$.

\begin{figure}[ptb]
\begin{tabular}{cc}
\epsfxsize=6.5cm  \epsffile{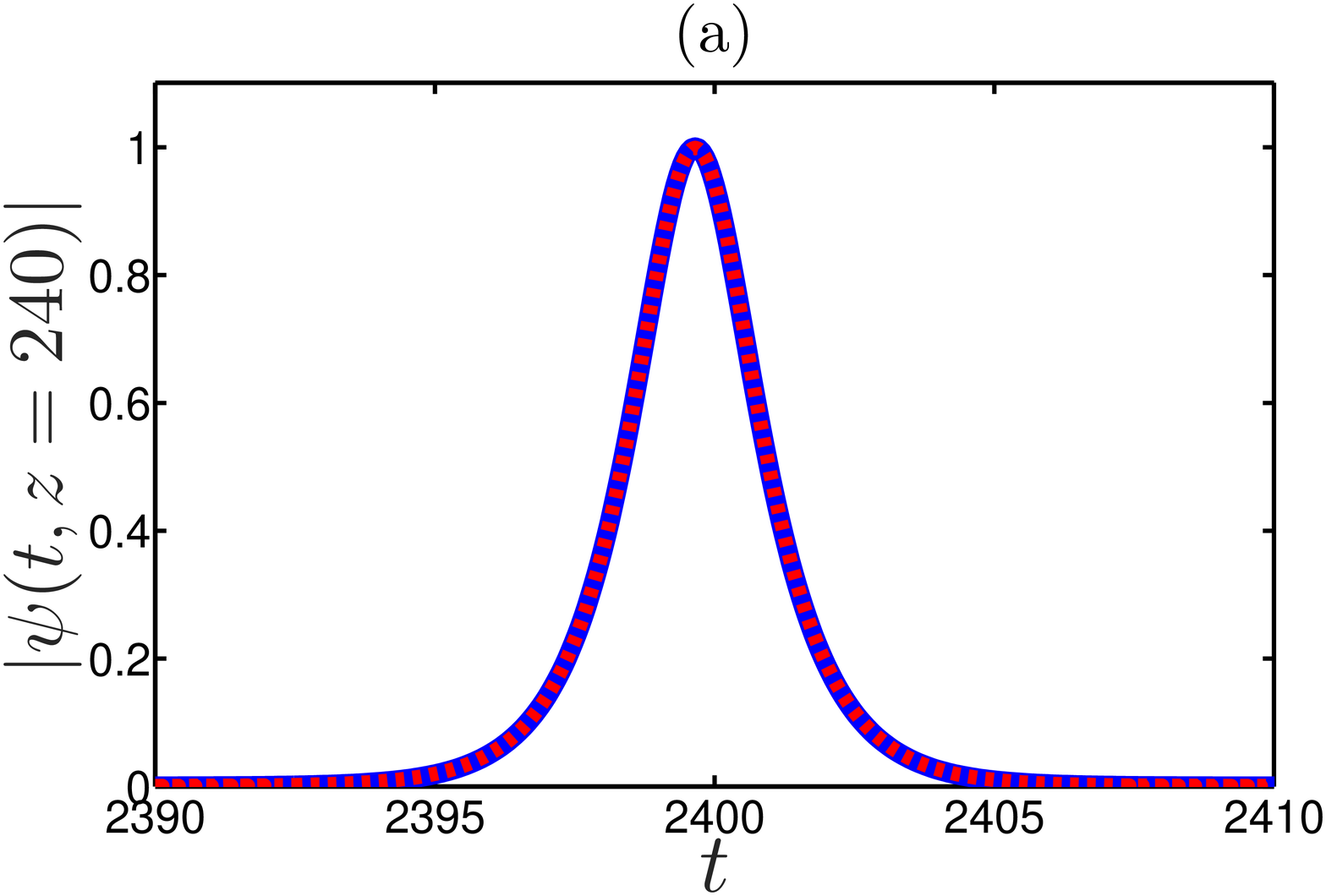} &
\epsfxsize=6.5cm  \epsffile{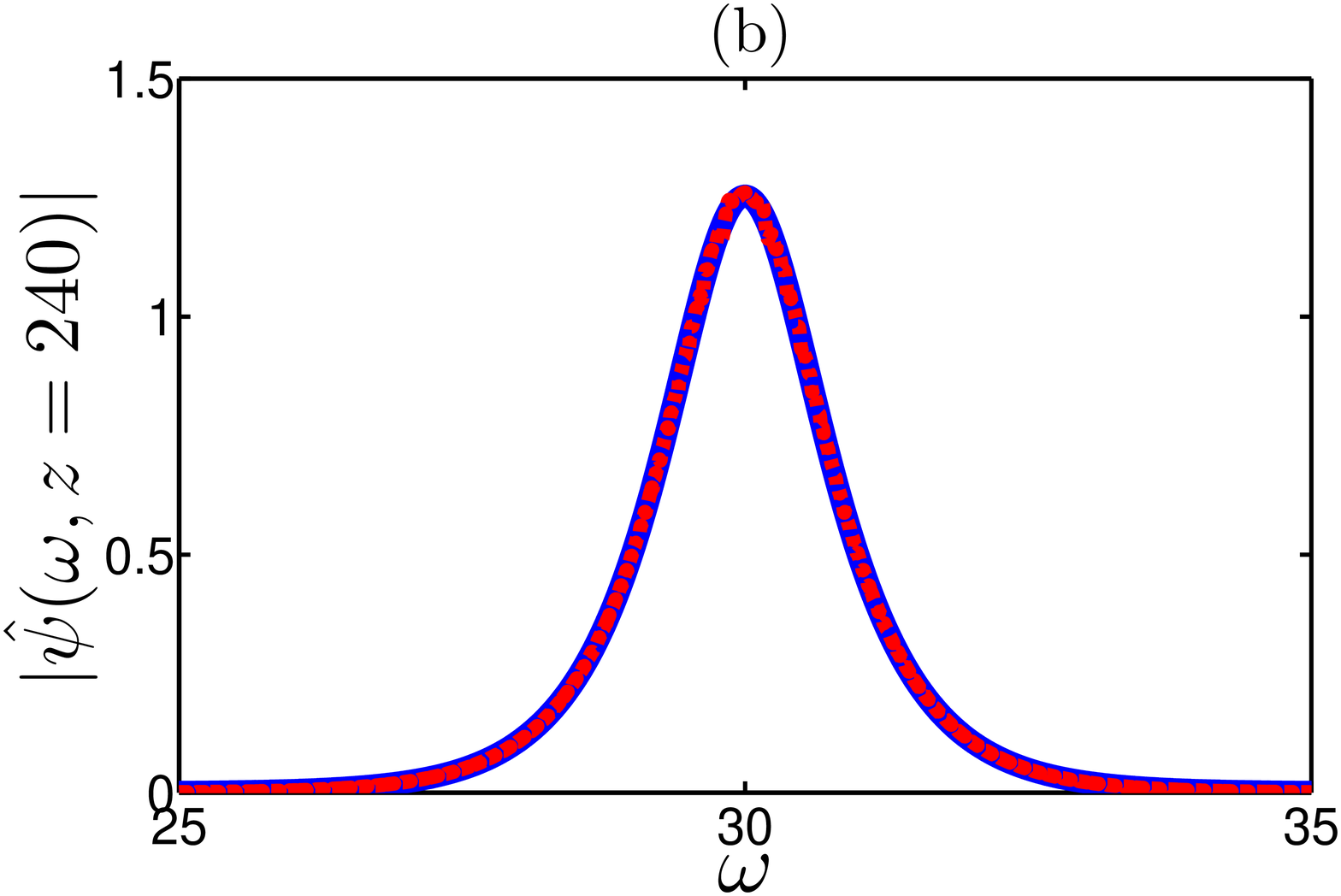} \\
\epsfxsize=6.5cm  \epsffile{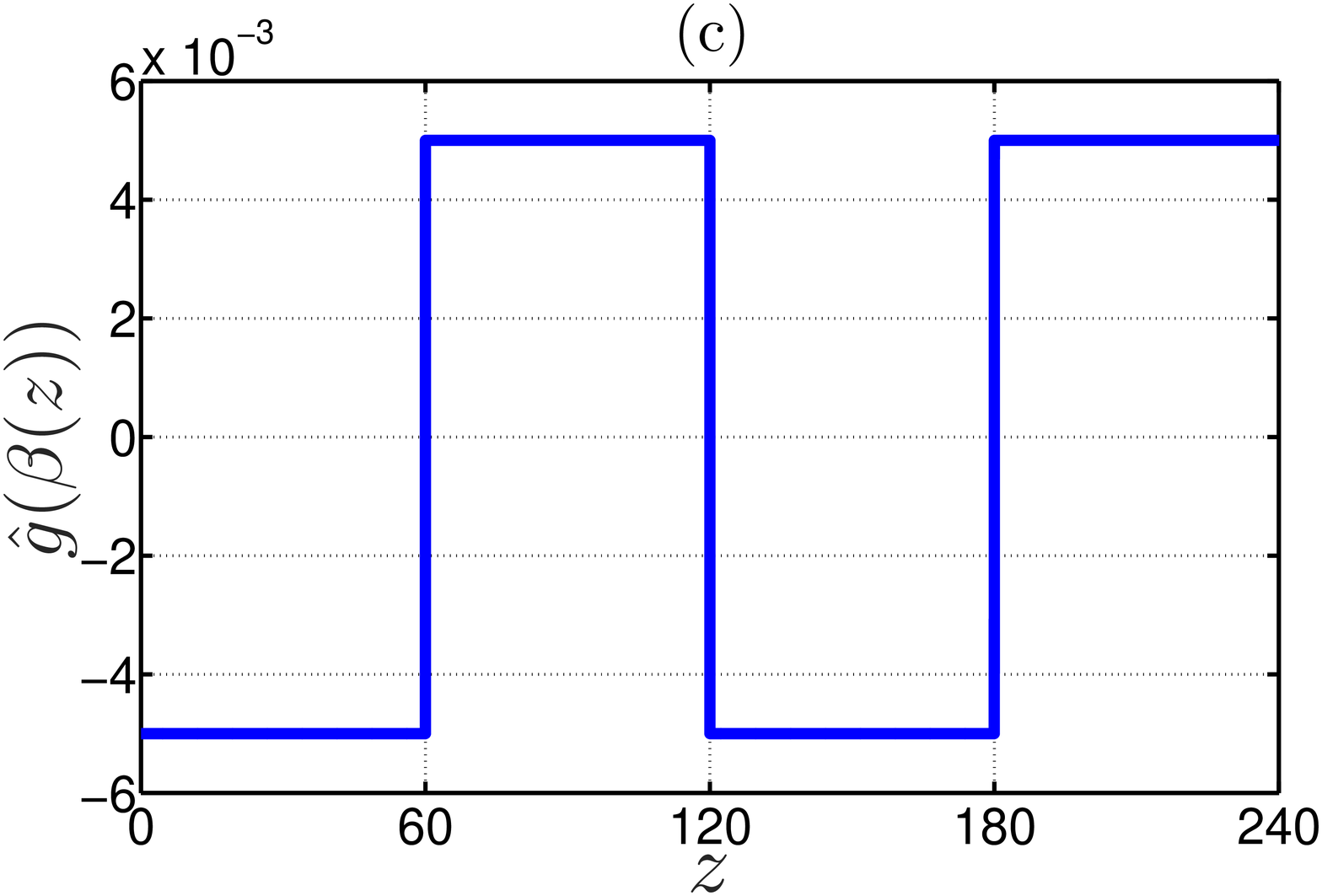}  &
\epsfxsize=6.5cm  \epsffile{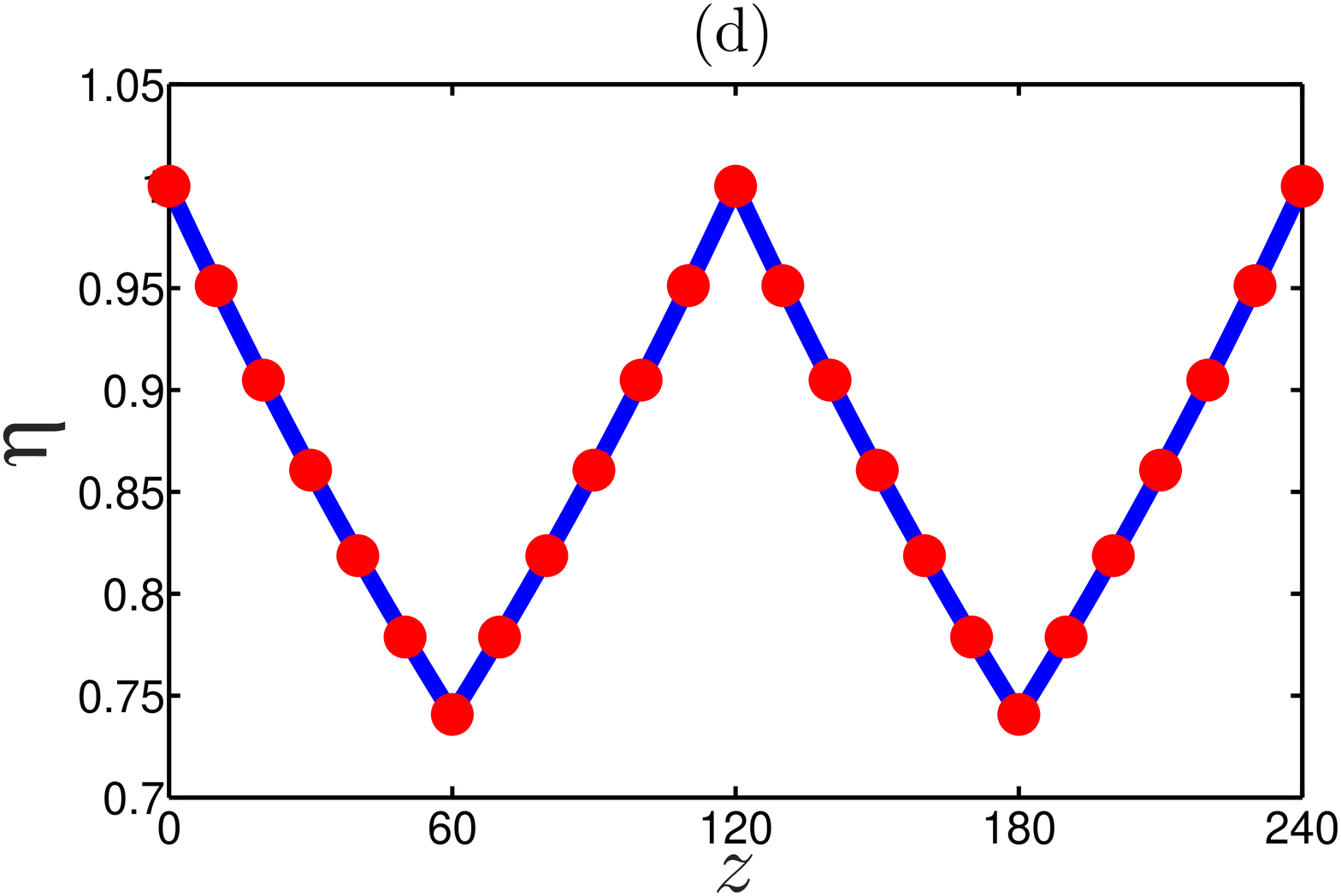} \\
\end{tabular}
\caption{(Color online) Multiple frequency shifting periodically in the presence of frequency dependent linear gain-loss for a single soliton ($\kappa=2$) with parameters in {\bf setup 5a}. (a)-(b) The soliton patterns in the time domain and in the frequency domain at the final propagation distance $z_{f}=240$, respectively. The red dashed and blue solid curves represent numerical results obtained from the simulation with Eq. (\ref{App_gl1}) and Eq. (\ref{App_gl2}) and its theoretical prediction, respectively. (c) The $z$ dependence of $\hat g(\beta(z))$ in Eq (\ref{App_gl2}). (d) The $z$ dependence of the soliton amplitude. The red circles represent the numerical soliton amplitude $\eta(z)$ measured by Eq. (\ref{App_gl1}) and Eq. (\ref{App_gl2}) while the blue solid curve corresponds to the theoretical prediction for $\eta(z)$ measured by Eq. (\ref{App_gl27}) and Eq. (\ref{App_gl28}).
}
 \label{fig9}
\end{figure}

\begin{figure}[ptb]
\begin{tabular}{cc}
\epsfxsize=7.0cm  \epsffile{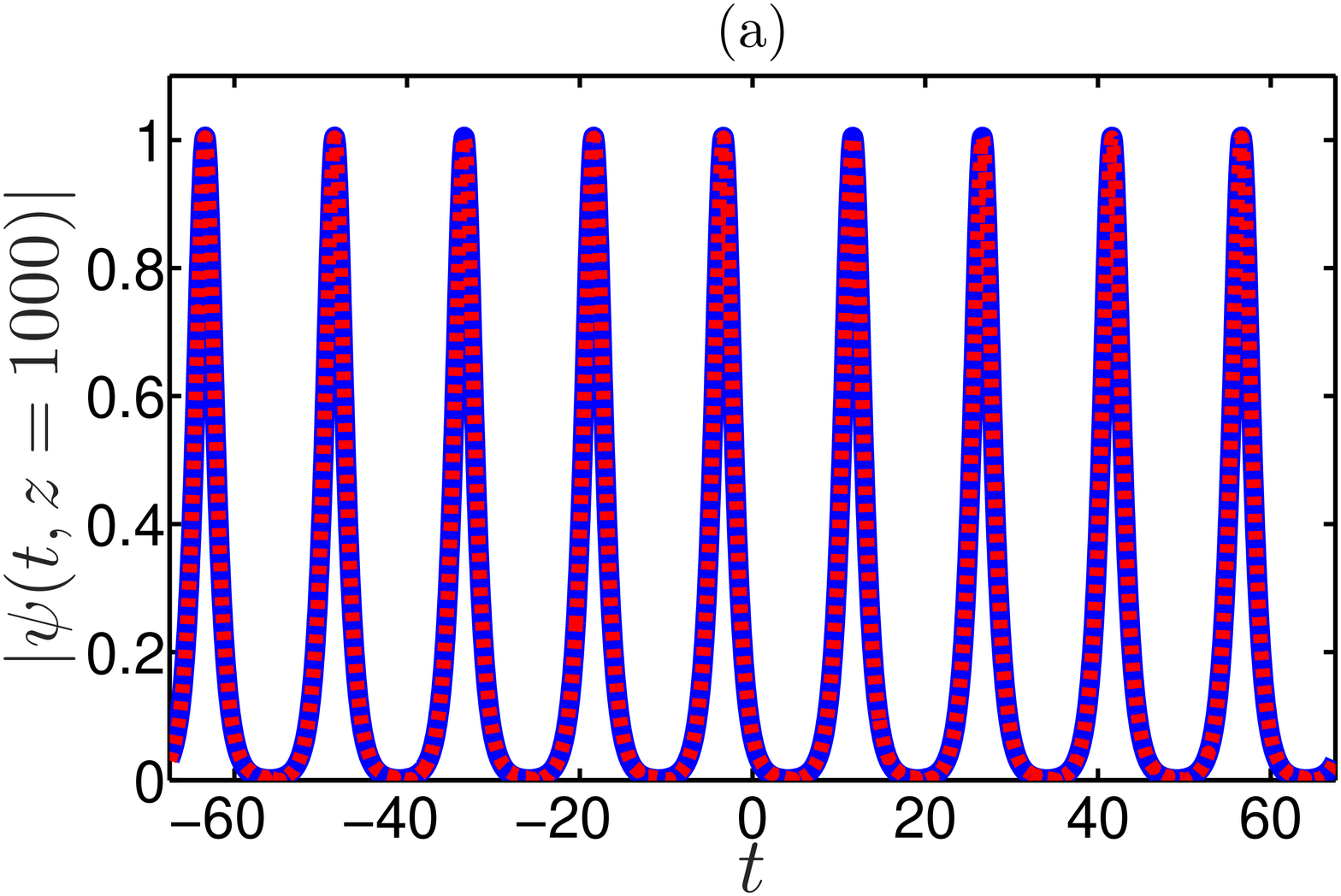} &
\epsfxsize=7.0cm  \epsffile{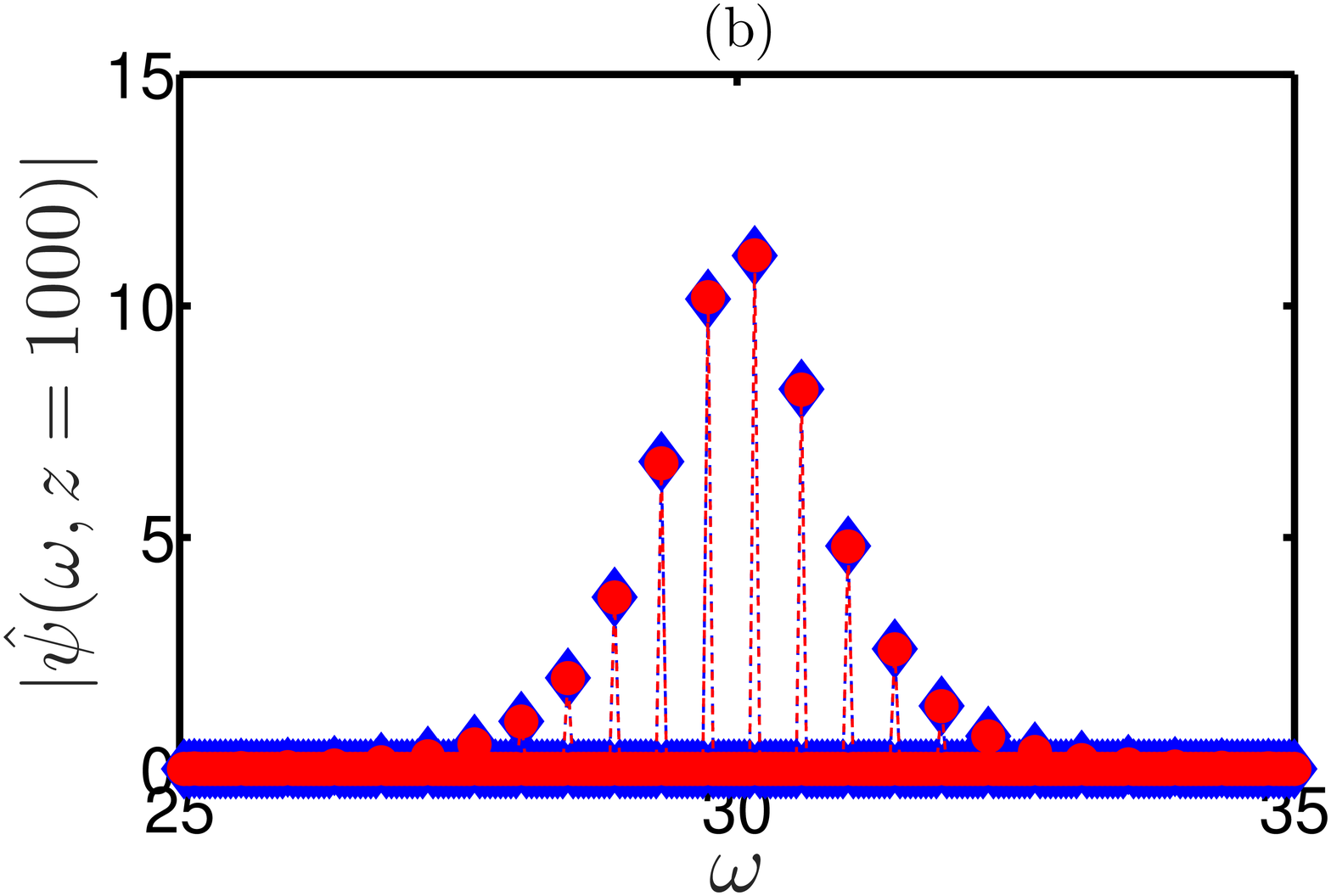} \\
\epsfxsize=7.0cm  \epsffile{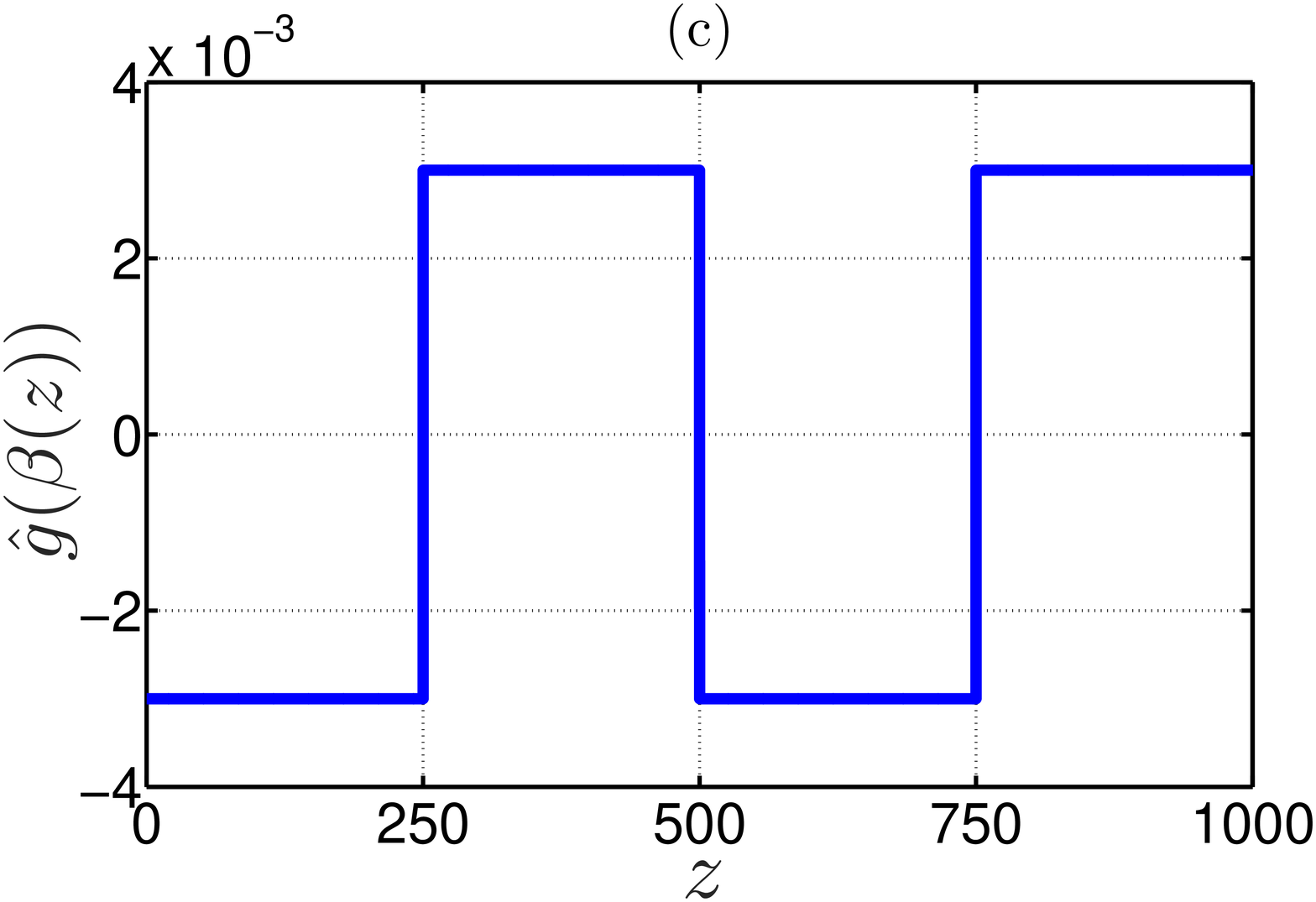} &
\epsfxsize=7.0cm  \epsffile{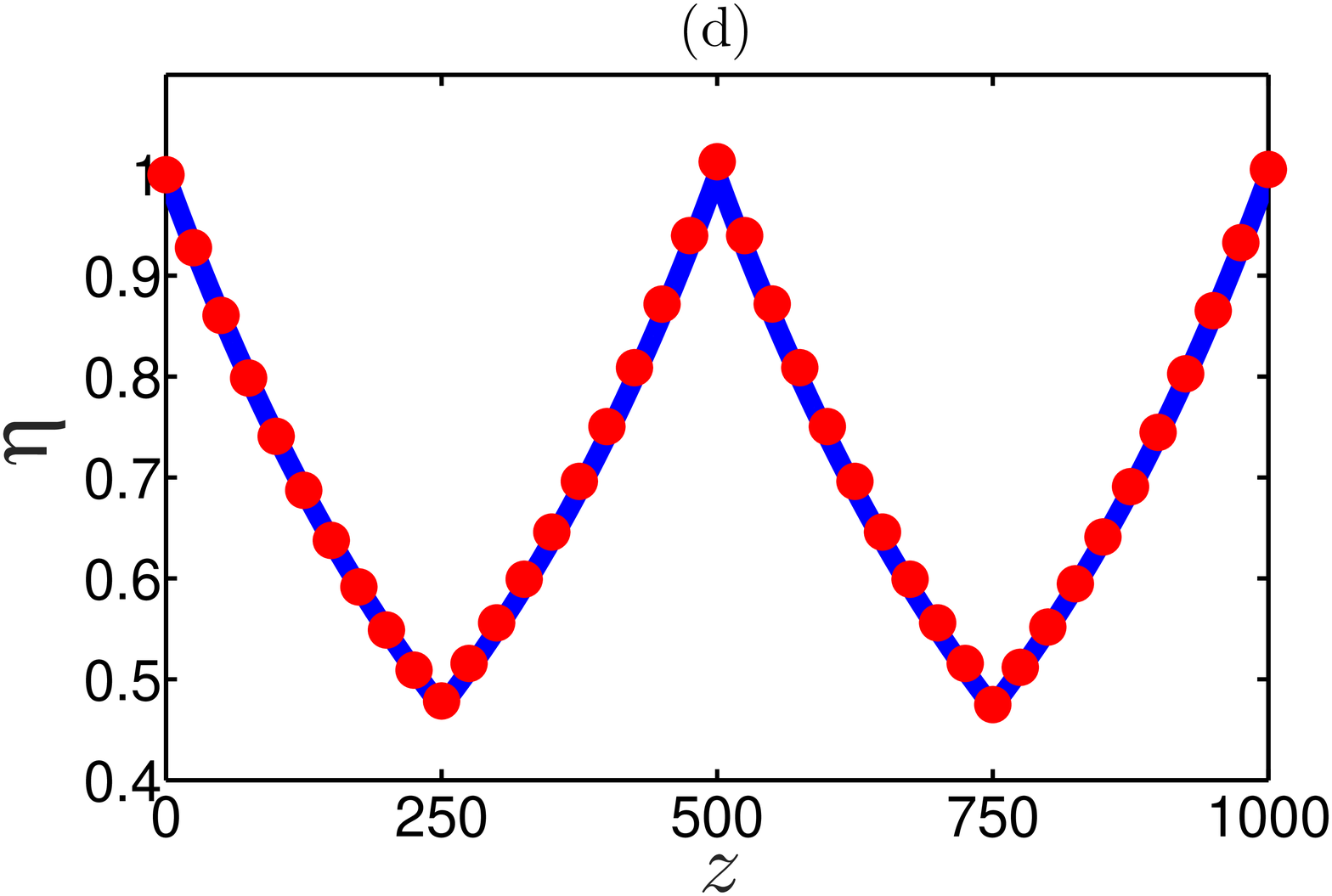} \\
\end{tabular}
\caption{(Color online) Multiple frequency shifting periodically in the presence of frequency dependent linear gain-loss for a sequence of solitons ($\kappa=2$) with parameters in {\bf setup 5b}. (a) The soliton patterns in the time domain at the end of period $z=1000$. The red dashed and blue solid curves correspond to $|\psi(t,z)|$ as obtained by the numerical simulation of Eq. (\ref{App_gl1}) and Eq. (\ref{App_gl2}) and its theoretical prediction. (b) The soliton patterns in the frequency domain at the distance $z=1000$. The red circles and blue diamonds represent $|\hat \psi (\omega ,z)|$ and its theoretical prediction.  (c) The $z$ dependence of $\hat g(\beta(z))$ in Eq (\ref{App_gl2}). (d) The $z$ dependence of the soliton amplitude. The red circles and blue solid curve represent the soliton amplitude $\eta(z)$ measured by Eqs. (\ref{App_gl1}) and (\ref{App_gl2}) and its theoretical prediction measured by Eqs. (\ref{App_gl27}) and (\ref{App_gl28}).
}
 \label{fig10}
\end{figure}

Second, we implement multiple periodic frequency shifting in the presence of frequency dependent linear gain-loss for a single soliton and for a sequence of solitons such that the solitons experience loss-gain $\kappa$ times during a certain propagation distance. 
The numerical simulations are implemented with Eq. (\ref{App_gl1}) and $\hat g(\omega)$ as in Eq. (\ref{App_gl2}) with $\kappa=2$.
It is useful to consider two numerical setups: {\bf setup 5a} for a single soliton and {\bf setup 5b} for a sequence of solitons.
The parameters of $\hat g(\omega)$ for {\bf setup 5a} are: $g_{0}=0.005$, $g_{L}=0.006$, $\Delta\beta=20$, $W=15$, $\rho=10$, and $\kappa=2$. This function $\hat g(\omega)$ is presented in Fig. \ref{fig5}(b). 
Other parameters are: $\beta(0) = - 3\Delta\beta /2 = - 30,\,\eta_{0} = 1,\,y(0) = 2400,\,\alpha(0)=0,\,\Delta z = 0.001,\,\Delta t = 0.0588,\,t_{\max} = 2415,\,t_{\min}=-t_{\max}$, and $z_{s_{k}}= 60k$ with $k=1,2,3$. That is, the frequency shift is implemented by $\Delta \beta$ at the propagation distances $z=60, 120, 180$. Solitons are experienced the processes of the linear loss-gain-loss-gain, respectively, over the propagation distance $0 \leq z \leq z_{f}=240$. The numerical simulation results are presented in Fig. \ref{fig9}. Figures \ref{fig9}(a) and \ref{fig9}(b) show the soliton patterns at the final propagation distance $z_{f}=240$ in the time domain and in the frequency domain, respectively.
The relative errors in measuring the soliton patterns $|\psi_{s1}(t,z)|$ and $|\hat\psi_{s1}(\omega,z)|$ at $z=z_{f}=240$ are $1.12\times 10^{-2}$ and $0.8\times 10^{-2}$, respectively.
Figure \ref{fig9}(c) presents the frequency dependent gain-loss function $\hat{g}(\beta(z))$ vs. $z$ used along the waveguide while Fig. \ref{fig9}(d) shows the amplitude dynamics $\eta(z)$.
The relative errors in measuring the amplitude $\eta(z)$ are less than $0.47\times 10^{-7}$, for $0 \leq z \leq z_{f}$.
The parameters of $\hat g(\omega)$ for {\bf setup 5b} are: $g_{0}=0.003$, $g_{L}=0.005$, $\Delta\beta=20$, $W=15$, $\rho=10$, and $\kappa=2$. Other parameters are: 
 $\beta(0)=-3\Delta \beta/2  = -30,\,\eta(0) = 1,\,y(0) = -5,\,\alpha(0)=0$, $L=15$, $T=15$, $J=4$, $\Delta z = 0.001,\,\Delta t = 0.0588,\,t_{\max} = 67.5,\,t_{\min}=-t_{\max}$, and $z_{s_{k}}=250k$ with $k=1,\,2,\,3$. That is, the multiple frequency shifting is implemented by $\Delta \beta$ at the propagation distances $z_{s_{k}}=250, \,500, \,750$.
The numerical simulation results are shown in Fig. \ref{fig10}. The soliton patterns at the final propagation distance $z_{f}=1000$ are presented in the time domain in Fig. \ref{fig10}(a) and in the frequency domain in Fig. \ref{fig10}(b).
The agreement between the
result of the simulations and the analytic prediction is
very good. In fact, the relative errors in measuring the soliton patterns $|\psi_{sq}(t,z)|$ and $|\hat \psi_{sq}(\omega,z)|$ at $z=z_{f}=1000$ are $0.86\times 10^{-2}$ and $0.32\times 10^{-2}$, respectively. 
Figure \ref{fig10}(c) presents $\hat{g}(\beta(z))$ vs. $z$ while Fig. \ref{fig10}(d) shows the amplitude dynamics $\eta(z)$.
The relative errors in measuring the amplitude dynamics are less than $1.3\times 10^{-2}$.

In summary, we obtain the very good agreement of the pulse pattern and of the amplitude dynamics between the simulation results and the theoretical predictions even at the very large propagation distance, for example, $z_{f}=1000$ for a sequence of solitions.
These confirm that the propagation of solitons with implementing multiple frequency shifting in the presence of frequency dependent linear gain-loss is stable over intermediate to long distances. It is emphasized that the soliton dynamics presented in Fig. \ref{fig10}(d) can be used for switching soliton dynamics or for transmission recovery. In these switching processes, one can turn off (or turn on) a sequence of solitons by guiding its amplitude below (or above) a threshold amplitude value $\eta_{th}$ (see \cite{PNH2017, NPT2015} for another approach to study the switching dynamics by using hybrid waveguides and stability analysis of the equilibrium states of the ODEs describing amplitude dynamics). The amplitude threshold value in Fig. \ref{fig10}(d) can be, for example, $\eta_{th}=0.5$.

%fig 8

%fig 9

\subsection{Numerical simulations for repeating soliton collisions in the presence of weak cubic loss}

%fig 10
\begin{figure}[ptb]
\begin{tabular}{cc}
\epsfxsize=14cm  \epsffile{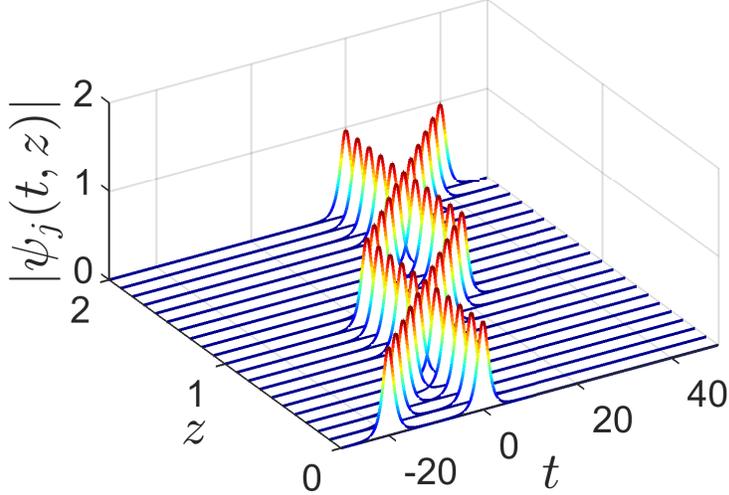} 
\end{tabular}
\caption{(Color online) Repeated two-soliton collisions in waveguides in the presence of weak cubic loss with Eq. (\ref{App_coll_2}).
}
 \label{fig11}
\end{figure}

%fig 11
\begin{figure}[ptb]
\begin{tabular}{cc}
\epsfxsize=6.5cm  \epsffile{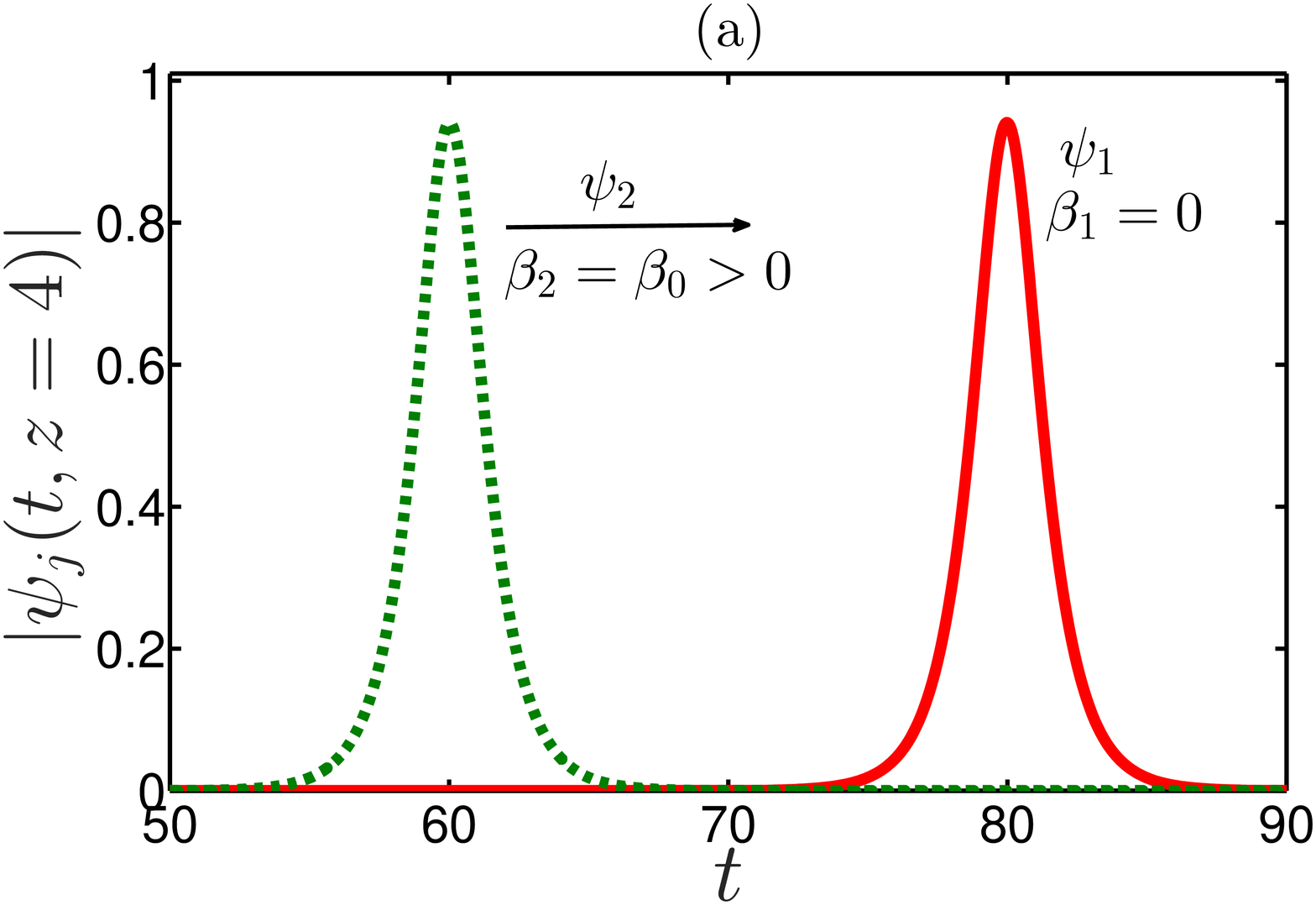} &
\epsfxsize=6.5cm  \epsffile{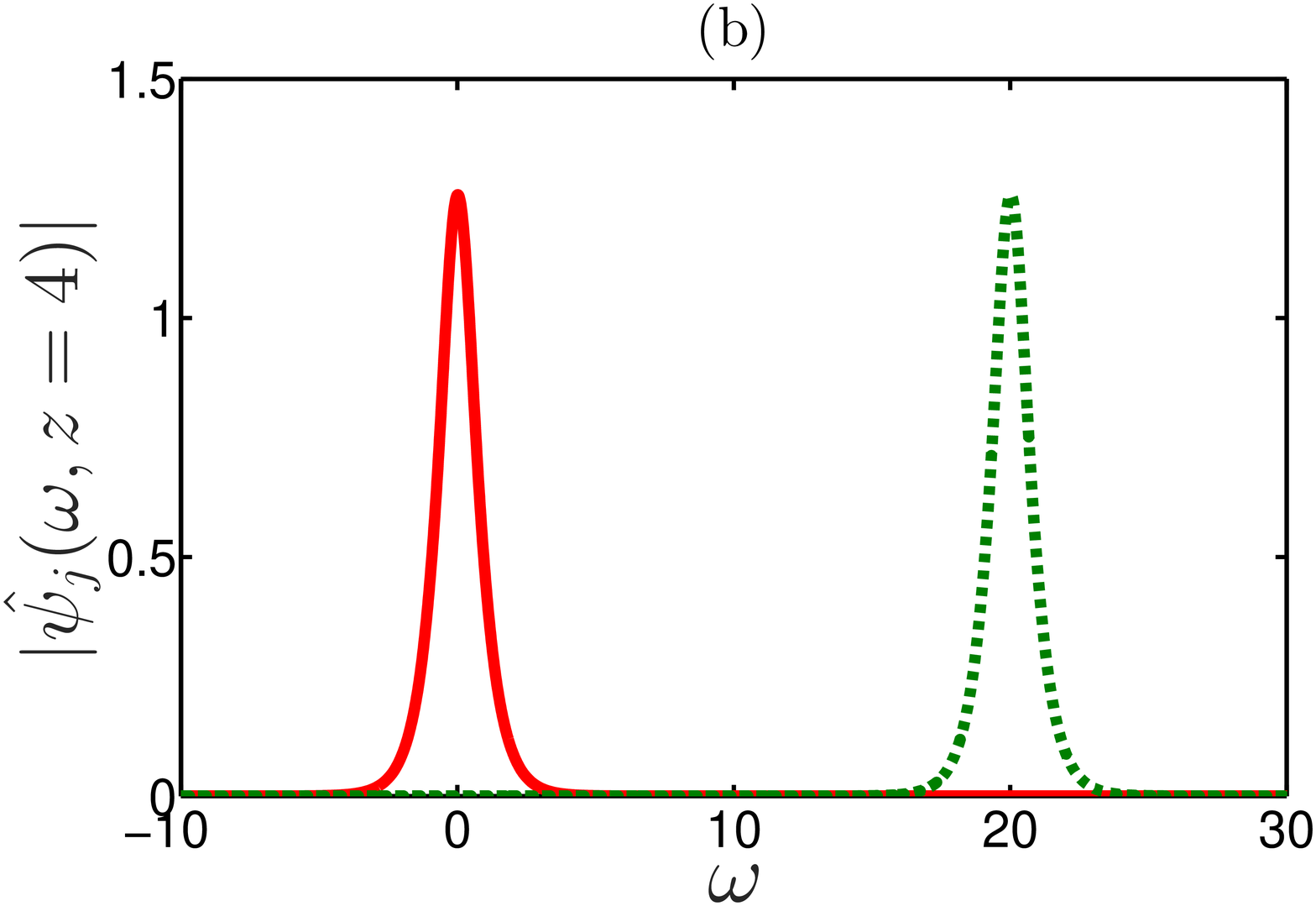} \\
\epsfxsize=6.5cm  \epsffile{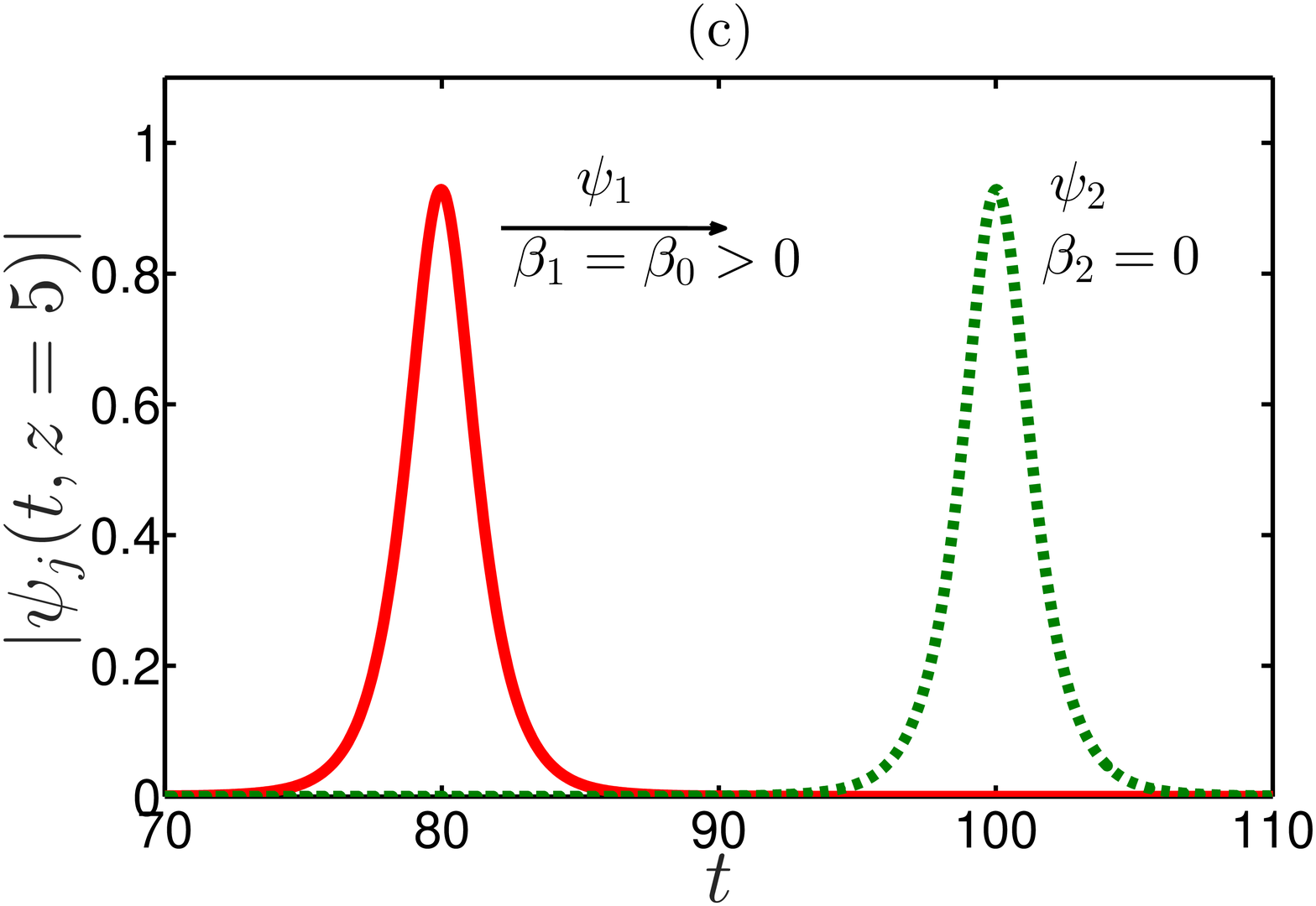} &
\epsfxsize=6.5cm  \epsffile{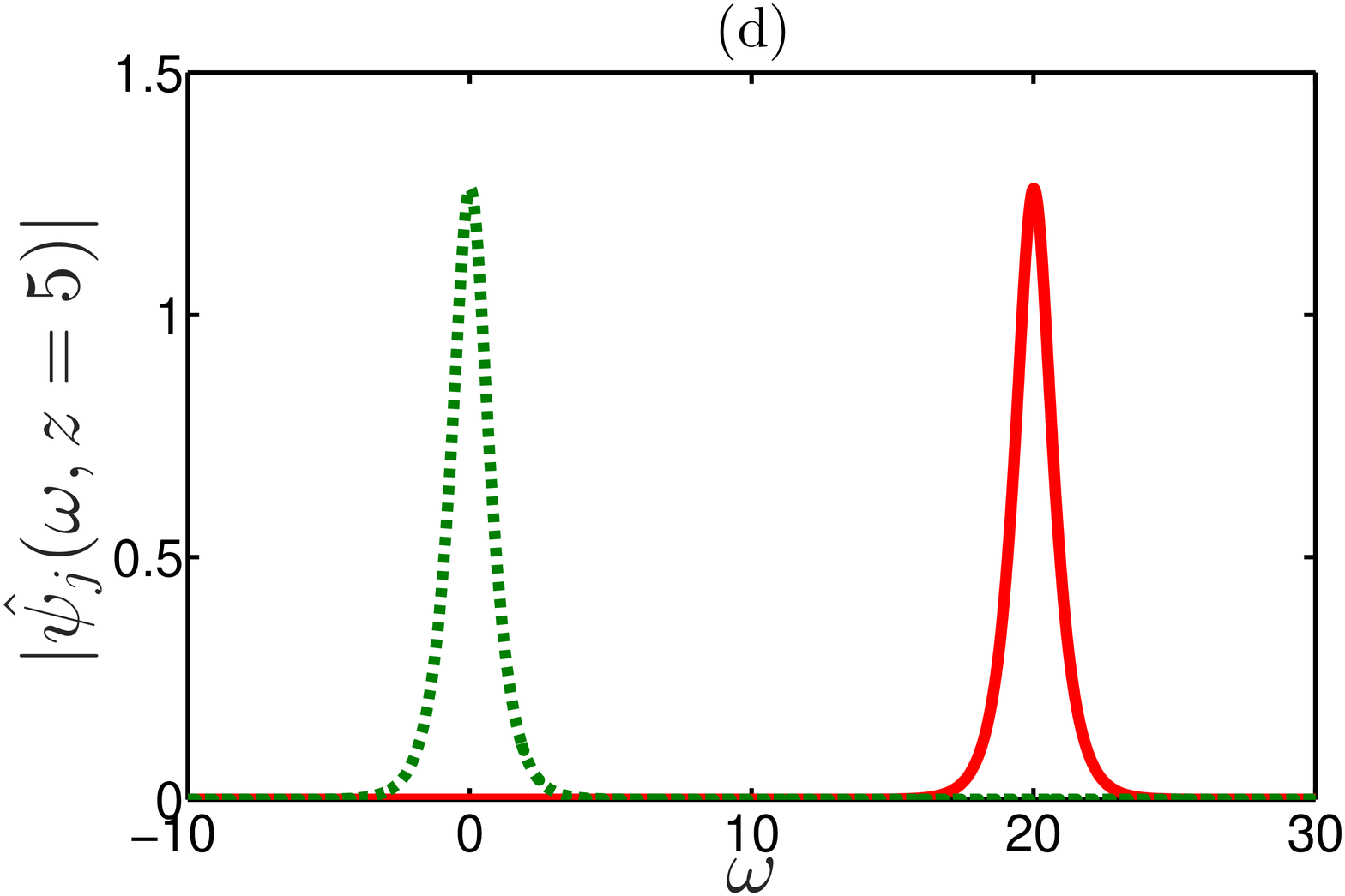} \\
\end{tabular}
\caption{(Color online) Repeated two-soliton collisions in waveguides in the presence of weak cubic loss with Eq. (\ref{App_coll_2}) for $5$ shifts of the frequency and $5$ collisions with the final propagation distance $z_{f}=5$. (a) The soliton patterns in the time domain at the distance $z=4$ (before the 5$^{th}$ collision). The red solid and green dashed curves correspond to $\left| {{\psi_j}(t,z)} \right|$, where $j=1,\,2$. (b) The soliton patterns in the frequency domain at the distance $z=4$.  The red solid and green dashed curves represent $\left| {{{\hat \psi_j }}(\omega,z)} \right|$. (c) The pulse patterns in the time domain at the final propagation distance $z_{f}=5$ (after the $5^{th}$ collision). The red solid and green dashed curves are the same as in (a). (d) The soliton patterns in the frequency domain at the distance $z=5$. The red solid and green dashed curves are the same as in (b).
}
 \label{fig12}
\end{figure}

%fig 12
\begin{figure}[ptb]
\begin{tabular}{cc}
\epsfxsize=6.8cm  \epsffile{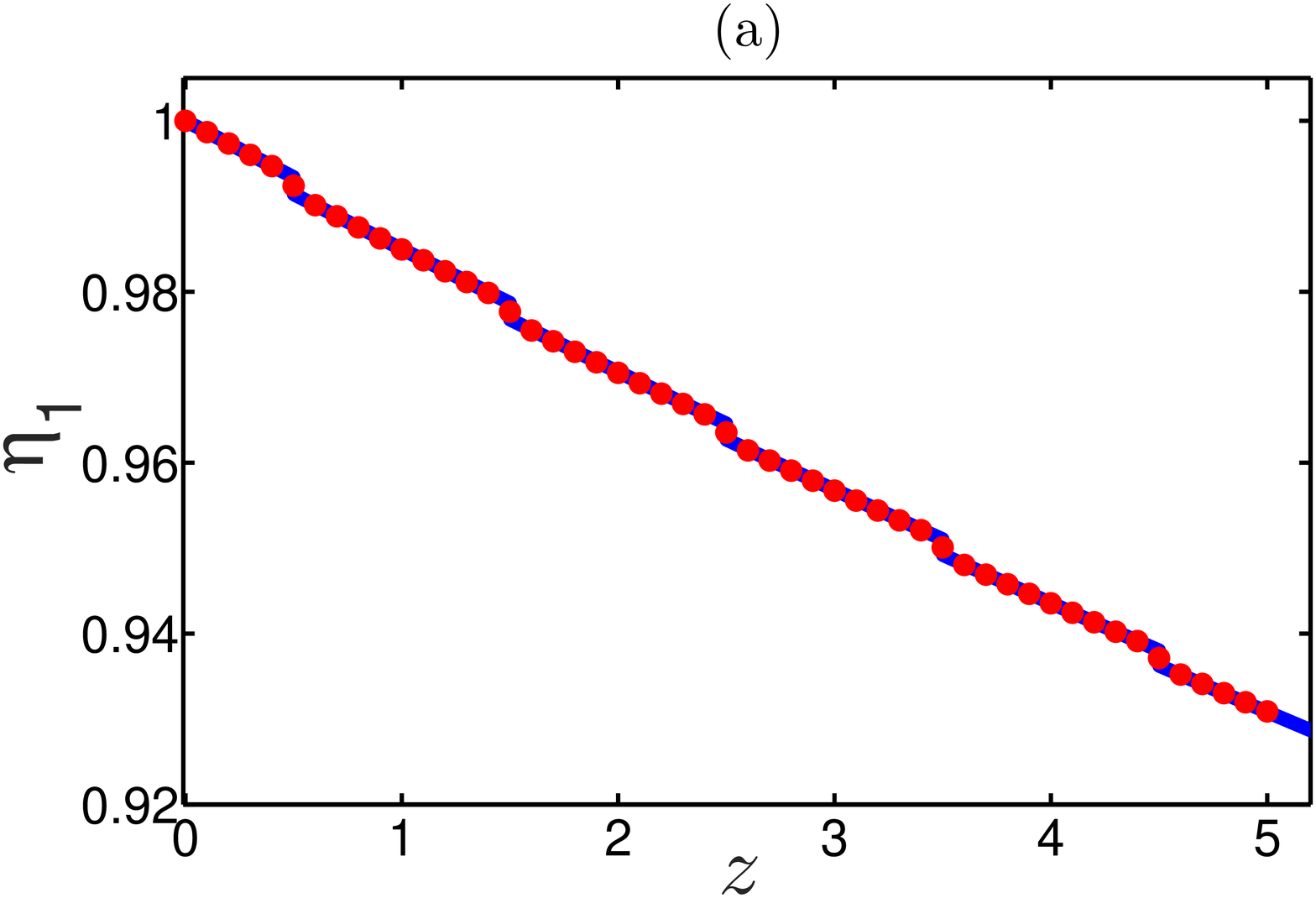} &
\epsfxsize=6.8cm  \epsffile{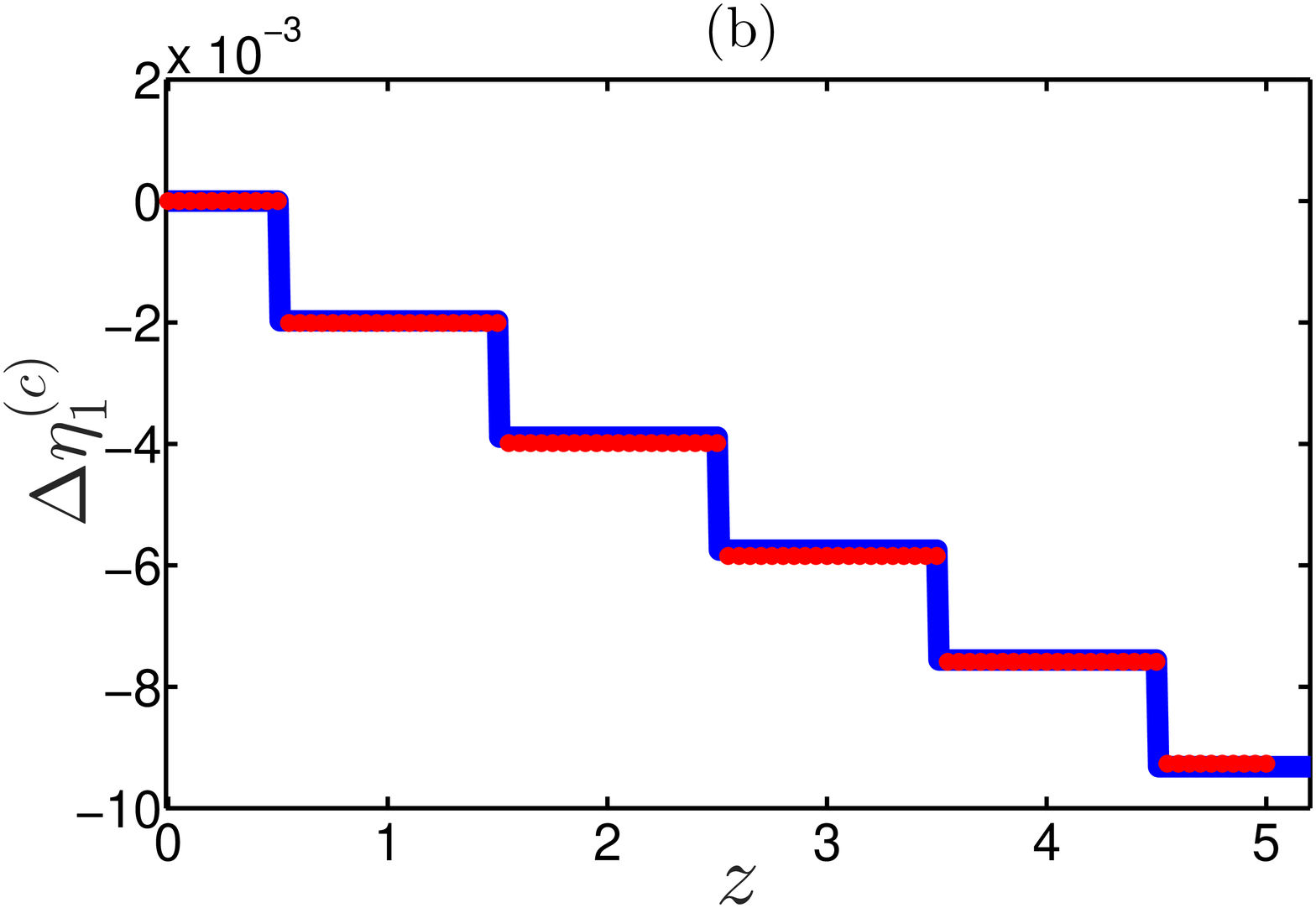}\\
\end{tabular}
\caption{(Color online) (a) The $z$ dependence of the soliton amplitude. The red circles and blue solid curve represent the amplitude $\eta_1(z)$ measured by the numerical simulation with Eq. (\ref{App_coll_2}) and the theoretical amplitude $\eta_1(z)$ obtained by Eq. (\ref{App_coll_13}). (b) The $z$ dependence of the total of the collision-induced amplitude shift $\Delta \eta_{1} ^{(c)} $. The red circles and blue solid curve represent the total of the collision-induced amplitude shift $ \Delta \eta_{1}^{(z_c)} $ measured by the numerical calculation with Eq. (\ref{App_coll_22}) and the theoretical calculation with Eq. (\ref{App_coll_18a}).
}
 \label{fig13}
\end{figure}

In this section, we present simulations for the use of the frequency shifting to repeated two-soliton collisions described as in section \ref{App-coll}. The value of the accumulative collision-induced amplitude shift is then numerically measured and is compared with its theoretical prediction.
It is useful to consider the numerical {\bf setup 6} with following parameters: $\epsilon_{3}=0.01,\,\beta_{1}(0) =  0,\,\beta_{2}(0) =  20,\,y_{1}(0) = 0,\,y_{2}(0)=-20,\,\eta_{1}(0)=\eta_{2}(0) = 1,\,\alpha_{1}(0)=\alpha_{2}(0)=0,\,\Delta z = 0.001,\,\Delta t = 0.0588,\,t_{\max} = 1500,\,t_{\min}=-t_{\max}$, and $\Delta \beta  = 20$. The numerical simulation results are presented in Figs. \ref{fig11}, \ref{fig12}, and \ref{fig13}. Figs. \ref{fig11} and \ref{fig12} illustrate the method and present the pulse patterns for repeated two-soliton collisions. Figure \ref{fig11} illustrates the two-soliton collisions occurring at the propagation distance $z_{c_{1}}=0.5$ and $z_{c_{2}}=1.5$. Figure \ref{fig13}(a) describes the amplitude dynamics $\eta_{1}(z)$ along the waveguides with the final propagation distance $z_{f}=5$. We perform 5 consecutive shifts of the frequency at the propagation distances $z_{s_{k}}=k$, where $k=1,2,\cdots , 5$. Therefore, due to changing the group velocity, there are a total of 5 consecutive soliton collisions at the distances $z_{c_{k}}=0.5k$, where $k=1,2,\cdots ,5$. As can be seen the amplitudes abruptly drop after each collision as in Fig. \ref{fig13}(a).
Figure \ref{fig13}(b) presents the $z$ dependence of the total of collision-induced amplitude shift $ \Delta \eta_{1} ^{(c)} $. The numerical value of the accumulative collision-induced amplitude shift in 5 collisions is $ \Delta \eta_{1} ^{(c)(num)} = -0.0093$ with the relative error, which is defined by $|\Delta \eta_{1} ^{(c)(num)} - \Delta \eta_{1} ^{(c)(th)}|/|\Delta \eta_{1} ^{(c)(th)}|$, of $0.57\times 10^{-2}$. Moreover, we also implement the numerical simulations for repeating soliton collisions with $70$ times of frequency shifting at propagation distances $z=k$. Thus, there are $70$ collisions at propagation distances $z=0.5k$, where $k=1,\,2,\cdots ,\,70$. The numerical measurement of the accumulative collision-induced amplitude shift is $ \Delta \eta_{1} ^{(c)(num)} = -0.0734$ with the relative error of $0.67\times 10^{-2}$. The small relative errors above confirm the robustness of the theoretical procedure for the frequency shifting and the theoretical analysis for repeating two-soliton collisions in the presence of weak cubic loss.

\section{Conclusions}
\label{conc}

In summary, we developed the theoretical procedures of the frequency shifting 
for a single soliton and in particular, for a sequence of solitons of the NLS equation and verified them by numerical simulations. 
The procedures are based on simple transformations of the FT of solitons and on the shape-preserving property of solitons. The frequency shifting procedure for a single soliton is based on shifting the FT of the soliton pattern in the frequency domain. The frequency shifting procedure for a sequence of solitons is based on {\it the decomposition method}. In this method, one can decompose $\Delta \beta=\Delta \beta_{1}+\Delta \beta_{2}$, where $\Delta\beta_{1}=2m\pi/T$, $m \in \mathbb{Z}$, and $-2\pi/T < \Delta\beta_{2} < 2\pi/T$, and implement independently the frequency shifting for each $\Delta \beta_{j}$, $j=1,2$, by using the different techniques for each one. To the best of our knowledge, the mathematical procedure of frequency shifting for a sequence of solitons by an arbitrary value has been successfully implemented and numerically verified for the first time in the current paper.

Furthermore, we demonstrated the use of frequency shifting procedures in two important applications: (1) stabilization of the soliton propagation in waveguides with frequency dependent linear gain-loss; (2) induction of repeated soliton collisions in waveguides with cubic loss. 
We showed that the amplitude dynamics of solitons in waveguides with frequency dependent linear gain-loss can be described by an ODE model in transmissions of a single soliton and of a sequence of solitons. This ODE model was used to validate the robust propagation of solitons experiencing the linear loss and experiencing the linear gain. The soliton dynamics studied in this paper can be useful for controlling and guiding solitons, for example, for switching soliton dynamics (see \cite{PNH2017, NPT2015} for another approach to stabilize and switch sequences of solitons by using hybrid waveguides and analyzing the ODE model for amplitude dynamics). In addition, the use of the frequency shifting was demonstrated to enable repeated two-soliton collisions in the presence of weak cubic loss and the theoretical prediction for the accumulative collision-induced amplitude shift was also measured.
The theoretical procedures and calculations were confirmed by 
numerical simulations with the NLS equation using the split-step Fourier method. Based on the results of numerical simulations, it can be concluded that the frequency procedures developed in the
current paper can indeed be successfully implemented to realize stable soliton propagations
and control of fast soliton collisions in nonlinear optical waveguides.

\appendix
\section{Implementations of the frequency shifting for a single soliton in simulations}
\label{App_A}

Let us describe implementations of the frequency shifting 
in numerical simulations for the propagation of a single NLS soliton. We assume that the 
soliton propagation is described by Eq. (\ref{single1})
with additional weak perturbation terms. 
It is emphasized that a similar procedure can be implemented with data obtained from experiments.

It is useful to denote by $\psi_{1}^{(num)}(t,z)$ the envelope of 
the electric field at the propagation distance $z$, obtained in the simulation (or in the experiment), 
and by $\hat\psi_{1}^{(num)}(\omega,z)$ the FT 
of $\psi_{1}^{(num)}(t,z)$ with respect to time.      
The first step in the implementation of the frequency 
shifting procedure is to determine $\Delta \beta^{(num)}$, 
which is the actual value of the frequency shift to be used in the numerical simulation (or in the experiment).
The value of $\Delta \beta^{(num)}$ is determined by using:  
$\Delta \beta^{(num)}=n\Delta\omega$, where $\Delta\omega$ 
is the spacing of the frequency grid in the simulation 
and $n$ is the integer, whose value is closest to $\Delta\beta/\Delta\omega$.  
The {\it new} (frequency shifted) FT of the envelope of the electric field, 
$\hat\psi_{1,n}^{(num)}(\omega,z)$, is then obtained by: 
\begin{eqnarray} &&
\hat\psi_{1,n}^{(num)}(\omega,z)=
\hat\psi_{1}^{(num)}(\omega - \Delta\beta^{(num)},z).
\label{App_A1}
\end{eqnarray} 
That is, $\hat\psi_{1,n}^{(num)}(\omega,z)$ 
is obtained by shifting $\hat\psi_{1}^{(num)}(\omega,z)$  $n$ grid points to 
the right  if $\Delta\beta^{(num)}>0$, and $n$ grid points to 
the left if $\Delta\beta^{(num)}<0$.     
The new field $\psi_{1,n}^{(num)}(t,z)$ is obtained by taking the 
inverse FT of $\hat\psi_{1,n}^{(num)}(\omega,z)$ with respect to $\omega$. 

In summary, the implementations of the frequency shifting procedure 
for a single soliton in simulations can be summarized by 3 steps:
\begin{enumerate} 
\item Determine $\Delta\beta^{(num)}$.
\item Determine $\hat\psi_{1,n}^{(num)}(\omega,z)$ by Eq. (\ref{App_A1}).  
\item Determine $\psi _{1,n}^{(num)}(t,z)$ by 
$\psi _{1,n}^{(num)}(t,z)=\mathcal {F}^{-1}(\hat\psi_{1,n}^{(num)}(\omega,z))$.
\end{enumerate}

{\it Discussion.} In implementing the frequency shifting procedure for a single soliton, one needs to take care of possible consequences of the finite size 
of the frequency interval $[\omega_{\min},\omega_{\max}]$. More specifically, suppose we employ steps (1) and (2) with $\Delta\beta^{(num)}>0$. 
Then the values of $\hat\psi_{1,n}^{(num)}(\omega ,z)$ 
are missing at $n$ grid points in the interval $[\omega_{\min},\omega_{\min}+\Delta\beta^{(num)})$ after employing the frequency shifting to the right in the frequency domain. 
Therefore, the values of $\hat\psi_{1,n}^{(num)}(\omega ,z)$ on the left tail of the pattern, i.e. for $\omega \in [\omega_{\min},\omega_{\min}+\Delta\beta^{(num)})$,  are determined by 
the extrapolation. 
One can carry out this extrapolation by fitting $\hat\psi_{1,n}^{(num)}(\omega,z)$ 
to the soliton part of the theoretical solution, given by Eq. (\ref{single4}), 
with parameters measured from simulations. 
That is, for $\omega \in [\omega_{\min},\omega_{\min}+\Delta\beta^{(num)})$, we fit $\hat\psi_{1,n}^{(num)}(\omega,z)$ to 
%\begin{eqnarray} &&
\[\hat\psi_{1,n}^{(th)}(\omega,z)= 
\left(\frac{\pi}{2} \right)^{1/2}
\frac{\exp[i\theta_{n}^{(num)}(z) - i\omega y^{(num)}(z)]}
{\cosh\left\{\pi \left[\omega - \beta_{n}^{(num)}(z)\right]/
\left[2\eta^{(num)}(z)\right]\right\}},
\]%\label{App_A2} 
%\end{eqnarray} 
where $\eta^{(num)}(z)$, $\beta_{n}^{(num)}(z)$, $y^{(num)}(z)$, 
and $\theta_{n}^{(num)}(z)$ are the values of the soliton's amplitude, 
new frequency, position, and new overall phase, which are determined 
from the simulation with $\beta_{n}^{(num)}(z)=$ $\beta^{(num)}(z)+\Delta\beta^{(num)}$ and $\theta_{n}^{(num)}(z) = \theta^{(num)}(z) + \Delta \beta^{(num)} y^{(num)}(z)$. Note that in a typical situation, the main body of $\hat\psi_{1,n}^{(num)}(\omega,z)$ is located at the central frequency $\beta_{n}^{(num)}(z)$ and are exponentially small for $\omega \in [\omega_{\min},\omega_{\min}+\Delta\beta^{(num)})$. Therefore, this extrapolation does not affect on the main body of the new soliton pattern.

\section{Implementations of the frequency shifting for a sequence of solitons in simulations}                
\label{App_B}

It is useful to denote by $\psi_{sq}^{(num)}(t,z)$ the envelope of 
the electric field at the distance $z$, obtained in the simulation (or in the experiment), 
and by $\hat\psi_{sq}^{(num)}(\omega,z)$ the FT 
of $\psi_{sq}^{(num)}(t,z)$ with respect to time.

{\it Procedure I - a naive frequency shifting procedure.} This procedure is similar to the frequency shifting procedure for a single soliton and can be summarized by 3 steps:
\begin{enumerate} 
\item Determine $\Delta\beta^{(num)}$ and measure $\eta^{(num)}(z)$, $\beta^{(num)}(z)$, $y^{(num)}(z)$, and $\theta^{(num)}(z)$ from $\hat\psi_{sq}^{(num)}(\omega,z)$.
\item Determine $\hat\psi_{sq,n}^{(num)}(\omega,z)$ by employing the transformation:
\[\hat \psi^{(num)}_{sq,n}(\omega,z) = \hat \psi^{(num)}_{sq}(\omega  - \Delta \beta^{(num)},z).
\]
Similarly to procedure I, the missing data of $\hat\psi_{sq,n}^{(num)}$ when implementing the transformation $\omega  \to \omega  -\Delta\beta^{(num)}$ can be extrapolated by its theoretical prediction: %$\hat\psi_{sq,n1}^{(th)}$ as in 
\[\!\!\!\!\!\!\!\!\!\!\!\!\!\!\!\!\!\!\!\!\!
\hat\psi^{(th)}_{sq,n}(\omega,z) = 
\left( \frac{\pi}{2} \right)^{1/2}
\frac{\exp[i\theta_{n}^{(num)}(z) - i\omega y^{(num)}(z)]}
{\cosh\left\{\pi \left[\omega - \beta_{n}^{(num)}(z)\right]/
\left[2\eta^{(num)}(z)\right]\right\}}
\sum\limits_{k=-J}^{J} e^{-ikT\omega},
\]
where  $\beta_{n}^{(num)}(z)=\beta^{(num)}(z)+\Delta\beta^{(num)}$ and $\theta_{n}^{(num)}(z) = \theta^{(num)}(z) + \Delta \beta^{(num)} y^{(num)}(z)$.
\item Determine $\psi_{sq,n}^{(num)}(t,z)$ by 
$\psi_{sq,n}^{(num)}(t,z)=\mathcal {F}^{-1}(\hat\psi_{sq,n}^{(num)}(\omega,z))$.
\end{enumerate}

{\it Procedure II - the frequency shifting of $V(\omega,z)$ and $U(\omega,z)$.}

First, one can numerically measure the envelope function $V^{(num)}(\omega ,z)$
and the fast oscillation function $W^{(num)}(\omega)$.
For this purpose, one needs to measure
$\theta^{(num)}(z)$, $\beta^{(num)}(z)$, $\eta^{(num)}(z)$, $y^{(num)}(z)$,
and $U^{(num)}(\omega,z)$ from $\hat \psi_{sq}^{(num)}(\omega,z)$,
where $\theta^{(num)}(z)$ can be measured by
$\arg [\hat \psi_{sq}^{(num)}(0, z) ]$
and $U^{(num)}(\omega,z) = \theta^{(num)}(z) - \omega y^{(num)}(z)$.
The normalized envelope function can be then measured:
%\begin{eqnarray} &&
\[V^{(num)}(\omega,z) = \tilde V^{(num)}(\omega,z)/(2J + 1),
\]%\label{App_B1}
%\end{eqnarray}
where the envelope function $\tilde V^{(num)}(\omega,z)$
is tangential to $|\hat \psi_{sq}^{(num)}(\omega,z )|$.
The function $\tilde V^{(num)}(\omega,z)$
can be measured by collecting the coordinates of all the points
on the graph of the numerical $|\hat \psi_{sq}^{(num)}(\omega,z)|$
where the curve of $|\hat \psi_{sq}^{(num)}(\omega,z)|$
is tangential to the function bounding $|\hat\psi^{(num)}_{sq}(\omega,z)|$ from above.
Once all these coordinates were collected,
an interpolation at other $\omega$-values is carried out by using Eq. (\ref{sequence3}) as follows:
\begin{eqnarray} &&
V^{(th)}(\omega, z) = (\pi /2)^{1/2} \sech\left\{ \pi \left[\omega  - \beta^{(num)}(z) \right]/\left[2\eta ^{(num)}(z) \right] \right\}.
\label{App_B1b}
\end{eqnarray} 
After measuring $V^{(num)}(\omega,z)$, the fast oscillating function $W^{(num)}(\omega)$ can be measured by 
\begin{eqnarray} &&
W^{(num)}(\omega ) = \hat \psi_{sq}^{(num)}(\omega,z){e^{-iU^{(num)}(\omega,z)}}/V^{(num)}(\omega ,z).
\label{App_B2}%Equation 20
\end{eqnarray}
We note that $V^{(num)}(\omega,z)$ is exponentially approaching to zero on both tails in the frequency domain. Therefore, the division in Eq. (\ref{App_B2}) can be numerically invalid when $V^{(num)}(\omega,z)$ is small enough. Let $L$ be length of the interval centered at $\beta^{(num)}(z)$ such that the division in Eq. (\ref{App_B2}) can be numerically valid, i.e., the division is implemented accurately with small relative errors. Based on extensively numerical tests with different values of $L$, the typical values of $L$ can be found as $15 \leq L \leq 40$. Thus, Eq. (\ref{App_B2}) can be invalid for $\omega<\beta^{(num)}(z)-L/2$ or $\omega>\beta^{(num)}(z)+L/2$. Therefore one needs to implement the extrapolation of $W^{(num)}(\omega)$ to its theoretical function $W^{(th)}(\omega )$ by using Eq. (\ref{sequence4}) for $\omega<\beta^{(num)}(z)-L/2$ or $\omega>\beta^{(num)}(z)+L/2$.

Second, one can implement the following transformations: $V^{(num)}_{n}(\omega, z) = V^{(num)}(\omega -\Delta\beta^{(num)}, z)$ and $U^{(num)}_{n}(\omega,z) = U^{(num)}(\omega-\Delta\beta^{(num)},z)$, where $\Delta\beta^{(num)}$ is the actually numerical frequency shift value.
Note that the theoretical predictions $V_{n}^{(th)}(\omega, z)$, defined by Eq. (\ref{App_B1b}), and $U_{n}^{(th)}(\omega,z)=$ $\theta_{n}^{(num)}(z)-\omega y^{(num)}(z)$ can be used to extrapolate the missing data of $V^{(num)}_{n}(\omega, z)$ and $U^{(num)}_{n}(\omega, z)$ when implementing the transformation $\omega  \to \omega  -\Delta \beta^{(num)}$. Here, $\beta_{n}^{(num)}(z)=\beta^{(num)}(z)+\Delta\beta^{(num)}$ and $\theta_{n}^{(num)}(z) = \theta^{(num)}(z) + \Delta \beta^{(num)} y^{(num)}(z)$ are used to measure $V^{(num)}_{n}(\omega, z)$ and $U^{(num)}_{n}(\omega, z)$.

Third, the new sequence of solitons in the frequency domain can be defined by
\begin{eqnarray} &&
\hat\psi^{(num)}_{sq,n}(\omega,z)= V^{(num)}_{n}(\omega,z) W^{(num)}(\omega) {e^{iU_{n}^{(num)}(\omega,z)}}.
\label{App_B3}%Equation 21
\end{eqnarray}

Finally, taking the inverse FT of $\hat\psi^{(num)}_{sq,n}(\omega,z)$ with respect to $\omega$, it yields $\psi_{sq,n}^{(num)}(t,z)$ with the new frequency of $\beta(z)+\Delta\beta^{(num)}$ as desired.

\begin{figure}[ptb]
\begin{center}
\begin{tabular}{cc}
\epsfxsize=9.0cm  \epsffile{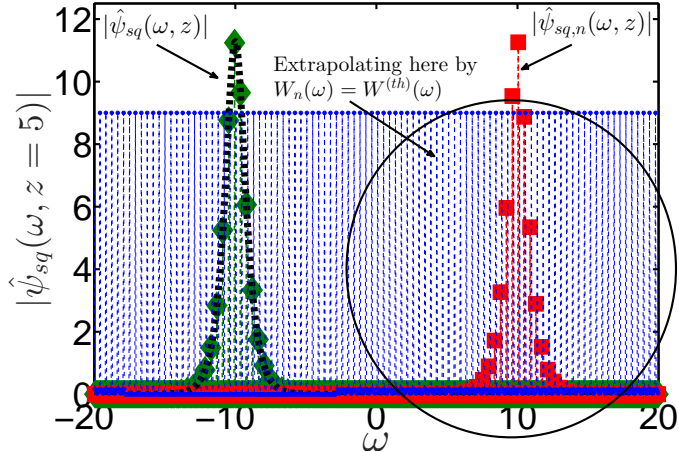} 
\end{tabular}
\end{center}
\caption{(Color online) The frequency shifting for a sequence of solitons by the procedure II with parameters as in {\bf setup 1}. The black dashed-dotted curve represents $\tilde V^{(num)} (\omega, z)$ as in \ref{App_B}. The blue circles represent $|W^{(num)}(\omega)|$ measured by Eq. (\ref{App_B2}). The green diamonds represent $| \hat \psi_{sq} (\omega ,z)|$ measured by the numerical simulation with Eq. (\ref{single1}) and the red squares represent the new pattern of $|\hat \psi_{sq,n}(\omega,z)|$ obtained by the procedure II.  
}
 \label{fig_appB1}
\end{figure}

Since the values of $W^{(num)}(\omega)$ can be only numerically measured in the interval $[\beta^{(num)}(z)-L/2,$ $\beta^{(num)}(z)+L/2]$ while other values are determined by extrapolations, the procedure II therefore has shortcomings as described in section \ref{2.2.2}.
To illustrate the shortcomings of this procedure, we perform the frequency shifting for a soliton sequence propagating periodically in a waveguide loop with Eq. (\ref{single1}) and parameters in {\bf setup 1}. Figure \ref{fig_appB1} depicts the soliton patterns in the frequency domain before and after shifting the frequency,  
$\left| {\hat \psi_{sq} (\omega ,z)} \right|$ and 
$\left| {\hat \psi_{sq,n} (\omega ,z)} \right|$, at the propagation distance $z=5$. The numerical frequency shifting value is $\Delta\beta^{(num)}=20.0131$.
The numerical measurements for the frequency before and after performing the frequency shifting are $\beta^{(num)}(z=5)=-10$ and $\beta_{n}^{(num)}(z=5)=10.0131$, respectively.
As can be seen, in the circled region in Fig. \ref{fig_appB1}, one needs to extrapolate $W(\omega)$ by using the theoretical prediction $W^{(th)}(\omega)$ for $\omega  > \beta^{(num)}(z) + L/2=-2.5$. 
As a result, extrapolated values of $W(\omega)$ are used for 
calculating $\hat\psi^{(num)}_{sq,n}(\omega,z)$ in the main body of $\hat\psi^{(num)}_{sq,n}(\omega,z)$. 
%fig2

{\it Procedure III - the decomposition procedure.} The implementations for this procedure include 5 main steps:  
\begin{enumerate}
\item Determine $\Delta\beta^{(num)}$ and decompose $\Delta \beta^{(num)}  =\Delta \beta_{1} + \Delta \beta_{2}$, where $\Delta \beta_{1}=2m\pi/T$ with $m$ is the nearest integer rounded by $\Delta\beta^{(num)} T/2\pi$ and $-2\pi/T < \Delta \beta_{2} < 2\pi/T$. 
\item Define $\hat \psi^{(num)}_{sq,n1}(\omega,z)$ from $\hat \psi^{(num)}_{sq}(\omega,z)$ by the {\it procedure I} with $\Delta\beta_{1}$.
\item Measure $V^{(num)}(\omega ,z)$ and $W^{(num)}(\omega)$ of $\hat \psi^{(num)}_{sq,n1}(\omega,z)$. 
\item Define $\hat\psi^{(num)}_{sq,n}(\omega,z)$ from $\hat \psi^{(num)}_{sq,n1}(\omega,z)$ by the {\it procedure II} with $\Delta\beta_{2}$.
\item Compute $\psi_{sq,n}^{(num)}(t,z)=\mathcal {F}^{-1}\left(\hat\psi_{sq,n}^{(num)}(\omega,z)\right)$. 
\end{enumerate}

\section{Split-step Fourier method}   
\label{App_C}

One method that has been used extensively to solve the pulse-propagation problem in nonlinear
dispersive media is the split-step Fourier method, which is a pseudo-spectral method. In this Appendix, we briefly describe the split-step Fourier method to implement the numerical simulations of the perturbed NLS equation \cite{Agrawal2001, Yang2010, Weideman1986}. It is useful to re-write the perturbed NLS equation in the form:
\begin{eqnarray} &&
\partial_{z}\psi = (L + N)\psi,
\label{App_C1} 
\end{eqnarray}
where $L$ is the linear and $N$ is the nonlinear operator. For example, if we solve the NLS with weak cubic loss: $i\partial_{z}\psi  + \partial_{t}^{2}\psi  + 2|\psi|^{2}\psi
= - i\epsilon_{3} |\psi|^{2}\psi$, then $L \psi = i\partial_{t}^{2}\psi,$ and $N \psi = (2i - \epsilon_{3})|\psi|^{2}\psi$. The theoretical exact solution of Eq. (\ref{App_C1}) at the propagation distance $z + h$ can be written as
%\begin{eqnarray} &&
\[\psi(z+h,t) = \exp{[h(L+N)]} \psi(z,t),
\]%\label{App_C4} 
%\end{eqnarray}
where $h=\Delta z$ is a small propagtion distance for the propagation of soliton from $z$ to $z+h$ in optical waveguide ($h$ is a $z$-step size).

In general, dispersion and nonlinearity act together along the length of the waveguide. In the split-step Fourier method, it is assumed that the linear and nonlinear effects act independently in the propagation of solitons over a small distance $h$. The propagation from $z$ to $z+h$ is carried out in two steps. In the first step, the nonlinearity acts alone and in the second step, dispersion acts alone. It thus implies two split equations:
\begin{eqnarray} &&
\partial_{z}\psi = L\psi,
\label{App_C5} 
\end{eqnarray}
and
\begin{eqnarray} &&
\partial_{z}\psi = N\psi.
\label{App_C6} 
\end{eqnarray}
The solution at the propagation distance $z+h$ of Eq. (\ref{App_C5}) and Eq. (\ref{App_C6}) can be written as $e^{hL}\psi(z,t)$ and $e^{hN}\psi(z,t)$, respectively. Using the Baker-Campbell-Hausdorff  formula for two non-commutative operators $A, B$ where $A=hL,B=hN$:
%\begin{eqnarray}
\[\exp (A)\exp (B) = \exp (A + B + \frac{1}{2}[A,B] + \frac{1}{{12}}[A - B,[A,B]] + ...),
\]%\label{App_C7}
%\end{eqnarray}
the error $E = \left| {\exp \left( {h\left( {L + N} \right)} \right) - \exp \left( {hL} \right)\exp \left( {hN} \right)} \right|$  is found to result from $\frac{1}{2}h^2 \left[ {N,L} \right]$, i.e, in the second order. The error can be reduced to the order ${\rm o}\left( h^{n} \right)$ if there is a set of real numbers $(c_l, c_2 .... , c_n)$ and $(d_l, d_2, ..., d_n)$ such that  
\begin{eqnarray}&&
\exp \left( {h\left( {L + N} \right)} \right) = \prod\limits_{i = 1}^n {\exp (c_i hL)\exp (d_i hN)} + {\rm o}\left( h^{n+1} \right).  \label{App_C8}
\end{eqnarray}
Using the results in \cite{Yoshida90}, it yields:
\begin{eqnarray} &&
c_{1}=\frac{1}{2(2-2^{1/3})},\,c_{2}=\frac{1-2^{1/3}}{2(2-2^{1/3})},\,c_{3}=c_{2},\,c_{4}=c_{1},
\nonumber \\ &&
d_{1}=\frac{1}{(2-2^{1/3})},\,d_{2}=\frac{-2^{1/3}}{(2-2^{1/3})},\,d_{3}=d_{1},\,d_{4}=0.
\label{App_C9} 
\end{eqnarray}
The split-step expressions in Eqs. (\ref{App_C8})-(\ref{App_C9}) are fourth-order accurate \cite{Yang2010, Yoshida90}.
The linear part $e^{c_{i}hL}$ can be numerically calculated by using the fast FT (FFT) \cite{trefethen2000}.  Note that using definition to compute a discrete FT of $n$ points takes $O(n^2)$ arithmetical operations, while an FFT can compute the same result in only $O(n \ln n)$ operations. This speeds up the computation compared with most finite difference schemes \cite{Agrawal2001, trefethen2000}. The nonlinear part $e^{d_{i}hN}$ can be solved by using Runge-Kutta fourth-order method. A sufficient condition for numerical stability is $\dfrac{\Delta z}{\Delta t^{2}} < \dfrac{1}{\pi}$ \cite{Yang2010, Lakoba2010}.

%\begin{acknowledgements}
\section*{Acknowledgements}

This work is funded by the Vietnam National Foundation for Science and Technology Development (NAFOSTED) under Grant No. 101.99-2015.29.
%\end{acknowledgements}

%\section*{References}

{}


\begin{thebibliography}{}
\bibitem{dimensions} The dimensionless distance $z$ in Eqs. (\ref{single1}) and Eq. (\ref{App_gl1}) is 
$z=X/(2L_{D})$, where $X$ is the dimensional distance, 
$L_{D}=\tau_{0}^{2}/|\tilde\beta_{2}|$ is the dimensional dispersion length,    
$\tau_{0}$ is the soliton width, and $\tilde\beta_{2}$ is the second-order
dispersion coefficient. The dimensionless time is
$t=\tau/\tau_{0}$, where $\tau$ is the time. 
$\psi=E/\sqrt{P_{0}}$, where $E$ is proportional to the 
electric field and $P_{0}$ is the peak power.   
The coefficient $g_{0}$ is related to the linear gain-loss coefficient $\tilde g$ by    
$g_{0}=2\tau_{0}^{2}\tilde g /|\tilde\beta_{2}|$, respectively.
\end{thebibliography}
\end{document}